\documentclass[11pt]{article}
\usepackage[dvipdfmx]{graphicx}   
\usepackage{amssymb}
\usepackage{array} 
\usepackage[authoryear]{natbib}
\usepackage{bibentry}
\usepackage{amsmath}
\usepackage{float}
\usepackage{verbatim}
\usepackage{color}
\usepackage{makeidx}
\usepackage{bm}
\usepackage{appendix}
\usepackage{lscape}

\DeclareMathOperator{\tr}{tr}
\DeclareMathOperator{\vecm}{vec}

\DeclareMathOperator{\Cov}{Cov}
\DeclareMathOperator{\Normal}{\operatorname{N}}
\DeclareMathOperator{\IG}{\operatorname{IG}}
\DeclareMathOperator{\IW}{\operatorname{IW}}

\DeclareMathOperator{\diag}{diag}
\newcommand{\bvec}[1]{\mbox{\boldmath $#1$}}
\DeclareMathOperator{\boldzero}{\bvec{0}}
\DeclareMathOperator{\boldone}{\bvec{1}}
\DeclareMathOperator{\boldy}{\bvec{y}}
\DeclareMathOperator{\boldV}{\mathbf{V}}

\DeclareMathOperator{\boldmu}{\bvec{\mu}}
\DeclareMathOperator{\boldalpha}{\bvec{\alpha}}
\DeclareMathOperator{\boldphi}{\bvec{\phi}}
\DeclareMathOperator{\boldPhi}{\mathbf{\Phi}}
\DeclareMathOperator{\boldeta}{\bvec{\eta}}
\DeclareMathOperator{\boldrho}{\bvec{\rho}}

\DeclareMathOperator{\boldtheta}{\bvec{\theta}}

\DeclareMathOperator{\boldsigma}{\bvec{\sigma}}
\DeclareMathOperator{\boldgamma}{\bvec{\gamma}}
\DeclareMathOperator{\boldSigma}{\mathbf{\Sigma}}
\DeclareMathOperator{\boldPsi}{\mathbf{\Psi}}
\DeclareMathOperator{\boldpsi}{\bvec{\psi}}

\DeclareMathOperator{\boldnu}{\bvec{\nu}}

\DeclareMathOperator{\boldh}{\bvec{h}}
\DeclareMathOperator{\bolde}{\bvec{e}}
\DeclareMathOperator{\boldR}{\mathbf{R}}
\DeclareMathOperator{\boldS}{\mathbf{S}}
\DeclareMathOperator{\boldG}{\mathbf{G}}
\DeclareMathOperator{\boldA}{\mathbf{A}}
\DeclareMathOperator{\boldb}{\mathbf{b}}

\DeclareMathOperator{\boldc}{\mathbf{c}}
\DeclareMathOperator{\boldd}{\mathbf{d}}

\DeclareMathOperator{\boldf}{\bvec{f}}

\DeclareMathOperator{\boldI}{\mathbf{I}}
\DeclareMathOperator{\boldx}{\bvec{x}}
\DeclareMathOperator{\boldX}{\mathbf{X}}
\DeclareMathOperator{\boldw}{\bvec{w}}
\DeclareMathOperator{\boldW}{\mathbf{W}}
\DeclareMathOperator{\boldg}{\bvec{g}}
\DeclareMathOperator{\boldm}{\bvec{m}}

\DeclareMathOperator{\boldB}{\mathbf{B}}

\DeclareMathOperator{\boldK}{\mathbf{K}}

\DeclareMathOperator{\boldbeta}{\bvec{\beta}}
\DeclareMathOperator{\boldepsilon}{\bvec{\epsilon}}

\DeclareMathOperator{\E}{E}
\DeclareMathOperator{\Var}{Var}

\DeclareMathOperator{\calF}{\mathcal{F}}

\DeclareMathOperator{\boldQ}{\mathbf{Q}}

\newcommand{\dtm}[1]{\lvert#1\rvert}

\begin{document}
\title{Dynamic factor, leverage and realized covariances \\ in multivariate stochastic volatility}
\author{Yuta Yamauchi\thanks{Graduate School of Economics, The University of Tokyo, Tokyo, Japan. E-mail:mchyuta@gmail.com.} \ and Yasuhiro Omori\thanks{Faculty of Economics, The University of Tokyo, Tokyo, Japan. E-mail:omori@e.u-tokyo.ac.jp. Phone: +81-3-5841-5516. Fax: +81-3-5841-5521.}}

\maketitle

\begin{abstract}
	In the stochastic volatility models for multivariate daily stock returns, it has been found that the estimates of parameters become unstable as the dimension of returns increases.
	To solve this problem, we focus on the factor structure of multiple returns and consider two additional sources of information: first, the realized stock index associated with the market factor, and second, the realized covariance matrix calculated from high frequency data. 
	The proposed dynamic factor model with the leverage effect and realized measures is applied to ten of the top stocks composing the exchange traded fund linked with the investment return of the S\&P500 index and the model is shown to have a stable advantage in portfolio performance. 

\end{abstract}

\newpage
\section{Introduction}

Portfolio management is one of the major purposes of constructing statistical models that estimate and predict time-varying variances and covariances of asset returns in financial econometrics. The weights on the financial assets are chosen to optimize the objective function for the portfolio based on the estimated models. The univariate stochastic volatility (SV) model is a popular statistical model than can well describe the dynamic stochastic process of the time-varying volatility of an asset return and there are various extensions to the multivariate SV model. 
As is often the case with the multivariate volatility model, the number of model parameters and dynamic latent variables increases as the number of assets increases, which leads to unstable and unreliable estimation results. Since a small number of market factors are often found to exist in empirical studies, this paper focuses on the factor multivariate stochastic volatility with the leverage effect (FMSV) model to overcome this problem. The FMSV model describes the factor structure of the dynamic covariance matrices among asset returns, and  has been investigated in the past literature  (e.g. \cite{PittShephard(99)}, \cite{aguilar2000bayesian}, \cite{chib2006analysis} , \cite{lopes2007factor}), and extended to incorporate the leverage effect (\cite{IshiharaOmori(17)}) which implies a decrease in  the stock return followed by the increase in its volatility (see e.g. \cite{Yu(05)}).

Even in such a parsimonious model, the information from daily returns is often insufficient to obtain accurate and stable estimates of the parameters and dynamic latent variables for multivariate SV models.
A common way to incorporate additional information on the volatilities and covolatilities of asset returns is to use high-frequency data that include information on intraday asset trades.
The realized stochastic volatility (RSV) models, for example, are this type of extension and are known to outperform models without realized measures at estimating model parameters, forecasting volatilities and portfolio performance (\cite{TakahashiOmoriWatanabe(09)}, \cite{HansenHuangShek(12)}, \cite{KoopmanScharth(13)}, \cite{TakahashiWatanabeOmori(16)}, \cite{ShirotaOmoriLopesPiao(17)}, \cite{KuroseOmori(19)}, \cite{YamauchiOmori(19)}).
However, it is not straightforward to extend the FMSV model using the realized covariance matrices in a similar manner, since the realized covariance matrices do not directly correspond to the factor loading matrix and the idiosyncratic volatilities that are not explained by the factors. This paper constructs the appropriate relationship between the realized covariance matrix and the true covariance matrix in the FMSV model, and  proposes a dynamic factor multivariate realized stochastic volatility with the leverage effect (FMRSV) model as an extension of both FMSV and RSV models. 

It should be mentioned that the estimate of the true covariance matrix using the realized volatilities and realized covolatilities is known to be biased due to market microstructure noise, nontrading hours, nonsynchronous trading and so forth. Although the biases can be removed by introducing the corresponding adjustment terms to the measurement equation in some RSV models, it turns out to be difficult in the multivariate SV model because of its factor structure. We instead show how to adjust the biases in the preprocessing step using the information of daily returns, thus extending the method of \cite{hansen2005forecast} for the univariate SV model.

Furthermore, we describe a novel method to estimate the relative weight of additional information from realized measures through the precision parameter.
 As we shall see in our empirical studies, the weight of  the realized measure equation is found to be large; and consequently, the FMRSV model estimates the leverage effect of latent factors and the correlation coefficients between assets with smaller absolute values than the FMSV model. In other words, without the additional information from realized covariances, we tend to overestimate the leverage effect and the strength of the linear relationship between asset returns. 




The rest of this paper is organized as follows.
Section 2 introduces the factor multivariate SV model with daily stock returns, realized factors and realized covariance matrices. Section 3 describes the estimation method using the Markov chain Monte Carlo simulation. In Section 4, we applied the proposed model to ten U.S. stock returns data.

\section{Factor multivariate stochastic volatility with realized measures}
\subsection{Dynamic factor and stochastic volatility}
First, we describe  the factor multivariate stochastic volatility (SV) model. Let $\boldy_t = (y_{1t},\ldots, y_{pt})'$ and $\bm{f}_t=(f_{1t},\ldots,f_{qt})'$ denote a $p \times 1$ stock return vector and a $q\times 1$ latent factor vector. As we shall see in our empirical studies, it is often the case that there is co-movement among stock returns (see Figure \ref{fig:realreturn}).
To model the co-movement, we assume that the return is the sum of the factor and idiosyncratic components as in \cite{ChibNardariShephard(02)} who considered the static factor without the leverage effect:
\begin{eqnarray}
\label{eq:fmsv-1}
	\boldy_t & =  & \boldB \boldf_t+ \boldV_{1t}^{1/2} \boldepsilon_{1t},  \quad t = 1,\ldots, T,
\\
	& & \bm{\epsilon}_{1t} \sim N(\bm{0},\mathbf{I}_p), \\
	& &   \boldV_{1t} = \mbox{diag}\left(\exp(h_{1t}), \ldots, \exp(h_{pt})\right),
\end{eqnarray}
where $\mathbf{B}$ is the $p\times q$ coefficient matrix of the factor, and $\mathbf{I}_p$ denotes the $p\times p$ identity matrix. Let us denote
\begin{align*}
	&\boldB = 
	\begin{bmatrix}
		\boldbeta_1' \\
		\vdots \\
		\boldbeta_p'
	\end{bmatrix}
	,\quad
	\boldbeta \equiv
	\vecm(\boldB')=
	\begin{bmatrix}
		\boldbeta_1 \\
		\vdots \\
		\boldbeta_p
	\end{bmatrix}.
\end{align*}
Further, we consider a dynamic process for the latent factor $\bm{f}_t$ (e.g. \cite{Han2006}).
Assume that it follows the first order stationary autoregressive process: 
\begin{eqnarray}
\label{eq:factor}
\boldf_t &=& \boldgamma + \boldpsi \odot (\boldf_{t-1} - \boldgamma) + \boldV_{2t}^{1/2} \boldepsilon_{2t}, \quad  t = 1,\ldots, T, 
\\
& & \bm{\epsilon}_{2t} \sim N(\bm{0},\mathbf{I}_q), \quad \boldf_0 \equiv \boldgamma,
\\
& &  \boldV_{2t} = \mbox{diag}\left(\exp(h_{p+1,t}), \ldots, \exp(h_{p+q,t})\right),
\end{eqnarray}
where $\odot$ denotes the Hadamard product, $\bm{\gamma}=(\gamma_1,\ldots,\gamma_q)'$ is a mean vector of $\bm{f}_t$ and $\bm{\psi}=(\psi_1,\ldots,\psi_q)'$ $(|\psi_i| < 1, i=1,\ldots,q)$ is an autoregressive coefficient vector. For the initial factor $\bm{f}_1$, we assume $\boldf_0 \equiv \boldgamma$ in Equation (\ref{eq:factor}) for simplicity. 
The log volatility $\bm{h}_t=(h_{1t},\ldots,h_{p+q,t})'$ is assumed to follow the first order stationary autoregressive process 
\begin{eqnarray}
\boldh_{t+1} &=& \boldmu + \boldphi \odot (\boldh_t - \boldmu) + \boldeta_t, \quad \bm{\eta}_{t} \sim N(\bm{0},\mathbf{\Sigma}_{\eta\eta}), \quad t = 1,\ldots, T-1, \\
&&\mathbf{\Sigma}_{\eta\eta}=\mbox{diag}\left(\boldsigma_\eta\right), \quad \boldsigma_\eta=(\sigma_{\eta,1}^2,\ldots,\sigma_{\eta,p+q}^2),\\
\boldh_{1} &=& \boldmu + \boldeta_0, \quad \bm{\eta}_{0} \sim N(\bm{0},\mathbf{\Sigma}_{h,0}), \quad \mathbf{\Sigma}_{h,0}=\mbox{diag}\left(\frac{\sigma_{\eta,1}^2}{1-\phi_1^2},\ldots,\frac{\sigma_{\eta,p+q}^2}{1-\phi_{p+q}^2}\right),
\end{eqnarray}
where $\bm{\mu}=(\mu_1,\ldots,\mu_{p+q})'$ is a mean vector of $\bm{h}_t$ and $\bm{\phi}=(\phi_1,\ldots,\phi_{p+q})'$ $(|\phi_i| < 1, i=1,\ldots,p+q)$ is an autoregressive coefficient vector.
To incorporate the leverage effects, we model the joint distribution of the error terms as follows:
\begin{align}
	&\begin{pmatrix}
		\boldepsilon_{1t} \\ 
		\boldepsilon_{2t} \\ 
		\boldeta_t
	\end{pmatrix} 
	\sim
	\Normal\left(\boldzero,
	\begin{bmatrix}
		\boldI_{p+q} & \boldSigma_{\epsilon\eta} \\
		\boldSigma_{\eta\epsilon}& \boldSigma_{\eta\eta}
	\end{bmatrix}
	\right),\hspace{.3cm}\boldSigma_{\epsilon\eta} =
	\mbox{diag}\left(\rho_1\sigma_{\eta,1},\ldots,\rho_{p+q}\sigma_{\eta,p+q}\right), 
\end{align}
where $\rho_i < 0$ implies that there is a leverage effect between $\epsilon_{it}$ and $h_{i,t+1}$. In empirical studies, the leverage effect is often found to exist only for the factor process, especially the first factor that represents the market factor (e.g. \cite{IshiharaOmori(17)}, \cite{YamauchiOmori(19)}). Thus, we assume that there is no leverage effect for those idiosyncratic components, i.e., $\rho_1=\ldots=\rho_p=0$,
\begin{align}
\label{eq:fmsv-2}
	&\hspace{1cm}\boldSigma_{\epsilon\eta} =
	\mbox{diag}\left(0,\ldots,0, \rho_{p+1}\sigma_{\eta,p+1},\ldots, \rho_{p+q}\sigma_{\eta,p+q}\right),
\end{align}
and denote $\bm{\rho}\equiv (\rho_{p+1},\ldots,\rho_{p+q})'$.
We define the FMSV model by Equations (\ref{eq:fmsv-1})--(\ref{eq:fmsv-2}).

\subsection{Realized factor and realized covariance matrix}
When there are many parameters and latent variables in the multivariate SV model, the parameter estimates of interest often become unstable and inaccurate. We overcome these difficulties by introducing additional measurement equations based on the realized measures. \\

%

\noindent
{\it Realized factor.} First, we introduce the realized measure for factor $\bm{f}_t$. In stock markets, it is usual that some major indices represent the market factor dynamics, such as the S\&P500 index return and sector index returns. We call them realized factor series in this paper and denote them by a $q \times 1$ observed realized factor vector $\bm{x}_t$. In practice, $q$ is expected to be small. Since the realized factors are considered to be correlated, we assume 
\begin{align}
\label{eq:fmrsv-1}
&\boldx_t = \boldA \boldf_t + \boldnu_t 
\quad \boldnu_t \sim \Normal (\boldzero, \boldSigma_\nu), \\
& \qquad
\boldSigma_\nu = \diag\left(\boldsigma_\nu), \quad \boldsigma_\nu=(\sigma^2_{\nu 1},\ldots,\sigma^2_{\nu q}\right),
\quad t = 1,\ldots, T, 
\end{align}
where  a loading matrix $\mathbf{A}$ is a $q\times  q$ lower triangular  matrix such that
\begin{align}
\boldA=
\begin{pmatrix}
		1      &     0  & \cdots & 0 \\ 
		a_{21} &     1  & \ddots & \vdots \\ 
		\vdots &\ddots  & \ddots & 0 \\ 
		a_{q1} &\cdots  & a_{q,q-1} & 1 
\end{pmatrix},
\end{align}
for identification reasons, and we let the lower triangular parameters of $\boldA$ be $\bm{\alpha}\equiv (\boldalpha'_2,\ldots,\boldalpha'_q)'$, where $\boldalpha_j=(\alpha_{j1},\ldots,\alpha_{j,j-1})'$ is a $(j-1) \times 1$ vector.
The first element of the realized factor $\bm{x}_t$ is chosen to represent the overall dynamics of the market such as the S\&P500.\\

\noindent
{\it Realized covariance matrix.}
Next, we consider the realized measures for $\bm{h}_t$. The high-frequency data are used to compute the realized covariance matrix $\boldW_t$ which is assumed to follow
\begin{align}
	\boldW_t \sim \IW (s_0, \{k_0 \Cov (\boldy_t \vert \boldh_t, \boldtheta)\}^{-1}),
	\quad \Cov(\boldy_t \vert \boldh_t, \boldtheta) = \boldB\boldV_{2t}\boldB' + \boldV_{1t},
\end{align}
where $s_0$ and $k_0$ are the constant hyperparameters, and 
the probability density function of $\boldW_t$ given the parameters and latent variables is
\begin{align}
	f(\boldW_t \vert \cdot) \propto \dtm{ \boldB\boldV_{2t}\boldB' + \boldV_{1t}}^{\frac{s_0}{2}}
	\times \dtm{\boldW_t}^{-\frac{s_0+p+1}{2}}
	\times \exp\left[-\frac{1}{2} \tr \left\{k_0  (\boldB\boldV_{2t}\boldB' + \boldV_{1t}\right)\}\boldW_t^{-1} ) \right].
\end{align}
How we set the hyperparameters $s_0$ and $k_0$ affects to what extent we incorporate the information of the realized covariance matrix. Thus, we introduce a new model parameter $\delta$ where $s_0 = \delta + p + 3$ and $k_0 = \delta + 2$. The expected values and covariances of $\mathbf{W}_t$ are given by
\begin{align}
& \E(\boldW_t) =
	\frac{
		k_0 \Cov (\boldy_t \vert \boldh_t \boldtheta)
	}{
		s_0 - (p+1)
	}
	= \Cov (\boldy_t \vert \boldh_t, \boldtheta),
\\
& \Var(w_{ii,t}) = \frac{2}{\delta} \sigma_{ii,t}^2,
\quad \Cov(w_{ij,t},w_{kl,t}) = \frac{2\sigma_{ij,t}\sigma_{kl,t}+(\delta+2)(\sigma_{ik,t}\sigma_{jl,t}+\sigma_{il,t}\sigma_{jk,t})}{\delta(\delta+3)},
\label{eq:fmrsv-2}
\end{align}
where $w_{ij,t}$ and $\sigma_{ij,t}$ denote the $(i,j)$-th element of $\mathbf{W}_t$ and $\Cov (\boldy_t \vert \boldh_t \boldtheta)$, respectively. 
While keeping $\mathbf{W}_t$ the unbiased estimator of the true covariance matrix $\Cov (\boldy_t \vert \boldh_t \boldtheta)$, we control its variance using the precision parameter $\delta$. 
A large $\delta$ implies a small variance for the realized covariance matrix, and consequently we place more weight on the information of the realized covariance matrix. 
We define the FMRSV model by Equations (\ref{eq:fmsv-1})--(\ref{eq:fmrsv-2}).

\subsection{Bias correction of realized volatilities and correlations using daily returns}
The realized covariance matrices are known to have estimation biases due to market microstructure noise, nontrading hours, nonsynchronous trades and so forth. In the preprocessing step, we use the information of daily returns to correct these biases for volatilities and correlation matrices, respectively.
\\

\noindent
{\it Bias correction of realized volatilities.}
We first correct the bias of the variance following \cite{hansen2005forecast}. Let $s_i^2$ and $\tilde{w}_{ii,t}$ respectively denote the sample variance of the daily return and the realized volatility at time $t$ for the $i$-th stock return. Compute a constant $c_i$ such that
	\begin{eqnarray*}
		c_i = \frac{s_i^2}{\frac{1}{T}\sum_{t=1}^T\tilde{w}_{ii,t}}, \quad i=1,\ldots,p, \quad t=1,\ldots,T.
	\end{eqnarray*}
Define the bias-corrected realized volatility $w_{ii,t}$ as
	\begin{eqnarray*}
		w_{ii,t} = c_i \tilde{w}_{ii,t}, \quad i=1,\ldots,p,  \quad t=1,\ldots,T,
	\end{eqnarray*}
so that the average of bias-corrected realized volatilities is equal to the sample variance of daily returns.
For example, if we ignore the overnight returns and compute the open-to-close realized volatilities, we tend to underestimate the true volatilities. By using the close-to-close daily returns, we can correct the biases as above. In Section 4, we illustrate examples where we found most of the values of $c_i$'s are greater than one. \\

\noindent
{\it Bias correction of realized correlation matrices.}
Let $\mathbf{R}$ and $\tilde{\mathbf{R}}_t$ respectively denote the sample correlation matrix using daily returns and the sample correlation matrix using intraday returns at time $t$. We compute the bias-corrected realized correlation matrix $\mathbf{R}_t$ where it is guaranteed to be positive definite:
\begin{enumerate}
	\item Compute the spectral decompositions of $\mathbf{R}$ and $\tilde{\mathbf{R}}_t$ as
	\begin{eqnarray*}
		\mathbf{R} = \mathbf{P}\mathbf{\Lambda}\mathbf{P}', \quad
		\tilde{\mathbf{R}}_t = \mathbf{P}_t\mathbf{\Lambda}_t\mathbf{P}'_t,
	\end{eqnarray*}
	where the $i$-th column of $\mathbf{P}$ ($\mathbf{P}_t$) is the eigenvector of $\mathbf{R}$ ($\tilde{\mathbf{R}}_t$), and $\mathbf{\Lambda}$ ($\mathbf{\Lambda}_t$) is the diagonal matrix whose $i$-th diagonal element is the eigenvalue corresponding to the $i$-th columns of $\mathbf{P}$ ($\mathbf{P}_t$). We define the logarithms of  $\mathbf{R}$ and $\tilde{\mathbf{R}}_t$ as 
	\begin{eqnarray*}
		\mathbf{LR} = \mathbf{P}\left(\log\mathbf{\Lambda}\right)\mathbf{P}', \quad
		\tilde{\mathbf{LR}}_t = \mathbf{P}_t\left(\log\mathbf{\Lambda}_t\right)\mathbf{P}'_t,
	\end{eqnarray*}
	where $\log\mathbf{\Lambda}$ ($\log\mathbf{\Lambda}_t$) denotes a diagonal matrix whose $i$-th diagonal element is a logarithm of the $i$-th diagonal element of $\mathbf{\Lambda}$ ($\mathbf{\Lambda}_t$).
	\item Compute a constant matrix $\mathbf{C}$ such that
	\[
	\mathbf{C} = \mathbf{LR} -\frac{1}{T}\sum_{t=1}^T \tilde{\mathbf{LR}}_t,
	\]
	and define the bias-corrected log correlation matrix $\mathbf{LR}_t$ as
	\[
	\mathbf{LR}_t = \tilde{\mathbf{LR}}_t +\mathbf{C}.
	\]
	\item Compute the bias-corrected correlation matrix $\mathbf{R}_t$ from $\mathbf{LR}_t$ as follows. The convergence of the algorithm is usually very fast as discussed in \cite{archakova2018new}.
	\begin{enumerate}
		\item Set $k=0$.
		\item Let $\bm{x}_k$ denote the vector whose $i$-th element is the $i$-th diagonal element of $\mathbf{LR}_t$. 
		\item Compute the spectral decomposition of $\mathbf{LR}_t$ such that
		\[
		\mathbf{LR}_t =\mathbf{Q}_t\mathbf{D}_t\mathbf{Q}_t',
		\]
		where the $i$-th column of $\mathbf{Q}_t$ is the eigenvector of $\mathbf{LR}_t$, and $\mathbf{D}_t$ is the diagonal matrix whose $i$-th diagonal element is the eigenvalue corresponding to the $i$-th columns of $\mathbf{Q}_t$. Then 
		$\exp(\mathbf{LR}_t)$ is
		\[
		\exp(\mathbf{LR}_t) =\mathbf{Q}_t\exp(\mathbf{D}_t)\mathbf{Q}_t',
		\]
		Let $\bm{\Delta}$ denote the vector whose $i$-th element is the logarithm of the $i$-th diagonal element of $\exp(\mathbf{LR}_t)$.
		\item Update $\bm{x}_{k+1}=\bm{x}_k -\bm{\Delta}$. Replace the diagonal elements ($\bm{x}_k$) of $\mathbf{LR}_t$ with $\bm{x}_{k+1}$.
		\item Set $k \leftarrow k+1$ and return to (b). Repeat until convergence is achieved.
		\item Replace the nondiagonal elements of the identity matrix by those of $\exp(\mathbf{LR}_t)$ and save the resulting matrix as $\mathbf{R}_t$.
	\end{enumerate}
\item 
Compute $\boldW_t$ using the bias-corrected realized volatilities ($w_{ii,t}$'s) and the bias-corrected realized correlation matrices ($\boldR_t$'s).
\end{enumerate}
In empirical studies, it is often pointed out that the realized correlations depend on the data sampling frequency. That  is, the correlations computed from the high frequency data tend to be smaller due to the market microstructure noise than those of the daily returns. This is well known to exist in the stock market and is known as the Epps effect (\cite{Epps(79)}, \cite{YamauchiOmori(19)}).  In fact, in Figure \ref{fig:corr_bias} of Section 4, some of the realized correlations are shown to have such biases.

\section{Markov chain Monte Carlo estimation}
\subsection{Prior distributions for parameters}
Since there are many parameters and latent variables in our FMRSV model, it is difficult to evaluate the likelihood and to implement the maximum likelihood estimation. Thus taking a Bayesian approach, we estimate the parameters and conduct statistical inference using a Markov chain Monte Carlo simulation. 

First, we set the prior distributions of parameters $\boldtheta  \equiv (\boldalpha, \boldbeta, \boldmu, \boldgamma, \boldphi, \boldpsi, \boldrho, \boldsigma_\eta, \boldsigma_\nu, \delta)$.
We assume the prior distributions of $\boldalpha_j$, $\boldbeta_i$, $\boldmu$, and $\boldgamma$ to follow multivariate normal distributions. The prior distributions of $\boldphi$, $\boldpsi$ and $\boldrho$ are assumed to follow beta distributions. 
We assume that $\boldsigma_\eta$ and $\boldsigma_\nu$ follow independent inverse gamma distributions. We assume $\delta$ follows a noninformative improper prior distribution. In summary, the following prior distributions are assumed:
\begin{align*}
	& \boldmu \sim \Normal(\boldm_\mu,\boldS_\mu), \quad \boldgamma \sim \Normal(\boldm_\gamma,\boldS_\gamma), \\
	&\boldbeta_{i} \sim \Normal(\boldm_{\beta_i}, \boldS_{\beta_i}), \quad i=1,\ldots,p, \quad \boldalpha_j \sim \Normal(\boldm_{\alpha_j}, \boldS_{\alpha_j}), \quad j=2,\ldots, q,\\
	&\frac{1+\phi_i}{2} \sim \text{Beta}(a_\phi,b_\phi), \quad \sigma_{\eta,i}^2 \sim \IG\left(\frac{n_\eta}{2}, \frac{d_\eta}{2}\right), \quad i=1,\ldots,p+q,\\
	&\frac{1+\psi_j}{2} \sim \text{Beta}(a_\psi,b_\psi),\quad \frac{1+\rho_{p+j}}{2} \sim \text{Beta}(a_\rho,b_\rho), \quad \sigma_{\nu,j}^2 \sim \IG\left(\frac{n_\nu}{2}, \frac{d_\nu}{2}\right), \quad j=1,\ldots,q,
\\
	&\pi(\delta) \propto I(\delta>0). 
\end{align*}
\subsection{Markov chain Monte Carlo algorithm}
Let $\bm{f}=(\bm{f}_1',\ldots,\bm{f}_T')'$, $\bm{h}=(\bm{h}_1',\ldots,\bm{h}_T')'$, $\bm{x}=(\bm{x}_1',\ldots,\bm{x}_T')'$, $\bm{y}=(\bm{y}_1',\ldots,\bm{y}_T')'$ and $\mathbf{W}=\{\mathbf{W}_t\}_{t=1}^T$. Furthermore, let  $\boldtheta_{\backslash \boldalpha}$ denote $\boldtheta$ excluding $\boldalpha$.
We implement the Markov chain Monte Carlo simulation as follows:
\begin{enumerate}
	\item
	Initialize $\boldh$, $\boldf$ and $\boldtheta$.
	\item 
	Generate $\boldh | \boldtheta, \boldf, \boldx, \boldy, \boldW$.
	\item 
	Generate $\boldf | \boldtheta, \boldh, \boldx, \boldy, \boldW$.
	\item 
	Generate $\boldalpha | \boldtheta_{\backslash \boldalpha}, \boldh, \boldf, \boldx, \boldy, \boldW$.
	\item 
	Generate $\boldbeta | \boldtheta_{\backslash \boldbeta}, \boldh, \boldf, \boldx, \boldy, \boldW$.
	\item 
	Generate $\boldmu | \boldtheta_{\backslash \boldmu}, \boldh, \boldf, \boldx, \boldy, \boldW$.
	\item 
	Generate $\boldgamma | \boldtheta_{\backslash \boldgamma}, \boldh, \boldf, \boldx, \boldy, \boldW$.
	\item 
	Generate $\boldphi | \boldtheta_{\backslash \boldphi}, \boldh, \boldf, \boldx, \boldy, \boldW$.
	\item 
	Generate $\boldpsi | \boldtheta_{\backslash \boldpsi}, \boldh, \boldf, \boldx, \boldy, \boldW$.
	\item 
	Generate $\boldrho | \boldtheta_{\backslash \boldrho}, \boldh, \boldf, \boldx, \boldy, \boldW$.
	\item 
	Generate $\boldsigma_{\eta} | \boldtheta_{\backslash \boldsigma_{\eta}}, \boldh, \boldf, \boldx, \boldy, \boldW$.
	\item 
	Generate $\boldsigma_{\nu} | \boldtheta_{\backslash \boldsigma_{\nu}}, \boldh, \boldf, \boldx, \boldy, \boldW$.
		\item 
	Generate $\delta | \boldtheta_{\backslash \delta}, \boldh, \boldf, \boldx, \boldy, \boldW$.
	\item 
	Return to Step 2.
\end{enumerate}
Let $\bm{h}_t^{(1)}=(h_{1t},\ldots,h_{pt})'$ and $\bm{h}_t^{(2)}=(h_{p+1,t},\ldots,h_{p+q,t})'$, and let $\bm{h}^{(1)}=\{\bm{h}^{(1)}_t\}_{t=1}^T$ and $\bm{h}^{(2)}=\{\bm{h}^{(2)}_t\}_{t=1}^T$.
We describe the generations of $\bm{\beta}$ and $\bm{h}^{(1)}$ below.  See Appendix \ref{sec:generate-h2} for the generation of $\bm{h}^{(2)}$, and the supplementary material for other steps.

\subsubsection{Generation of $\boldbeta$}
The logarithm of the conditional posterior density of $\boldbeta$ given the other parameters and latent variables is
\begin{align}
\lefteqn{\log \pi(\boldbeta \vert \cdot)} & \nonumber\\
&= \mbox{const} +\frac{s_0}{2} \sum_{t=1}^Tg_t(\boldbeta_i)  -\frac{k_0}{2} \tr \left( \sum_{t=1}^{T} \boldB \boldV_{2t}\boldB' \boldW_t^{-1} \right)-\frac{1}{2} \sum_{t=1}^{T} (\boldy_t - \boldB \boldf_t)'\boldV_{1t}^{-1}(\boldy_t - \boldB \boldf_t) .
\end{align}
where $g_t(\boldbeta_i) \equiv \log\dtm{\boldB \boldV_{2t}\boldB' + \boldV_{1t}}$. 
Since the logarithm of the determinant component, $g_t(\boldbeta_i)$, cannot be transformed to some well-known density form, we can construct a proposal distribution for the Metropolis-Hastings (MH) algorithm without this term and adjust it by the acceptance probability in the MH algorithm. However, it results in inefficient sampling and we need to approximate $g_t(\boldbeta_i)$ using some known density to improve the sampling efficiency. 
In order to approximate it using the normal density, we consider Taylor expansion around $\hat{\bm{\beta}}_i$, the mode of the conditional posterior density,
\begin{align*}
	g_t(\boldbeta_i) \approx g_t(\hat{\boldbeta}_i) +\boldg_t'
	(\boldbeta_i - \hat{\boldbeta}_i) - \frac{1}{2} (\boldbeta_i - \hat{\boldbeta}_i)'\boldG_t^{-1}(\boldbeta_i - \hat{\boldbeta}_i).
\end{align*}
where
\begin{align*}
	\boldg_t = \left[\frac{\partial g_t(\boldbeta_i)}{\partial \boldbeta_i} \right]_{\boldbeta_i = \hat{\boldbeta}_i},\quad
	\boldG_t^{-1} = -\left[\frac{\partial^2 g_t(\boldbeta_i)}{\partial \boldbeta_i \partial \boldbeta_i'} \right]_{\boldbeta_i = \hat{\boldbeta}_i}.
\end{align*}
It can be shown that
\begin{align*}
	&\frac{\partial g_t(\boldbeta_i)}{\partial \boldbeta_i}
	= 2 \boldV_{2t} \boldB' (\boldB \boldV_{2t}\boldB' + \boldV_{1t})^{-1} \bolde_i, \\
	&\frac{\partial^2 g_t(\boldbeta_i)}{\partial \boldbeta_i \partial \boldbeta_i'}
	= 2 d_{ii}
	\left\{
	\boldV_{2t} - \boldV_{2t} \boldB' (\boldB\boldV_{2t}\boldB' + \boldV_{1t})^{-1} \boldB \boldV_{2t}
	\right\}
	- \frac{1}{2} \left[\frac{\partial g_t(\boldbeta_i)}{\partial \boldbeta_i}\right]
	\left[\frac{\partial g_t(\boldbeta_i)}{\partial \boldbeta_i}\right]'.
\end{align*}
where $\bolde_i$ denotes a $p \times 1$ vector with the $i$-th element equal to one and zero otherwise, and $d_{ii}$ is the $(i,i)$-th element of $(\boldB \boldV_{2t}\boldB' + \boldV_{1t})^{-1}$ (the proof is given by Proposition 1 of Appendix \ref{sec:prop1}).
%
%
Further, let $w_{t}^{ij}$ denote the $(i,j)$-th element of $\boldW_t^{-1}$. 
Noting that
	\begin{eqnarray}
	\tr(\boldB \boldV_{2t}\boldB' \boldW_t^{-1}) &=& \vecm(\boldB')' (\boldW_t^{-1} \otimes \boldV_{2t}) \vecm(\boldB'), \nonumber\\
	&=& \mbox{const}+\boldbeta_i'(w_{t}^{ii}\boldV_{2t})\boldbeta_i
		+ 2\boldbeta_i'(\boldV_{2t}\sum_{i\neq j}w_{t}^{ij}\boldbeta_j),\\
	\tr (\boldf_t'\boldB' \boldV_{1t}^{-1} \boldB \boldf_t)
		&=& \tr(\boldB \boldf_t \boldf_t'\boldB' \boldV_{1t}^{-1} ) \nonumber \\
	&=& \mbox{const}+\boldbeta_i'\{\exp(-h_{it})\boldf_t \boldf_t'\}\boldbeta_i,\\
	\tr \{\boldy_t'\boldV_{1t}^{-1} \boldB\boldf_t \}
		&=& \tr \{\boldf_t \boldy_t'\boldV_{1t}^{-1} \boldB \}
		= \vecm(\boldB')'\vecm \{\boldf_t \boldy_t'\boldV_{1t}^{-1}\}, \nonumber\\
	&=& \mbox{const}+\boldbeta_i' \{y_{it}\exp(-h_{it}) \boldf_t\},
\end{eqnarray}
we obtain the normal approximation for the conditional posterior density
\begin{align}
\log \pi(\boldbeta \vert \cdot)
\approx \mbox{const}-\frac{1}{2}(\boldbeta - \hat{\boldm}_{\beta_i})'\hat{\boldSigma}_{\beta_i}(\boldbeta - \hat{\boldm}_{\beta_i}) + r(\boldbeta_i),
\end{align}
where
\begin{align*}
	\hat{\boldm}_{\beta_i}
	&= \hat{\boldSigma}_{\beta_i}\left[ \sum_{t=1}^{T}\frac{s_0}{2}\boldG_t^{-1} \left(\hat{\boldbeta}_i + \boldG_t\boldg_t\right) - k_0 \sum_{t=1}^{T}\boldV_{2t}\left(\sum_{i\neq j}w_{t}^{ij}\boldbeta_j\right) + \sum_{t=1}^{T} y_{it}\exp(-h_{it})\boldf_t +\boldS_{\beta_i}^{-1}\boldm_{\beta_i}
	\right] \\
	\hat{\boldSigma}_{\beta_i}
	&= \left[
	\frac{s_0}{2}\sum_{t=1}^{T}\boldG_t^{-1}
	+ k_0\sum_{t=1}^{T}w_t^{ii}\boldV_{2t}
	+ \sum_{t=1}^{T} \exp(-h_{it})\boldf_t\boldf_t'
	+ \boldS_{\beta_i}^{-1}
	\right]^{-1},\\
	&r(\boldbeta_i) = 
	\frac{s_0}{2}
	\sum_{t=1}^{T}
	\left\{
	g_t(\boldbeta_i) - \boldg_t'(\boldbeta_i - \hat{\boldbeta}_i) + \frac{1}{2} (\boldbeta_i -\hat{\boldbeta}_i)'\boldG_t^{-1}(\boldbeta_i - \hat{\boldbeta}_i)
	\right\}.
\end{align*}
When the current value is $\boldbeta^o_i$, we generate $\boldbeta_i^n$ from $\Normal(\hat{\boldm}_{\beta_i}, \hat{\boldSigma}_{\beta_i})$ and accept $\boldbeta_i^{n}$ with probability $\min\{1,\exp(r(\boldbeta^n_i)-r(\boldbeta_i^o))\}$.

\subsubsection{Generation of $\bm{h}^{(1)}$}
The log conditional posterior density of $\boldh^{(1)}$ given other latent variables and parameters is
\begin{eqnarray}
\lefteqn{\log \pi(\boldh^{(1)} \vert \cdot)} & & \nonumber \\
	 &=& \mbox{const}+
	 \frac{s_0}{2}\sum_{t=1}^{T}\log\dtm{\boldB \boldV_{2t}\boldB' + \boldV_{1t}}
	-\frac{k_0}{2}\sum_{t=1}^{T} \mbox{tr}(\boldV_{1t}\boldW_{t}^{-1})\nonumber\\
	&&
	-\frac{1}{2}\sum_{t=1}^{T}\left\{\log\dtm{\boldV_{1t}}+(\boldy_t - \boldB \boldf_t)'\boldV_{1t}^{-1}(\boldy_t - \boldB \boldf_t)\right\}-\frac{1}{2} \{\boldh_{1} - \boldmu)'\boldSigma_{h,0}^{-1}(\boldh_{1} - \boldmu)\nonumber
	 \\
	&&-\frac{1}{2} \sum_{t=1}^{T-1}\left\{\boldh_{t+1} - (\boldI - \boldPhi)\boldmu - \boldPhi\boldh_{t}\right\}'\boldSigma_{\eta\eta}^{-1}\{\boldh_{t+1} - (\boldI - \boldPhi)\boldmu - \boldPhi\boldh_{t}\}
	 \nonumber \\
	& = & 
	\mbox{const} -\sum_{i=1}^p\frac{(1-\phi_i^2)(h_{i1}-\mu_i)^2}{2\sigma_{\eta,i}^2}-\sum_{t=1}^{T-1}\sum_{i=1}^p\frac{\{h_{i,t+1}-h_{it}-(1-\phi_i)\mu_i\}^2}{2\sigma_{\eta,i}^{2}} 
	+\sum_{t=1}^{T}l_t,
\hspace{3mm}\mbox{}
\end{eqnarray}
where
\begin{eqnarray*}
l_t = \frac{s_0}{2}\log\dtm{\boldB \boldV_{2t}\boldB' + \boldV_{1t}}
	-\frac{k_0}{2} w_{t}^{ii}\exp(h_{it}) 
	-\frac{1}{2}\sum_{i=1}^p\left\{h_{it}+(y_{it}- \bm{\beta}_i'\bm{f}_t)^2\exp(-h_{it})\right\},
\end{eqnarray*}
and $w_t^{ii}$ denotes the $(i,i)$-th element of $\mathbf{W}_t^{-1}$.
Although it is simple and easy to implement a single move sampler that generates a single state variable $h_{it}$ ($i=1,\ldots,p$, $t=1,\ldots,T$) at a time, it would result in inefficient sampling. That is, it is well known to generate highly autocorrelated samples when state variables are highly correlated as in stochastic volatility models. A multi-move sampler, which generates a block of state variables using stochastic knots (e.g. \cite{shephard1997likelihood}, \cite{watanabe2004multi}), is one of the most efficient ways to generate latent state variables with high autocorrelations. In the multi-move sampler, first, we divide $(h_{i,1},\ldots,h_{i,T})$ into $K$ blocks 
$(h_{i,s_k},\ldots,h_{i,s_{k+1}-1})$, where $k=1,\ldots,K$ with $1 = s_1 < s_2 < \ldots < s_{K+1} = T+1$.
Then we approximate the nonlinear Gaussian state space model using the linear Gaussian state space model to sample from the conditional posterior distribution of the state variables for each block. To  sample the state variables $(h_{i,s},\ldots,h_{i,s+m})$ from their conditional posterior distribution efficiently, we consider sampling the corresponding disturbances $(\eta_{i,s-1},\ldots,\eta_{i,s+m-1})$. The log posterior density of $(\eta_{i,s-1},\ldots,\eta_{i,s+m-1}),$ ($i = 1,\ldots, p$) is
\begin{align}
\log f(\eta_{i,s-1},\ldots,\eta_{i,s+m-1}\vert \cdot)&= \mbox{const} -\frac{1}{2}\sum_{t=s-1}^{t=s+m-1}\eta_{it}^2+ \sum_{t=s}^{s+m}l_{it}& \nonumber \\
	&
	 -\frac{\left\{h_{i,s+m+1} - \phi_i h_{i,s+m} - (1-\phi_i)\mu_i\right\}^2}{2\sigma_{\eta,i}^{2}}I(s+m<T),
\end{align}
where
\begin{align*}
	& \hspace{5mm} l_{it}=\frac{s_0}{2}\log\dtm{\boldB \boldV_{2t}\boldB' + \boldV_{1t}} 
	-\frac{k_0}{2} w_{t}^{ii}\exp(h_{it}) 
	-\frac{1}{2}\left\{h_{it}+(y_{it}- \bm{\beta}_i'\bm{f}_t)^2\exp(-h_{it})\right\},
\end{align*}
and $I(A)$ is an indicator function such that $I(A)=1$ if $A$ is true and 0 otherwise.
By Taylor expansion of $l_{it}$ around the conditional mode, $\hat{h}_{it}$, the approximated conditional posterior density $f^*(\eta_{i,s-1},\ldots,\eta_{i,s+m-1})$ of the block is given by
\begin{align}
\log f^*(\eta_{i,s-1},\ldots,\eta_{i,s+m-1}) 
	&= \mbox{const}-\frac{1}{2}\sum_{t=s-1}^{t=s+m-1}\eta_{it}^2
	+\sum_{t=s}^{t=s+m}
	\left[
	\hat{l}_{it} + (h_{it} - \hat{h}_{it})\hat{l}_{it}'
	+\frac{1}{2} (h_{it} - \hat{h}_{it})^2 \hat{l}_{it}''
	\right], \nonumber \\
	&\hspace{3mm}-\frac{\left\{h_{i,s+m+1} - \phi_i h_{i,s+m} - (1-\phi_i)\mu_i\right\}^2}{2\sigma_{\eta,i}^{2}}I(s+m<T).
\end{align}
where $\hat{l}_{it}'$ and $\hat{l}_{it}''$ are
\begin{align*}
	&l_{it}' = \frac{s_0}{2} d_{ii}\exp(h_{it})
	-\frac{k_0}{2} w_{t}^{ii}\exp(h_{it}) 
	+\frac{1}{2} \left\{-1+(y_{it} - \bm{\beta}_i'\boldf_t)^2\exp(-h_{it})\right\},
	\\
	&l_{it}'' = \frac{s_0}{2} d_{ii}\exp(h_{it})-\frac{s_0}{2} d_{ii}^2\exp(2h_{it})
	-\frac{k_0}{2} w_{t}^{ii}\exp(h_{it}) 
	- \frac{1}{2} (y_{it} - \bm{\beta}_i'\boldf_t)^2\exp(-h_{it}),
\end{align*}
respectively, evaluated at $h_{it}=\hat{h}_{it}$ using 
\begin{align*}
\frac{\partial \log\dtm{\boldB \boldV_{2t}\boldB' + \boldV_{1t}}}{\partial h_{it}}
&= d_{ii} \exp(h_{it}), \quad
\frac{\partial^2 \log\dtm{\boldB \boldV_{2t}\boldB' + \boldV_{1t}}}{\partial h_{it}^2}
= d_{ii} \exp(h_{it}) - d_{ii}^2\exp(2h_{it}),
\end{align*}
for $i=1,\ldots,p$ and $t=1,\ldots,T$ (the proof is given by Proposition 2 of Appendix \ref{sec:prop2}
). 
To construct the approximated linear Gaussian state space model from which we sample a proposal of $(\eta_{i,s-1},\ldots,\eta_{i,s+m-1})$, we define the auxiliary variables $\hat{y}_{it}$ and $v_{it}$ as follows.
For $t = s,\ldots,s+m-1$ or $t = s + m = T$,
\begin{align}
	&\hat{y}_{it} = \hat{h}_{it} + v_{it} \hat{l}_{it}', \quad
	v_{it} = -\hat{l}_{it}^{\prime\prime -1},
\end{align}
	and, for $t= s + m < T$,
\begin{align}
	&\hat{y}_{it} = v_{it} \left[
	\left\{
	\hat{l}_{it}' - \hat{l}_{it}''\hat{h}_{it}
	\right\}
	+ \phi_i \sigma_{\eta,i}^{-2}
	\left\{
	h_{t+1,i} - (1 - \phi_i)\mu_i
	\right\}
	\right], \quad
	v_{it} = \left(
	\phi_i^2\sigma_{\eta,i}^{-2}
	- \hat{l}_{it}''
	\right)^{-1}.
\end{align}
Then, consider the following linear Gaussian state space model,
\begin{eqnarray}
	\hat{y}_{it} &=& h_{it} + \epsilon_t, \quad \epsilon_t \sim \Normal(0,v_{it}), \\
	h_{i,t+1} &=& (1-\phi_i) \mu_i + \phi_i h_{it} +  \eta_{it}, \quad \eta_{it} \sim \Normal(0,\sigma_{\eta,i}).
\end{eqnarray}
Given $h_{i,s-1},(\hat{y}_{is},\ldots,\hat{y}_{i,s+m})$ and other parameters, we can generate the candidate of \\ $(\eta_{i,s-1},\ldots,\eta_{i,s+m-1})$ from the approximated density $f^*$ using a simulation smoother (e.g. \cite{deJongShephard(95)}, \cite{DurbinKoopman(02)}), and conduct the MH algorithm. That is, when we have the current samples $(\eta_{i,s-1}^{o},\ldots,\eta_{i,s+m-1}^{o})$, we accept the new samples $(\eta_{i,s-1}^{n},\ldots,\eta_{i,s+m-1}^{n})$ generated from $f^*(\eta_{i,s-1},\ldots,\eta_{i,s+m-1})$ with probability
\begin{align*}
	\min\left\{1,\frac{f(\eta_{i,s-1}^{n},\ldots,\eta_{i,s+m-1}^{n})f^*(\eta_{i,s-1}^{o},\ldots,\eta_{i,s+m-1}^{o})}{f(\eta_{i,s-1}^{o},\ldots,\eta_{i,s+m-1}^{o})f^*(\eta_{i,s-1}^{n},\ldots,\eta_{i,s+m-1}^{n})}\right\}.
\end{align*}
To obtain the conditional mode $(\hat{h}_{is},\ldots,\hat{h}_{i,s+m})$, we select some initial mode values such as the current state vector of  $(h_{is},\ldots,h_{i,s+m})$ and repeat the disturbance smoother several times (see e.g. \cite{shephard1997likelihood}, \cite{watanabe2004multi}).\\

\noindent
Remark. Parameter $K$ is chosen to obtain stable and efficient estimation results. In our empirical study, we used $K=470$, but, in general, a smaller $K$ could be used.
\section{Empirical studies}
\subsection{Data}
{\it Data}. In this section, we apply our proposed model to the daily returns of ten U.S. stocks with the bias-corrected realized covariance matrices. The ten series of stock returns are chosen from top stocks composing the exchange traded fund (ETF) that seeks to track the performance of a benchmark index that measures the investment return of the S\&P500 index\footnote{The ETF is Vanguard S\&P 500 ETF (VOO).}. 
%
They are 1: Apple Inc. (AAPL), 2: Microsoft Corp. (MSFT), 3: Amazon.com Inc. (AMZN), 4: JPMorgan Chase \& Co. (JPM), 5: Berkshire Hathaway Inc. Class B (BRKB), 6: Alphabet Inc. Class A (GOOGL), 7: Johnson \& Johnson (JNJ), 8: Proctor \& Gamble Co. (PG), 9: Exxon Mobil Corp (XOM), and 10: AT\&T (T). 
The federal funds rate is used as a risk-free asset.
The daily returns for the $i$-th stocks are defined as $y_{it} = 100 \times (\log p_{it} - \log p_{i, t-1})$, where $p_{it}$ is the closing price of the $i$-th asset at time $t$.
The (open-to-close) realized covariance matrices are computed as $\sum_{s=1}^{78}\bm{r}_{st}\bm{r}_{st}'$, where $\bm{r}_{st}$ is the $s$-th return vector during day $t$ at intervals of 5 minutes from 9:35 to 16:00\footnote{The intraday price data was obtained from Tick Data (\tt{http://www.tickdata.com}).}.
The sample period is from September 1, 2004 to December 31, 2013, and the number of observations is $T=2350$. \\

\begin{figure}[H]
	\centering
	\includegraphics[width=15.5cm]{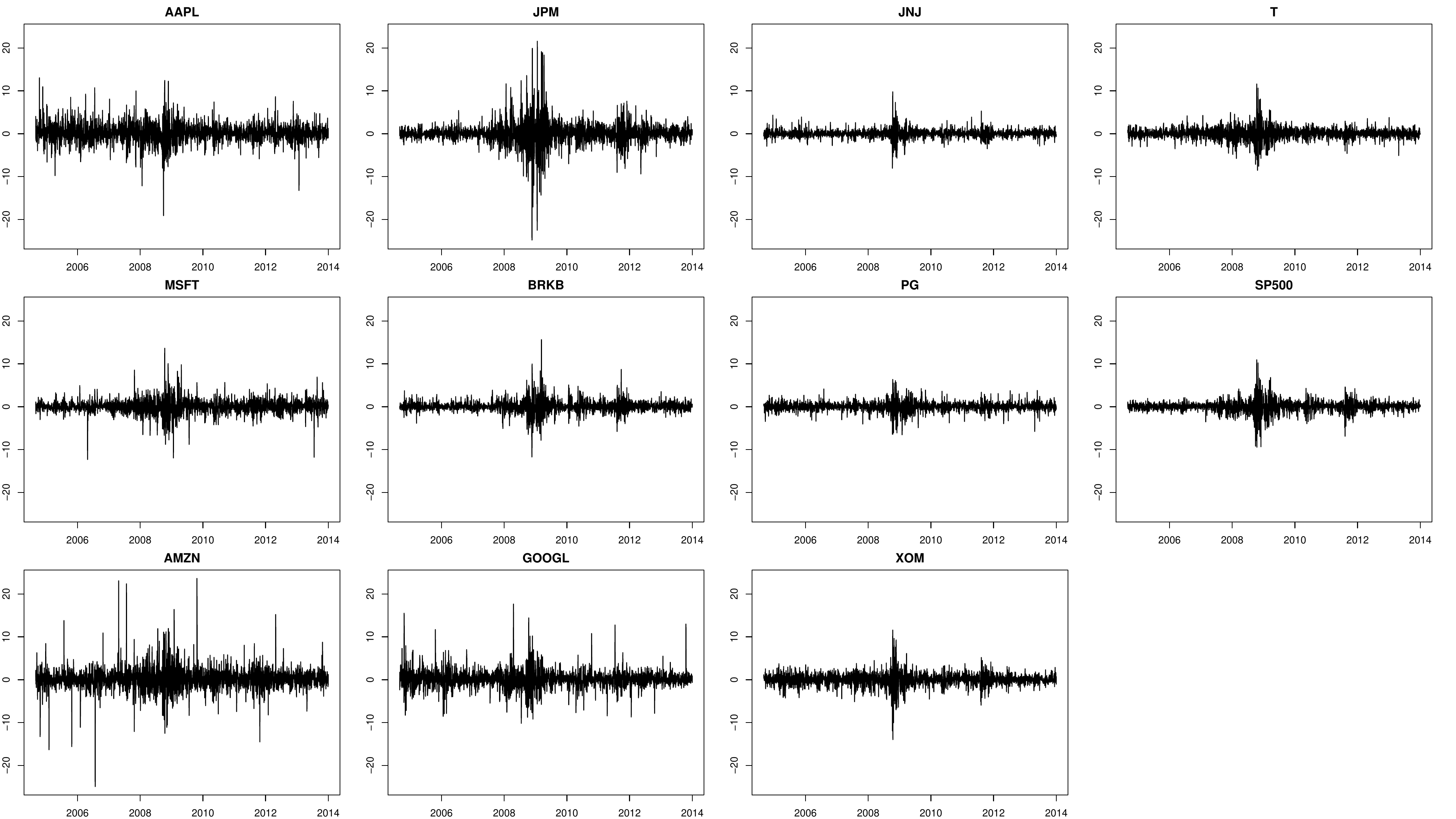}
	\caption{Time series plots of ten U.S. stock (close-to-close) returns and S\&P 500 return.}
	\label{fig:realreturn}
\end{figure}

\noindent
{\it Realized factor.} The time series plots of $y_{it}$'s are shown in Figure \ref{fig:realreturn}, which shows that there is a very high volatility period in 2008 ( the time of the global financial crisis when Lehman Brothers filed for Chapter 11 bankruptcy protection). We can see the co-movement among the ten stock returns and the S\&P500 index. Since the stock index is considered to represent the overall movement in the stock market, we use the S\&P500 index as the realized factor $x_t$ which corresponds to the U.S. stock market factor; and we set $q = 1$ and $\mathbf{A} = 1$ in this analysis.
We also considered the case  $q = 2$, but the second factor does not seem to exist, resulting in weakly identified parameter estimates. Instead, to illustrate the case $q=2$, we conducted the simulation study in Supplementary Material.

\begin{figure}[H]
	\centering
	\includegraphics[width=12cm]{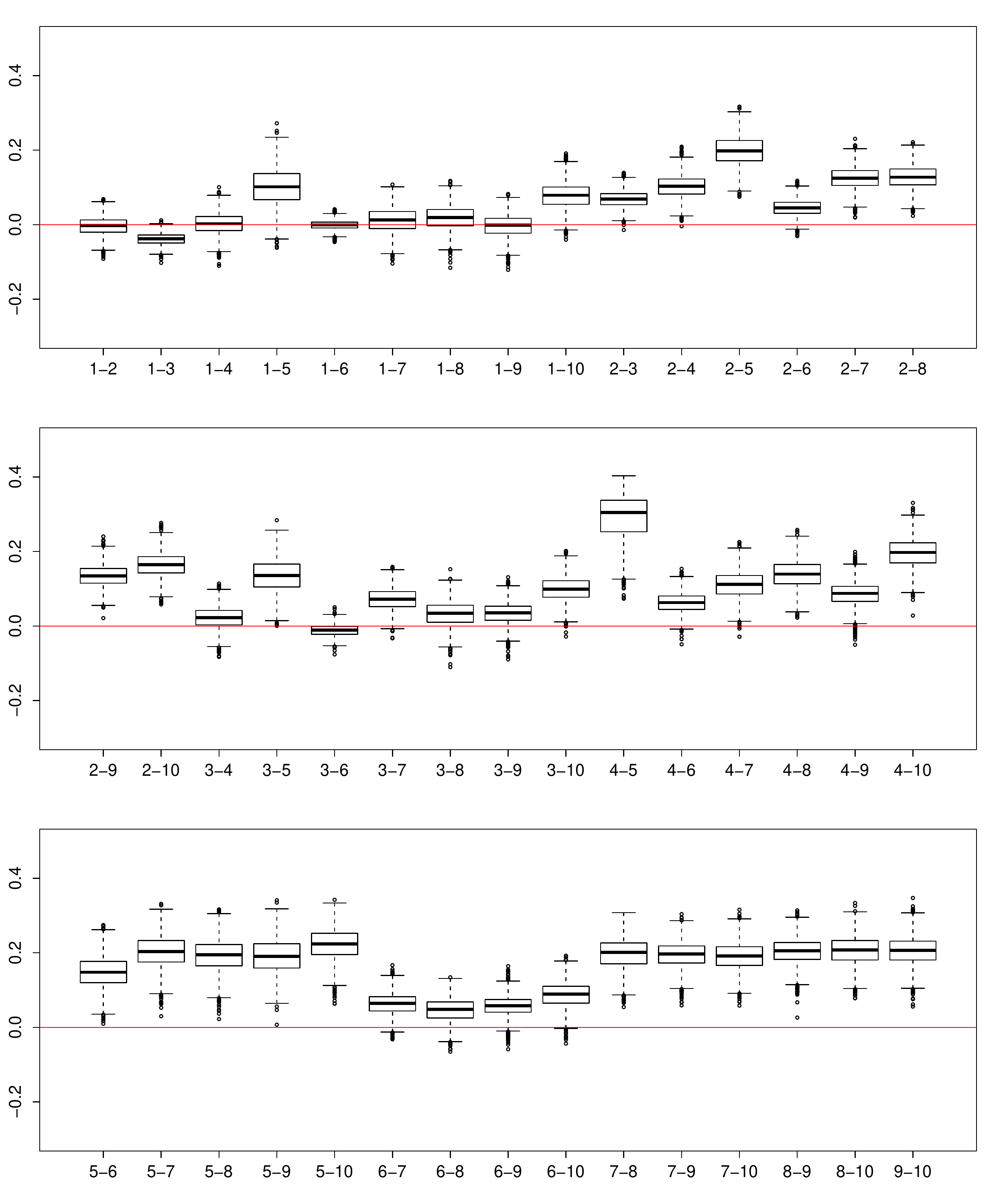}
	\caption{Boxplot of $\mathbf{R}_t-\tilde{\mathbf{R}}_t$. Differences between the bias-corrected and raw realized correlations.}
	\label{fig:corr_bias}
\end{figure}

\noindent
{\it Bias correction of the realized volatilities and correlations}. 
The bias correction vector $\bm{c}$ for the realized volatilities is obtained as
\begin{eqnarray*}
\bm{c} & = &  
(1.53,	1.39, 1.55, 1.37, 1.02, 1.50, 1.04, 1.06, 1.21, 0.99)',
\end{eqnarray*}
and the realized volatilities (except for T) appear to underestimate the true volatility due to ignoring the overnight returns. 
The bias correction matrix $\mathbf{C}$ for the log correlation matrices is omitted since it is difficult to interpret intuitively; instead, we show the boxplots of the differences between the bias-corrected realized correlation and the raw realized correlation in Figure \ref{fig:corr_bias}. Most of the bias-corrected realized correlations are found to be larger (except those for AAPL-AMZN (1-3) and AMZN-GOOGL (3-6)) than the realized correlations, indicating the existence of the Epps effect. 

\noindent
The prior distributions are assumed as follows:
\begin{align*}
& \mu_i \sim \Normal(0, 4), \quad 
\frac{1+\phi_i}{2} \sim \text{Beta}(20, 1.5), \quad \sigma_{\eta,i}^2 \sim \IG\left(\frac{0.1}{2}, \frac{0.1}{2}\right),\quad i=1,\ldots,11, \\
&\gamma_1 \sim \Normal(0, 1),\quad \frac{1+\psi_1}{2} \sim \text{Beta}(1,1),\quad \frac{1+\rho_{11}}{2} \sim \text{Beta}(1,1), \quad \sigma_{\nu,1}^2 \sim \IG\left(\frac{0.1}{2}, \frac{0.1}{2}\right),\\
& \beta_{i} \sim \Normal(0, 1),\quad i=1,\ldots, 10, \quad \pi(\delta) \propto I(\delta>0).
\end{align*}

%
\subsection{Estimation results}
The MCMC simulation is iterated to obtain 20,000 posterior samples after discarding 10,000 samples as the burn-in period for the FMSV model, and 10,000 posterior samples after discarding 2,000 samples as the burn-in period for the FMRSV model. Tables \ref{table:emp_param_norcov_2004} and \ref{table:emp_param_rcov_2004} show the posterior estimation results of the parameters for the FMSV model (without realized covariances) and the FMRSV model (with realized covariances), respectively. The inefficiency factors\footnote{The inefficiency factor is defined as $1+2\sum_{g=1}^\infty \rho(g)$, where $\rho(g)$ is the sample autocorrelation at lag $g$.  This is interpreted as the ratio of the numerical variance of the posterior mean from the chain to the variance of the posterior mean from hypothetical uncorrelated draws. The smaller the inefficiency factor is, the closer the MCMC sampling is to the uncorrelated sampling. The effective sample size is obtained as the posterior sample size divided by the inefficiency factor.} range from 1 to 114 (the effective sample sizes range from 175 to 20,000) for the FMSV model, and from 1 to 94  (the effective sample sizes are from 106 to 10,000) for the FMRSV model, which implies that our sampling algorithm works quite efficiently.

The conditional means  of the log volatilities ($\mu_j$) vary from $-1.105$ (JNJ) to 0.804 (AAPL) in the FMSV model and from $-1.047$ (JNJ) to 1.240 (AMZN) in the FMRSV model. Overall, these posterior means of the FMRSV model are larger than those of the FMSV model. In both models, $\mu_1$ (AAPL) and $\mu_3$ (AMZN) are the largest two, while $\mu_7$ (JNJ) is the smallest, as expected from Figure \ref{fig:realreturn}. 

The persistences of the log volatilities ($\phi_j$) are high from 0.619 (AMZN) to 0.982 (factor) in the FMSV model and from 0.640 (AMZN) to 0.927 (XOM) in the FMRSV model, where the estimates of the FMRSV model are relatively lower than those of the FMSV model. The persistences ($\phi_{11}$) of the log volatilities of the dynamic factor are higher than those for most of the stock returns. 
The factor loadings ($\beta_j$) are all positive, ranging from 0.470 (JNJ) to 1.260 (JPM) in the FMSV model and from 0.521 (JNJ) to 1.126 (JPM) in the FMRSV model, suggesting co-movement between stock returns. Among the factor loadings,  the estimates of AMZN and JPM are found to be the largest. 
The posterior probability that the factor mean ($\gamma_1$) is positive is greater than 0.975 since the 95\% credible interval is above 0. This implies that the expected market return is positive during this sample period. The autoregressive coefficient ($\psi_1$) is estimated to be negative but close to zero.

The leverage effect of the factor, denoted by $\rho_{11}$, is estimated to be $-0.609$ in the FMSV model, and $-0.234$ in the FMRSV model. It is expected to be negative with the posterior probability greater than 0.975, suggesting the existence of the leverage effect. The absolute value of the posterior estimate is found to be much smaller in the FMRSV model. 
%

%
%
%
%
%
\normalsize

\scriptsize
\begin{table}[H]
	\begin{center}
		\begin{tabular}{lrcrlrcr}
			\hline\hline
			\multicolumn{1}{l}{Par.}&\multicolumn{1}{c}{Mean}&\multicolumn{1}{c}{95\% interval}&\multicolumn{1}{c}{IF}&\multicolumn{1}{c}{Par.}&\multicolumn{1}{c}{Mean}&\multicolumn{1}{c}{95\% interval}&\multicolumn{1}{c}{IF}\tabularnewline
			\hline
			$\mu_{1}$&$0.804$&$[0.653,0.960]$&$6$&$\phi_{1}$&$0.823$&$[0.759,0.879]$&$68$\tabularnewline
			$\mu_{2}$&$-0.281$&$[-0.433,-0.124]$&$5$&$\phi_{2}$&$0.799$&$[0.729,0.858]$&$71$\tabularnewline
			$\mu_{3}$&$0.767$&$[0.645,0.893]$&$8$&$\phi_{3}$&$0.619$&$[0.503,0.708]$&$70$\tabularnewline
			$\mu_{4}$&$0.067$&$[-0.453,0.575]$&$1$&$\phi_{4}$&$0.976$&$[0.960,0.989]$&$64$\tabularnewline
			$\mu_{5}$&$-0.478$&$[-0.856,-0.103]$&$1$&$\phi_{5}$&$0.971$&$[0.954,0.985]$&$59$\tabularnewline
			$\mu_{6}$&$0.292$&$[0.117,0.467]$&$4$&$\phi_{6}$&$0.832$&$[0.775,0.880]$&$67$\tabularnewline
			$\mu_{7}$&$-1.105$&$[-1.272,-0.936]$&$5$&$\phi_{7}$&$0.864$&$[0.811,0.908]$&$65$\tabularnewline
			$\mu_{8}$&$-0.777$&$[-0.927,-0.622]$&$5$&$\phi_{8}$&$0.858$&$[0.801,0.904]$&$79$\tabularnewline
			$\mu_{9}$&$-0.398$&$[-0.761,-0.028]$&$2$&$\phi_{9}$&$0.977$&$[0.961,0.989]$&$62$\tabularnewline
			$\mu_{10}$&$-0.423$&$[-0.623,-0.220]$&$3$&$\phi_{10}$&$0.932$&$[0.893,0.961]$&$87$\tabularnewline
			$\mu_{11}^{\dagger}$&$-0.382$&$[-0.867,0.042]$&$2$&$\phi_{11}^{\dagger}$&$0.982$&$[0.972,0.990]$&$41$\tabularnewline
			$\beta_{1}$&$0.934$&$[0.880,0.990]$&$3$&$\sigma_{\eta,1}$&$0.572$&$[0.470,0.684]$&$87$\tabularnewline
			$\beta_{2}$&$0.835$&$[0.798,0.873]$&$6$&$\sigma_{\eta,2}$&$0.656$&$[0.546,0.778]$&$90$\tabularnewline
			$\beta_{3}$&$1.146$&$[1.088,1.205]$&$5$&$\sigma_{\eta,3}$&$0.896$&$[0.791,1.024]$&$81$\tabularnewline
			$\beta_{4}$&$1.260$&$[1.209,1.311]$&$5$&$\sigma_{\eta,4}$&$0.277$&$[0.216,0.350]$&$110$\tabularnewline
			$\beta_{5}$&$0.687$&$[0.646,0.727]$&$6$&$\sigma_{\eta,5}$&$0.243$&$[0.189,0.305]$&$101$\tabularnewline
			$\beta_{6}$&$0.865$&$[0.823,0.907]$&$5$&$\sigma_{\eta,6}$&$0.649$&$[0.548,0.756]$&$88$\tabularnewline
			$\beta_{7}$&$0.470$&$[0.445,0.495]$&$4$&$\sigma_{\eta,7}$&$0.497$&$[0.407,0.594]$&$89$\tabularnewline
			$\beta_{8}$&$0.516$&$[0.486,0.546]$&$4$&$\sigma_{\eta,8}$&$0.472$&$[0.386,0.569]$&$102$\tabularnewline
			$\beta_{9}$&$0.860$&$[0.828,0.891]$&$3$&$\sigma_{\eta,9}$&$0.186$&$[0.142,0.241]$&$109$\tabularnewline
			$\beta_{10}$&$0.678$&$[0.645,0.711]$&$4$&$\sigma_{\eta,10}$&$0.301$&$[0.234,0.389]$&$114$\tabularnewline
			$\psi_{1}^{\dagger}$&$-0.059$&$[-0.102,-0.016]$&$4$&$\sigma_{\eta,11}^{\dagger}$&$0.194$&$[0.159,0.234]$&$88$\tabularnewline
			$\gamma_{1}^{\dagger}$&$0.051$&$[0.022,0.081]$&$4$&$\sigma_{\nu,1}^{\dagger}$&$0.136$&$[0.110,0.161]$&$78$\tabularnewline
			&&&&$\rho_{11}^{\dagger}$&$-0.609$&$[-0.705,-0.489]$&$44$\tabularnewline
			\hline
	\end{tabular}\end{center}
	\caption{FMSV model. The posterior means, 95 \% credible intervals and inefficiency factors (IFs) of the parameters for ten U.S. stocks returns.}
	\label{table:emp_param_norcov_2004}\newcolumntype{.}{D{.}{.}{-1}}\newcolumntype{.}{D{.}{.}{-1}}
$\dagger$: parameters related to the market factor.
\end{table}

\begin{table}[H]
\newcolumntype{.}{D{.}{.}{-1}}
\begin{center}
	\begin{tabular}{lrcrlrcr}
		\hline\hline
		\multicolumn{1}{l}{Par.}&\multicolumn{1}{c}{Mean}&\multicolumn{1}{c}{95\% interval}&\multicolumn{1}{c}{IF}&\multicolumn{1}{c}{Par.}&\multicolumn{1}{c}{Mean}&\multicolumn{1}{c}{95\% interval}&\multicolumn{1}{c}{IF}\tabularnewline
		\hline
		$\mu_{1}$&$0.898$&$[0.794,1.002]$&$1$&$\phi_{1}$&$0.781$&$[0.752,0.808]$&$3$\tabularnewline
		$\mu_{2}$&$0.113$&$[0.0368,0.190]$&$1$&$\phi_{2}$&$0.769$&$[0.736,0.801]$&$6$\tabularnewline
		$\mu_{3}$&$1.240$&$[1.178,1.303]$&$2$&$\phi_{3}$&$0.640$&$[0.603,0.676]$&$6$\tabularnewline
		$\mu_{4}$&$0.355$&$[0.219,0.489]$&$1$&$\phi_{4}$&$0.815$&$[0.791,0.840]$&$3$\tabularnewline
		$\mu_{5}$&$-0.643$&$[-0.794,-0.489]$&$1$&$\phi_{5}$&$0.849$&$[0.824,0.872]$&$4$\tabularnewline
		$\mu_{6}$&$0.615$&$[0.503,0.729]$&$1$&$\phi_{6}$&$0.820$&$[0.794,0.846]$&$4$\tabularnewline
		$\mu_{7}$&$-1.047$&$[-1.134,-0.960]$&$1$&$\phi_{7}$&$0.803$&$[0.773,0.834]$&$6$\tabularnewline
		$\mu_{8}$&$-0.804$&$[-0.890,-0.718]$&$1$&$\phi_{8}$&$0.799$&$[0.768,0.829]$&$5$\tabularnewline
		$\mu_{9}$&$-0.280$&$[-0.451,-0.106]$&$1$&$\phi_{9}$&$0.927$&$[0.909,0.945]$&$6$\tabularnewline
		$\mu_{10}$&$-0.551$&$[-0.661,-0.443]$&$1$&$\phi_{10}$&$0.845$&$[0.819,0.870]$&$5$\tabularnewline
		$\mu_{11}^{\dagger}$&$-0.278$&$[-0.433,-0.127]$&$4$&$\phi_{11}^{\dagger}$&$0.900$&$[0.881,0.919]$&$6$\tabularnewline
		$\beta_{1}$&$0.787$&$[0.770,0.804]$&$20$&$\sigma_{\eta,1}$&$0.557$&$[0.535,0.579]$&$6$\tabularnewline
		$\beta_{2}$&$0.858$&$[0.843,0.875]$&$34$&$\sigma_{\eta,2}$&$0.420$&$[0.399,0.443]$&$11$\tabularnewline
		$\beta_{3}$&$1.100$&$[1.078,1.123]$&$24$&$\sigma_{\eta,3}$&$0.533$&$[0.513,0.553]$&$12$\tabularnewline
		$\beta_{4}$&$1.126$&$[1.105,1.147]$&$33$&$\sigma_{\eta,4}$&$0.613$&$[0.592,0.636]$&$8$\tabularnewline
		$\beta_{5}$&$0.606$&$[0.595,0.618]$&$36$&$\sigma_{\eta,5}$&$0.566$&$[0.542,0.592]$&$12$\tabularnewline
		$\beta_{6}$&$0.808$&$[0.792,0.825]$&$26$&$\sigma_{\eta,6}$&$0.499$&$[0.478,0.521]$&$10$\tabularnewline
		$\beta_{7}$&$0.521$&$[0.512,0.532]$&$36$&$\sigma_{\eta,7}$&$0.417$&$[0.394,0.441]$&$13$\tabularnewline
		$\beta_{8}$&$0.557$&$[0.546,0.568]$&$35$&$\sigma_{\eta,8}$&$0.416$&$[0.395,0.438]$&$11$\tabularnewline
		$\beta_{9}$&$0.796$&$[0.783,0.810]$&$37$&$\sigma_{\eta,9}$&$0.310$&$[0.290,0.331]$&$16$\tabularnewline
		$\beta_{10}$&$0.694$&$[0.682,0.706]$&$35$&$\sigma_{\eta,10}$&$0.415$&$[0.393,0.437]$&$10$\tabularnewline
		$\psi_{1}^{\dagger}$&$-0.0677$&$[-0.107,-0.0281]$&$4$&$\sigma_{\eta,11}^{\dagger}$&$0.362$&$[0.343,0.382]$&$16$\tabularnewline
		$\gamma_{1}^{\dagger}$&$0.0902$&$[0.0604,0.121]$&$2$&$\sigma_{\nu,1}^{\dagger}$&$0.132$&$[0.108,0.159]$&$58$\tabularnewline
		$\delta$&$16.478$&$[16.243,16.692]$&$94$&$\rho_{11}^{\dagger}$&$-0.234$&$[-0.288,-0.181]$&$8$\tabularnewline
		\hline
\end{tabular}\end{center}
\caption{FMRSV model. The posterior means, 95 \% credible intervals and inefficiency factors (IFs) of the parameters for ten U.S. stocks returns.}
\label{table:emp_param_rcov_2004}\newcolumntype{.}{D{.}{.}{-1}}
$\dagger$: parameters related to the market factor.
\end{table}

\normalsize
The posterior mean of the precision parameter $\delta$ is estimated to be large at approximately 16.5 in the FMRSV model. This result suggests that the distribution of the realized covariances is concentrated around the expected value of the true covariance matrix (under our factor model), and that the model fit is good for the measurement equation of the realized covariance matrix. 

\begin{figure}[H]
	\centering
	\includegraphics[width=15cm]{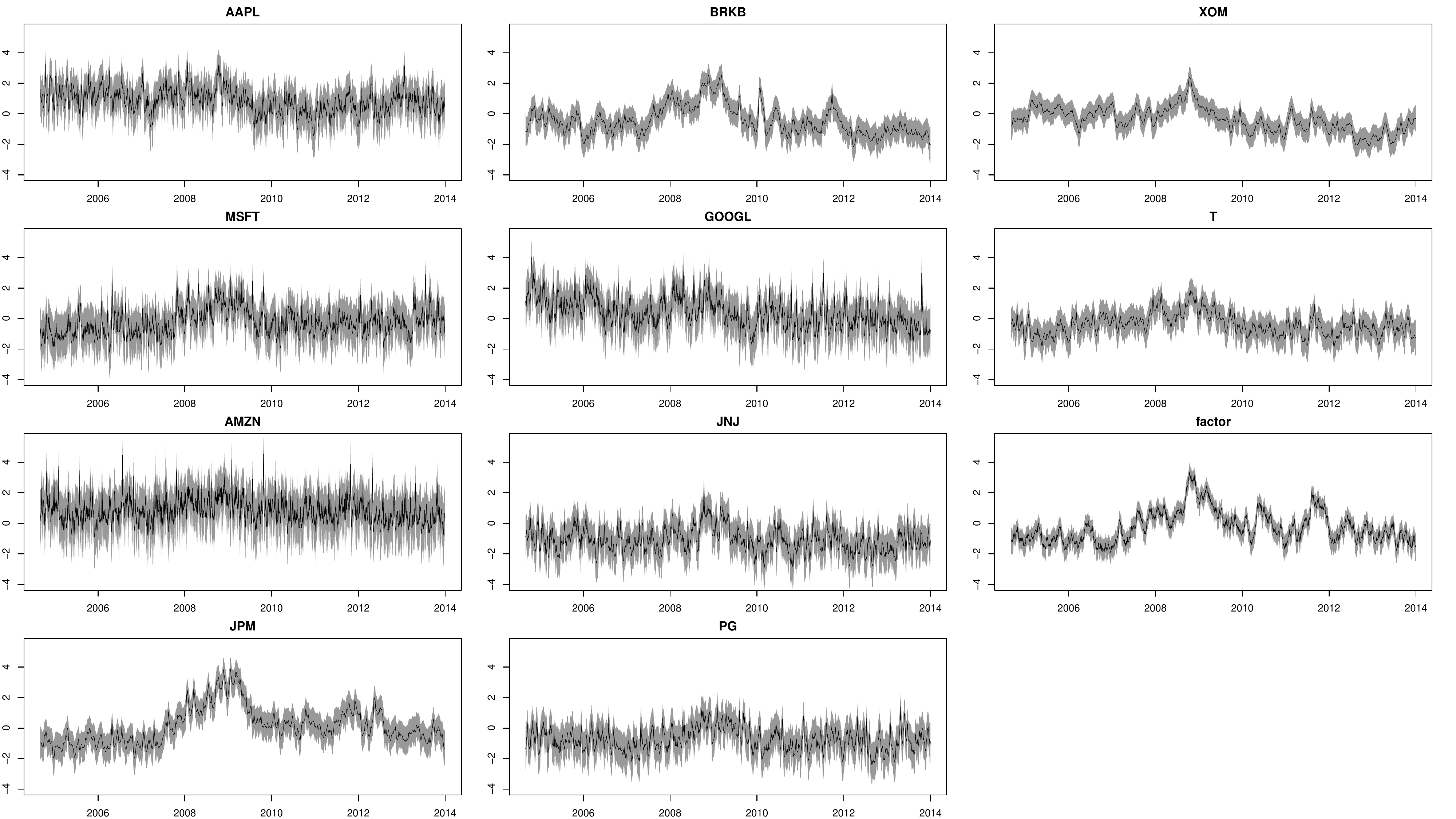}
	\caption{The posterior means (solid lines) and 95\% credible intervals (shaded areas) of each estimated log volatility of the stocks and the latent factor in the FMSV model.}
	\label{fig:vol1}
\end{figure}

\begin{figure}[H]
	\centering
	\includegraphics[width=15cm]{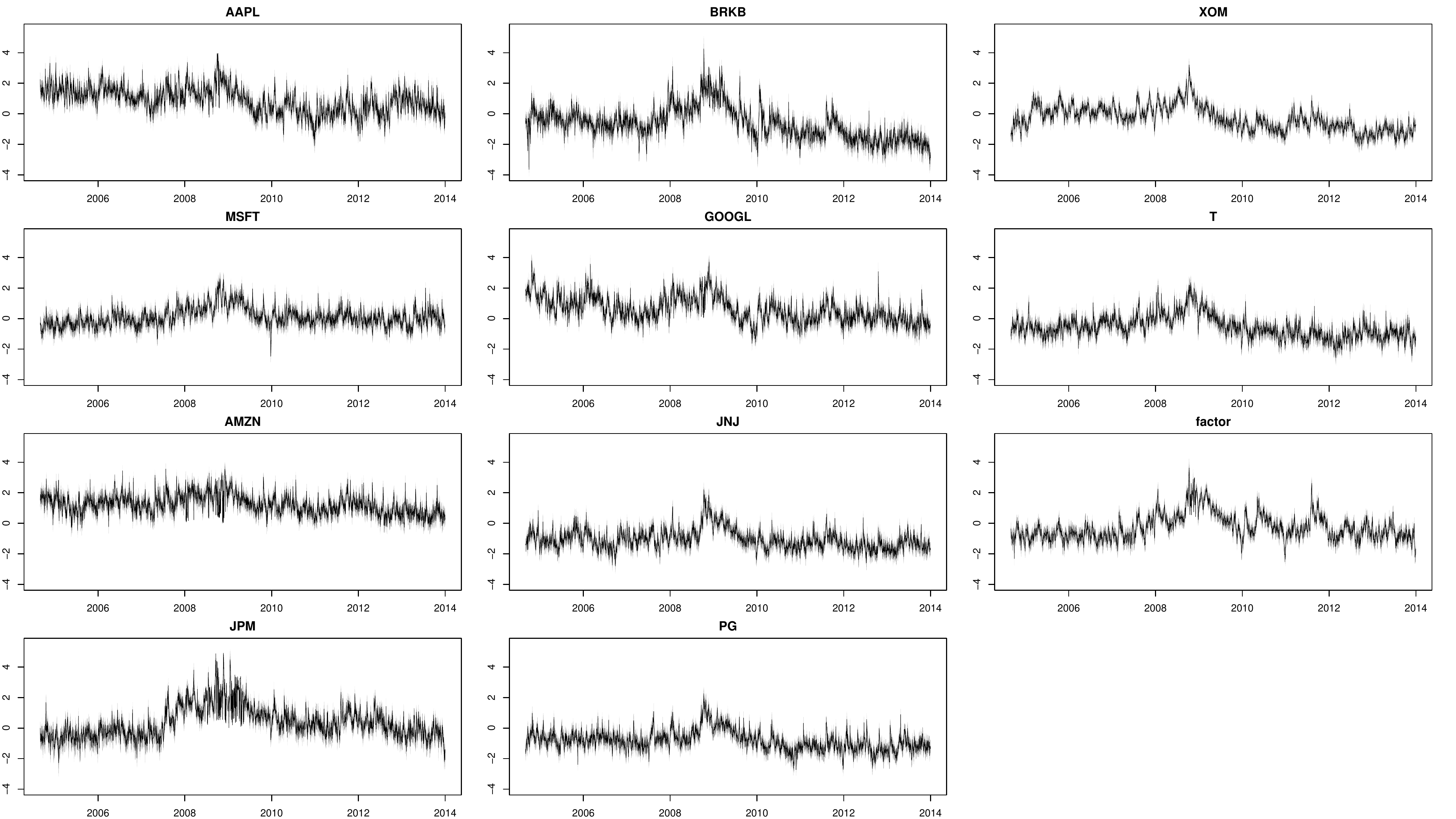}
	\caption{The posterior means (solid lines) and 95\% credible intervals (shaded areas) of the estimated idiosyncratic log volatilities of the stocks and the latent factor in the FMRSV model.}
	\label{fig:vol2}
\end{figure}

\begin{figure}[H]
	\centering
			\includegraphics[width=7.5cm]{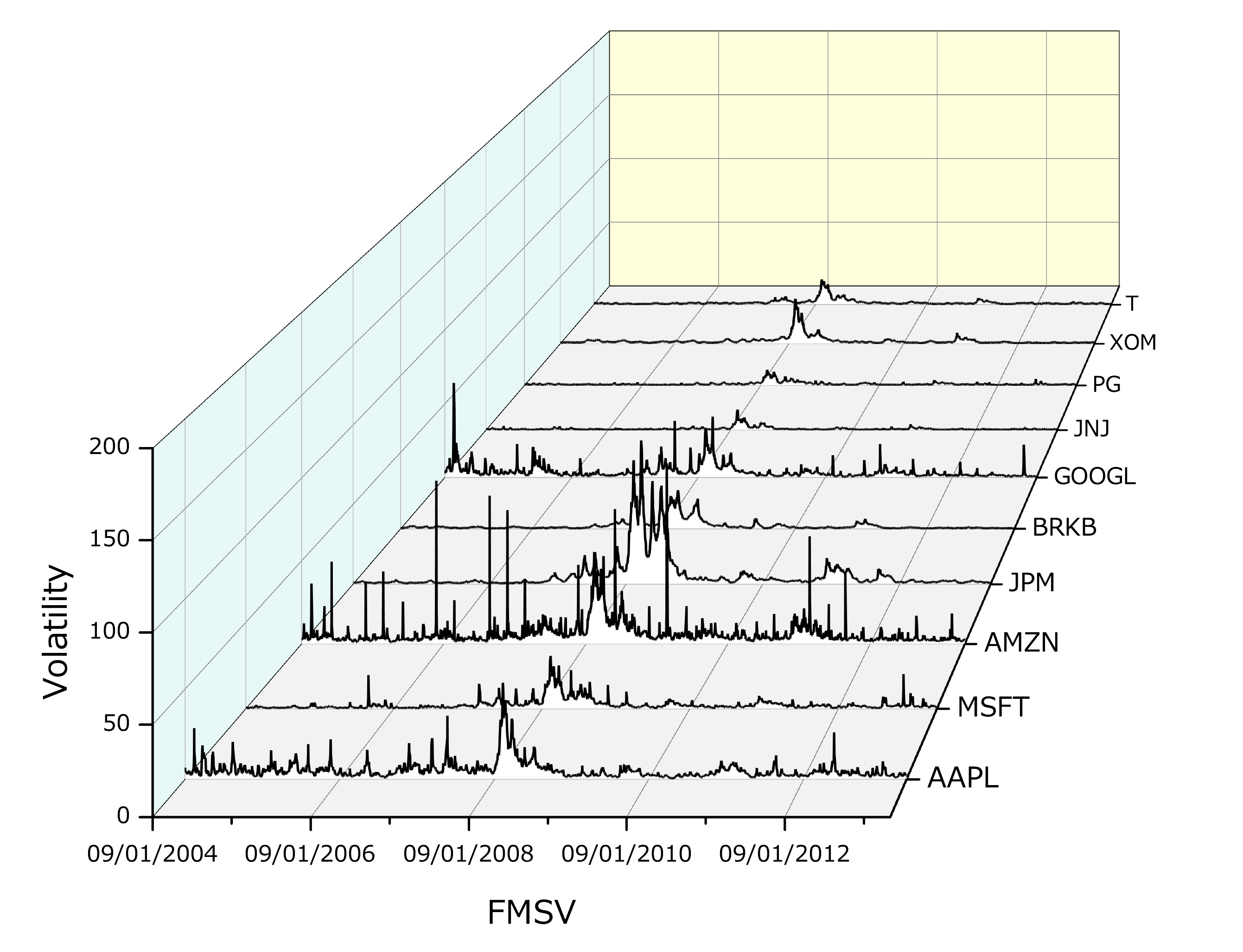}
			\includegraphics[width=7.5cm]{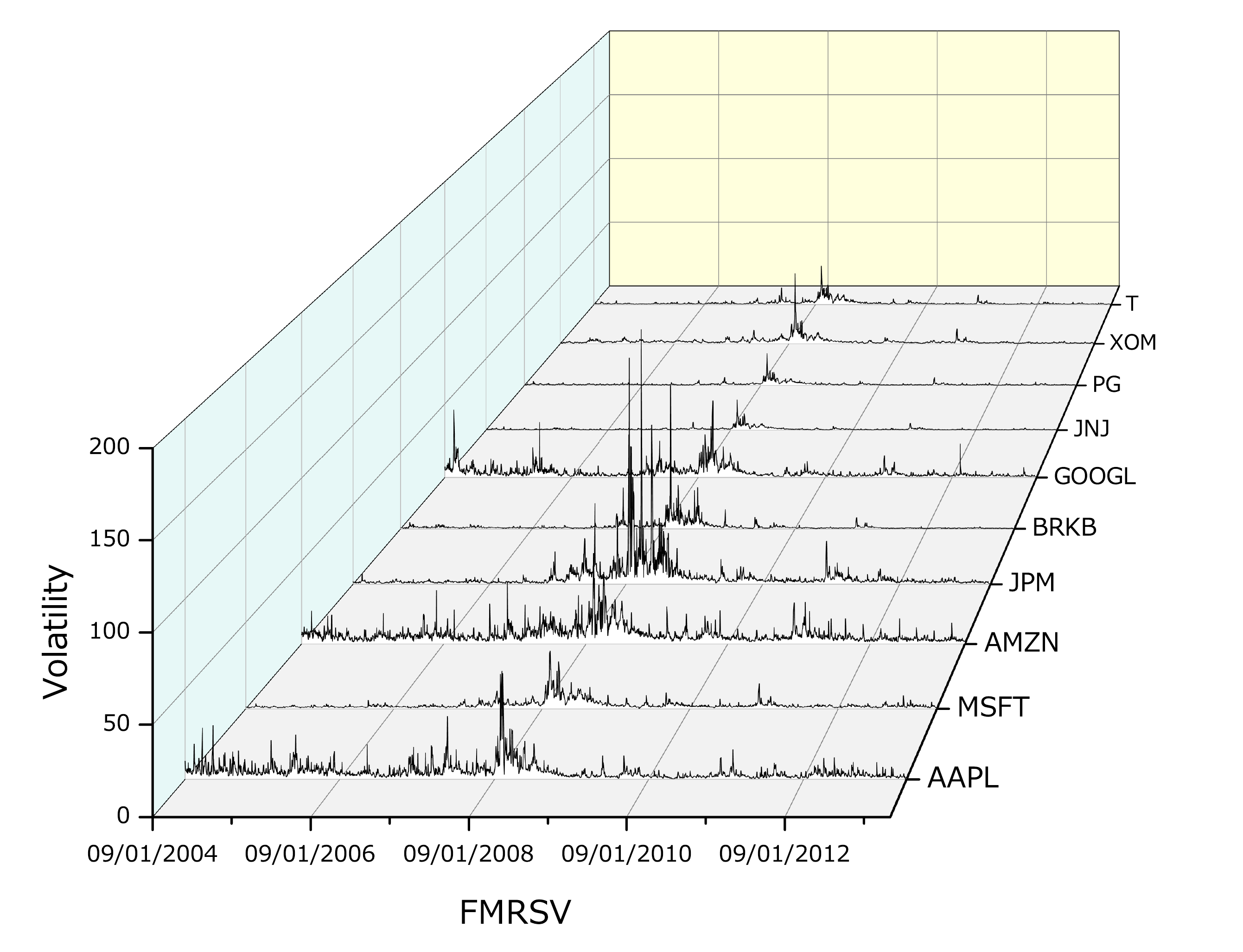}
	\caption{The posterior means of the estimated volatilities. }
	Left: FMSV model. Right: FMRSV model. 
        \label{fig:vol3}
\end{figure}

Figures \ref{fig:vol1} and \ref{fig:vol2} are the time series plots of the log idiosyncratic volatilities in the two models. The credible intervals are narrower and more stable in the FMRSV model than those in the FMSV model. Similar results are found for the  dynamic correlations (see the supplementary material). 
This finding implies that the additional information of the realized covariances enables us to estimate the true volatilities and correlations more accurately. 
Moreover, Figure \ref{fig:vol3} shows the posterior means of the estimated volatilities for ten stock returns. For example, the estimates of the return volatilities of AMZN in the FMSV model are overall smooth with many large irregular jumps throughout the sample period, while those in the FMRSV model are large only around the time of the global financial crisis with small irregular jumps. These results also confirm the usefulness of the additional information of realized covariances.

\begin{figure}[H]
	\centering
	\includegraphics[width=15cm]{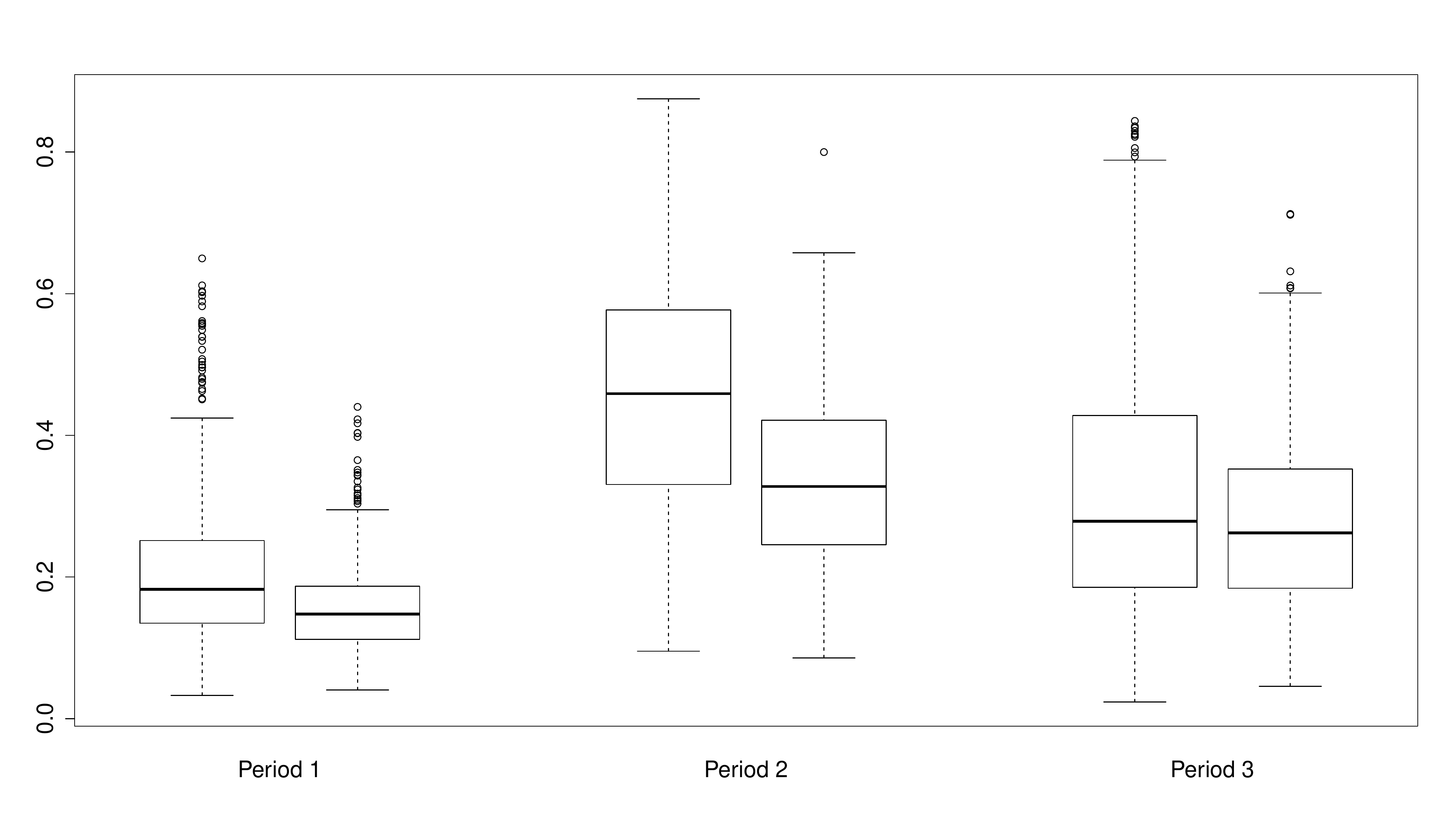}
	\caption{Boxplots of the posterior means of the correlation between AAPL and MSFT.}
	1: 9/1/2004--10/5/2007.  2: 10/8/2007--11/9/2010. 3: 11/10/2010--12/31/2013.\\
	Left: FMSV model. Right: FMRSV model. 
	\label{fig:corrbox}
\end{figure}

In Figure \ref{fig:corrbox}, we compare boxplots of the correlation coefficients between AAPL and MSFT during the three periods: (1) period 1: from September 1, 2004, to October 5, 2007, (2) period 2: from October 8, 2007, to November 9, 2010, and (3) period 3: from November 10, 2010, to December 31, 2013. The estimated correlations are high for both models during period 2, which includes the global financial crisis. Overall, the correlations in the FMSV model are estimated to be larger and more dispersed than those of the FMRSV models (the boxplots of the difference between the correlation estimates of the FMSV and FMRSV models are shown in the supplementary material). This indicates that we  overestimate the correlations among stock returns, especially during volatile markets when we do not use the information of the realized covariances.

\begin{figure}[H]
	\centering
	\begin{tabular}{ccc}
	\begin{minipage}{0.7\hsize}
		\centering
		\includegraphics[width=\textwidth]{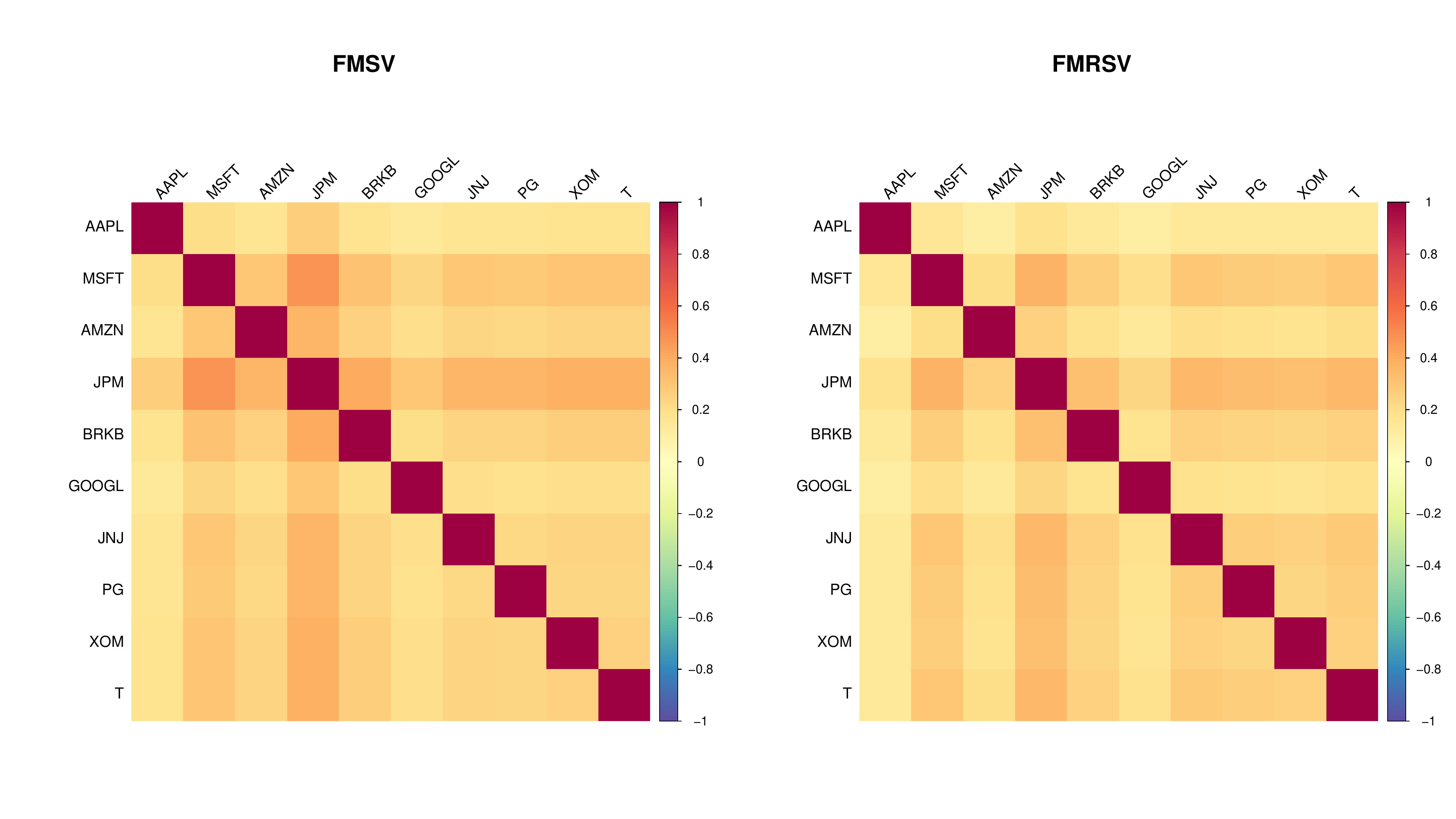}
	\end{minipage}\\
	\begin{minipage}{0.7\hsize}
		\centering
		\includegraphics[width=\textwidth]{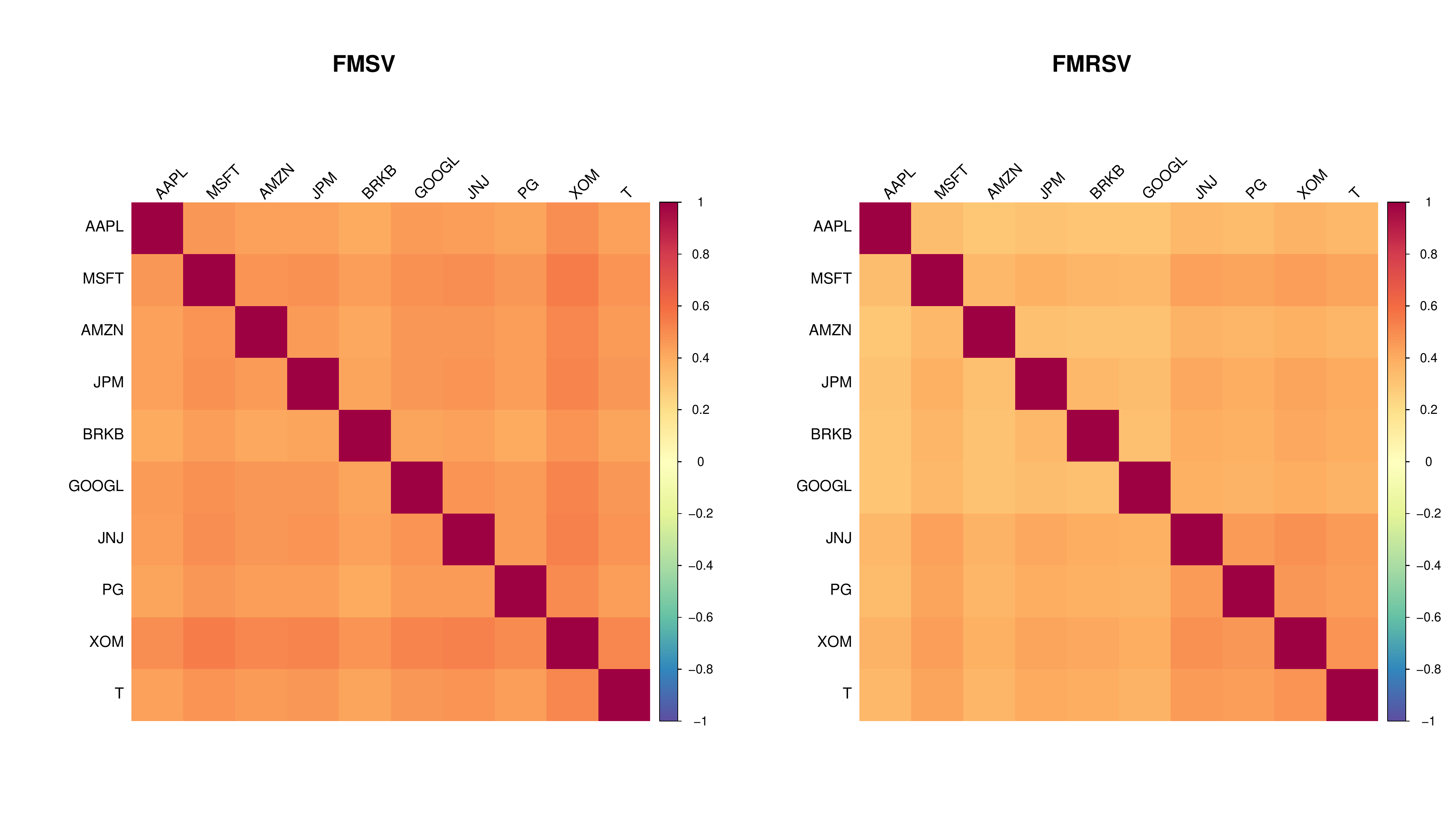}
	\end{minipage}\\
		\begin{minipage}{0.7\hsize}
		\centering
		\includegraphics[width=\textwidth]{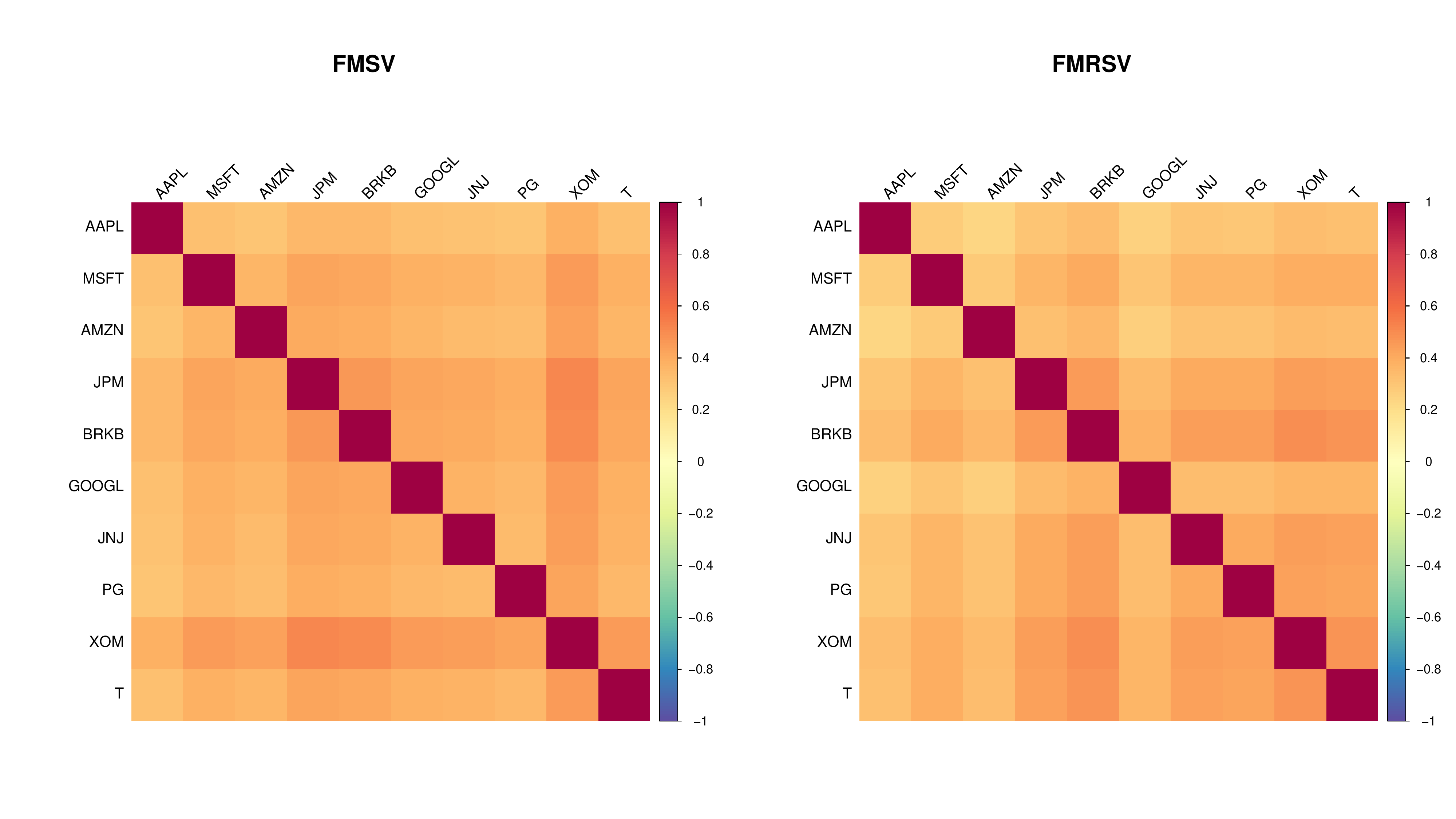}
	\end{minipage}
	\end{tabular}
	\caption{Heatmaps of the posterior means of the correlation coefficients.}
	Top: 9/1/2004--10/5/2007. Middle: 10/8/2007--11/9/2010. \\ Bottom: 11/10/2010--12/31/2013.
	Left: FMSV model. Right: FMRSV model. 
	\label{fig:corrmap}
\end{figure}

The heatmaps of the posterior means of all correlations between the ten stocks for the above three periods are also shown in Figure \ref{fig:corrmap}. All posterior means of the correlations are found to be positive and suggest the existence of a common market factor. For the FMSV model, in period 1, MSFT and JPM have larger correlations with other stock returns; while in periods 2 and 3, XOM seems to have the largest correlations. However, for the FMRSV model, we do not observe such significant differences among correlations.
As is also indicated by the box plots of the correlations between AMZN and MSFT in the three periods, most correlations are larger in the FMSV model than in the FMRSV model for all periods. During period 2, which includes the global financial crisis, all correlations are the highest, suggesting the co-movement of all stock returns through the market factor.

\newpage


\subsection{Comparison of portfolio performance}
In addition to the estimation results, we also compare the portfolio performance for the FMSV and FMRSV models. The portfolio return at time $t+1$ is defined as 
\begin{eqnarray}
r_{p,t+1} = \boldw_t'\boldy_{t+1} + (1- \boldw_t'\boldone_p)r_f,
\end{eqnarray}
where  $\boldw_t$ is a $p \times 1$ portfolio weight vector for the stock return $\bm{y}_{t+1}$, $\bm{1}_p$ denotes a $p\times 1$ vector with all elements equal to one, and $r_f$  is the risk-free asset return. The weight $\boldw_t$ is unrestricted (allowing short selling) and chosen to optimize the objective function based on the portfolio strategy, as follows. 

The conditional mean and conditional variance of $r_{p,t+1}$ given the information set $\calF_t$ are
\begin{align*}
\mu_{p,t+1}
&\equiv \E[\bm{r}_{p,t+1} \vert \calF_t]
= \boldw_t'\boldm_{t+1 \vert t} + (1- \boldw_t'\boldone_p)r_f, \\
\sigma^2_{p,t+1} &\equiv
\Var[\bm{r}_{p,t+1} \vert \calF_t]
=\boldw_t' \boldSigma_{t+1 \vert t} \boldw_t,
\end{align*}
where 
\begin{align*}
\boldm_{t+1 \vert t}
& \equiv
\E[\boldy_{t+1} \vert \calF_t]=\mathbf{B}\{\bm{\gamma}+\bm{\psi}\odot (\bm{f}_t-\bm{\gamma})\},
\\
\boldSigma_{t+1 \vert t} & \equiv
\Var[\boldy_{t+1} \vert \calF_t]
= \mathbf{B}\mathbf{V}_{2,t+1}\mathbf{B}' + \mathbf{V}_{1,t+1}.
\end{align*}
This paper considers the portfolio strategy to minimize the conditional expected variance $\sigma^2_{p,t+1}$ given the target conditional expected return $\mu_{p,t+1} = \mu^{*}_{p}$. The solution of the weight is given by 
\begin{align*}
& \hat{\boldw}_t = \boldSigma_{t+1 \vert t}^{-1} (\boldm_{t+1 \vert t} - r_f \boldone_p) \frac{\mu_p^* - r_f}{(\boldm_{t+1 \vert t} - r_f \boldone_p)' \boldSigma_{t+1 \vert t}^{-1}(\boldm_{t+1 \vert t} - r_f \boldone_p)}
\end{align*}
where we obtain the estimates of $\bm{m}_{t+1|t}$ and $\mathbf{\Sigma}_{t+1|t}$ via a one-step ahead forecast using a rolling estimation.

To investigate the effect of including the realized covariances as the additional information and the leverage effect, we compare the portfolio performance using the following three models.
\begin{enumerate}
	\item
	FMSV model: Factor multivariate stochastic volatility model with the leverage effect, but without realized covariances.
	\item 
	FMRSV-NL model: Factor multivariate stochastic volatility model without the leverage effect, but with realized covariances.
	\item 
	FMRSV model: Factor multivariate stochastic volatility model, with the leverage effect and realized covariances.
\end{enumerate}
Two different forecast periods with  100 one-day ahead forecasts are considered using the rolling estimation with the number of observations equal to 2250:
\begin{itemize}
	\item Period  I.  From August  9, 2013, to December 31, 2013.
	\item Period II.  From August  9, 2019, to December 31, 2019.
\end{itemize}
Period I includes the time of the global financial crisis in the estimation period.
The rolling forecast and estimation are implemented as follows.
\begin{enumerate}
	\item[]\hspace{-7mm}Step 1. 
	First, we estimate parameters using the first $2250$ observations from  September 1, 2004, to August 8, 2013. and forecast the mean, the volatility and the correlation of the multiple stock returns for August 9, 2013. Use them to obtain the optimal weights of the assets for the above portfolio strategies  where the federal funds rate is used as the risk-free asset return $r_f$.

	\item[]\hspace{-7mm}Step 2. 
	Next, we drop the first observation (September 1, 2004) from the sample period and add a new observation (August 9, 2013). The new sample period is from September 2, 2004, to August 9, 2013. We estimate the parameters using these observations and forecast the mean, the volatility and the correlation for August 10, 2013. Then, we use them to obtain the optimal weights in a similar manner.
	\item[]\hspace{-7mm}Step 3.  We iterate these rolling forecasts until December 31, 2013, to obtain the 100 one-day ahead forecasts and corresponding weights.
\end{enumerate}
Table \ref{table:portfolio_cum_obj_2004} displays the cumulative realized variances of three models in the two periods. The FMRSV-NL and FMRSV models show the best performance in periods I and II (except for $\mu^*_p = 0.007$ in Period I), respectively among these models, suggesting that the introduction of realized covariances improves the prediction of the conditional means and covariances of stock returns.
On the other hand, the effect of introducing the leverage depends on the forecasting period. The performance of the FMRSV-NL model is the best in period I, but the worst in period II. The performance of the FMRSV model is overall good and stable in both periods.

\small
\begin{table}[H]
	\centering
	\begin{tabular}{lrrr}
		\hline\hline
		~~&$\mu_p^{*} = 0.004$&$\mu_p^{*} = 0.01$&$\mu_p^{*} = 0.02$\\
		~~FMSV            &$0.291$&$1.959$&$8.027$\\
		~~FMRSV-NL        &$\textbf{0.131}$&$\textbf{0.882}$&$\textbf{3.616}$\\
		~~FMRSV         &$0.229$&$1.550$&$6.362$\\
		\hline
	\end{tabular}
	\vspace{2mm}
	\\
	Period I (8/9/2013--12/31/2013)
	\vspace{5mm}\\
	\begin{tabular}{lrrr}
		\hline\hline
		~~&$\mu_p^{*} = 0.007$&$\mu_p^{*} = 0.015$&$\mu_p^{*} = 0.03$\\
		~~FMSV            &$\textbf{0.115}$&$3.049$&$19.481$\\
		~~FMRSV-NL        &$0.135$&$3.285$&$20.722$\\
		~~FMRSV         &$0.126$&$\textbf{3.011}$&$\textbf{18.918}$\\
		\hline
	\end{tabular}
	\vspace{2mm}
	\\
	Period II (8/9/2019--12/31/2019)\\
	\caption{The cumulative realized variances computed as $\sum_{t=T-100}^{T-1}\hat{w}_t'\boldSigma_{t+1}\hat{w}_t$ where $\boldSigma_{t+1}$ is evaluated using the bias-corrected realized covariances at time $t+1$.}
	\label{table:portfolio_cum_obj_2004}
\end{table}
\normalsize
\vspace{-5mm}
\begin{figure}[H]
	\centering
	\includegraphics[width=8.25cm]{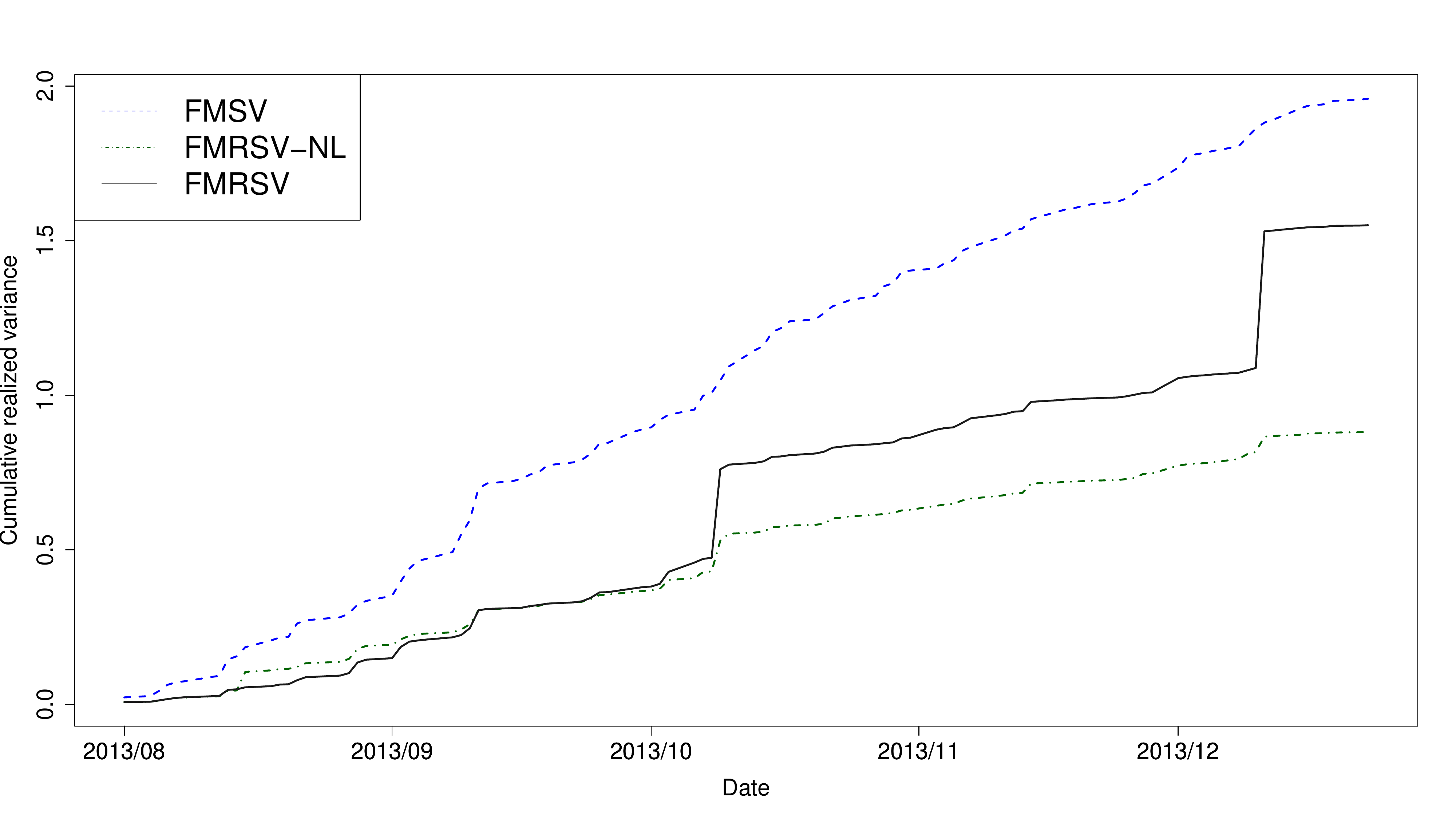}
	\\
	Period I (8/9/2013--12/31/2013)\\
	\includegraphics[width=8.25cm]{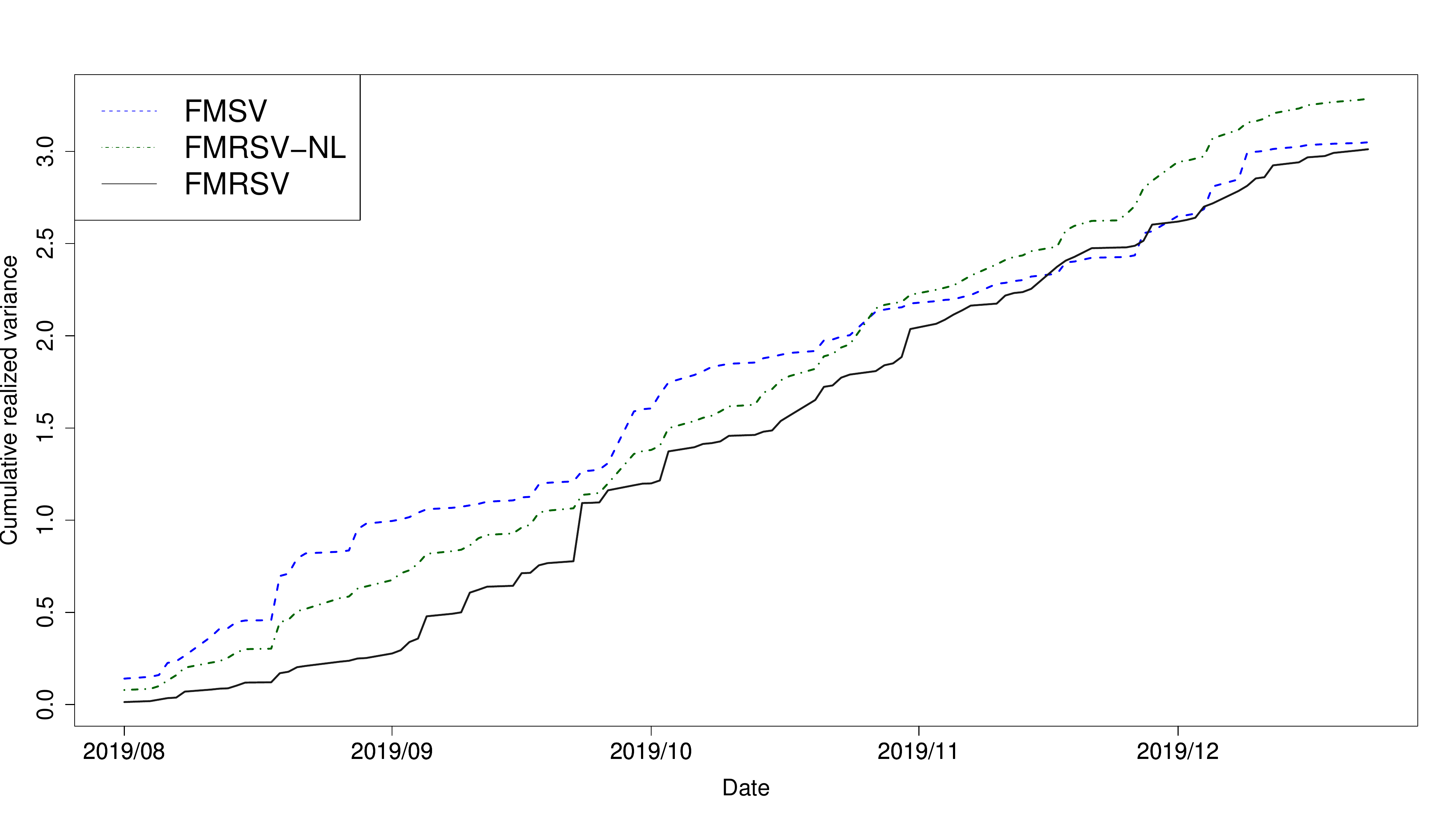}
\\
	Period II (8/9/2019--12/31/2019)
	\caption{The cumulative realized variances with $\mu^*_p = 0.01$ (Period I) and $\mu^*_p = 0.015$ (Period II).
	FMRSV: solid black. FMRSV-NL: dashed green. FMSV: dotted blue. }
	\label{fig:cumr_maxr_2004}
\end{figure}

Figure \ref{fig:cumr_maxr_2004} shows the time series plots of the cumulative realized variances for the three models. The FMRSV-L and FMRSV models outperform the FMSV model in periods I and II respectively, implying that the information of the realized covariances improves the portfolio performance consistently. The performance of the FMRSV-NL model seems to depend on the forecast period. For example,  it outperforms the FMRSV model after November 2013, while it underperforms the FMSV and FMRSV models  after November 2019.

\begin{figure}[H]
	\centering
	\includegraphics[width=7.5cm]{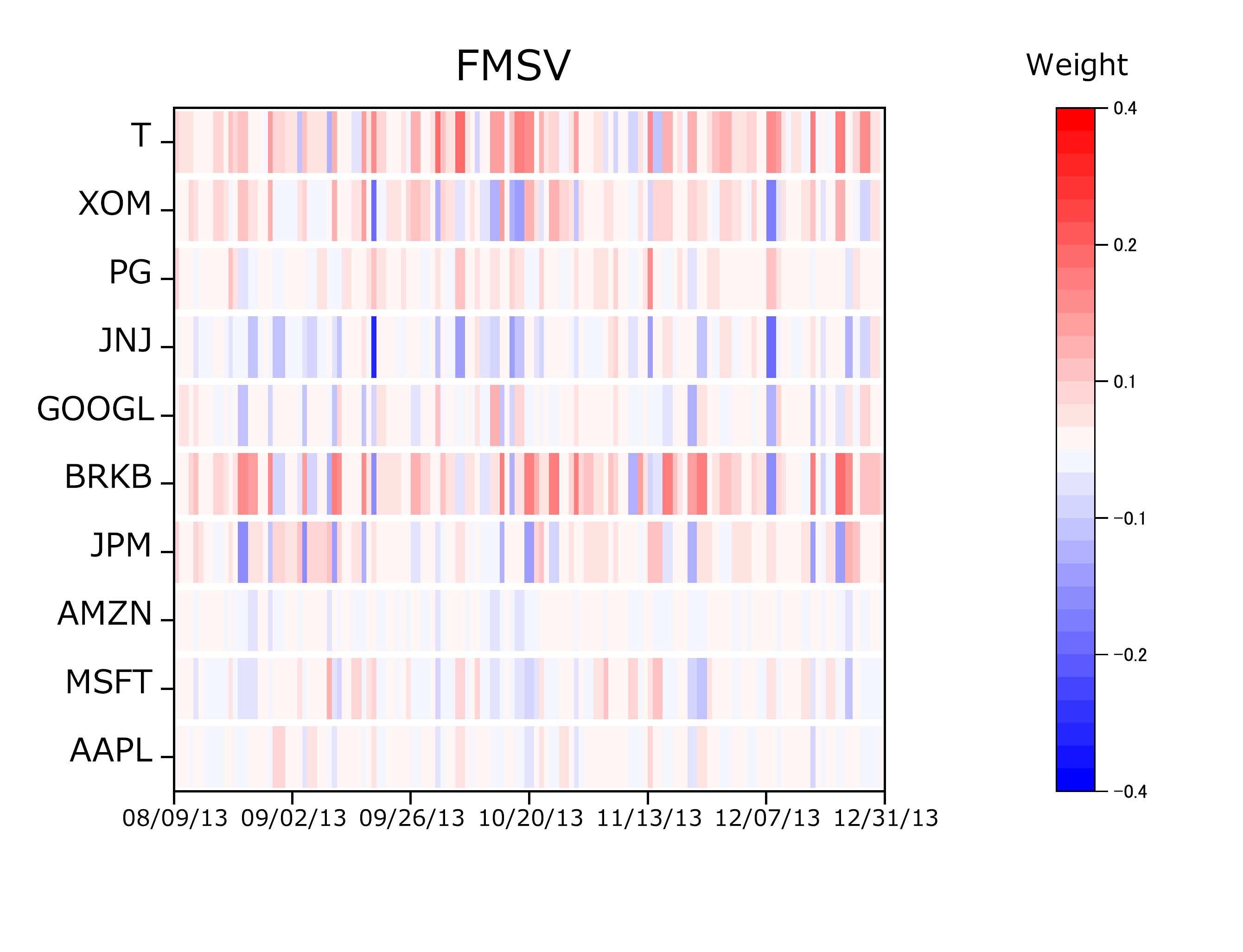}
	\includegraphics[width=7.5cm]{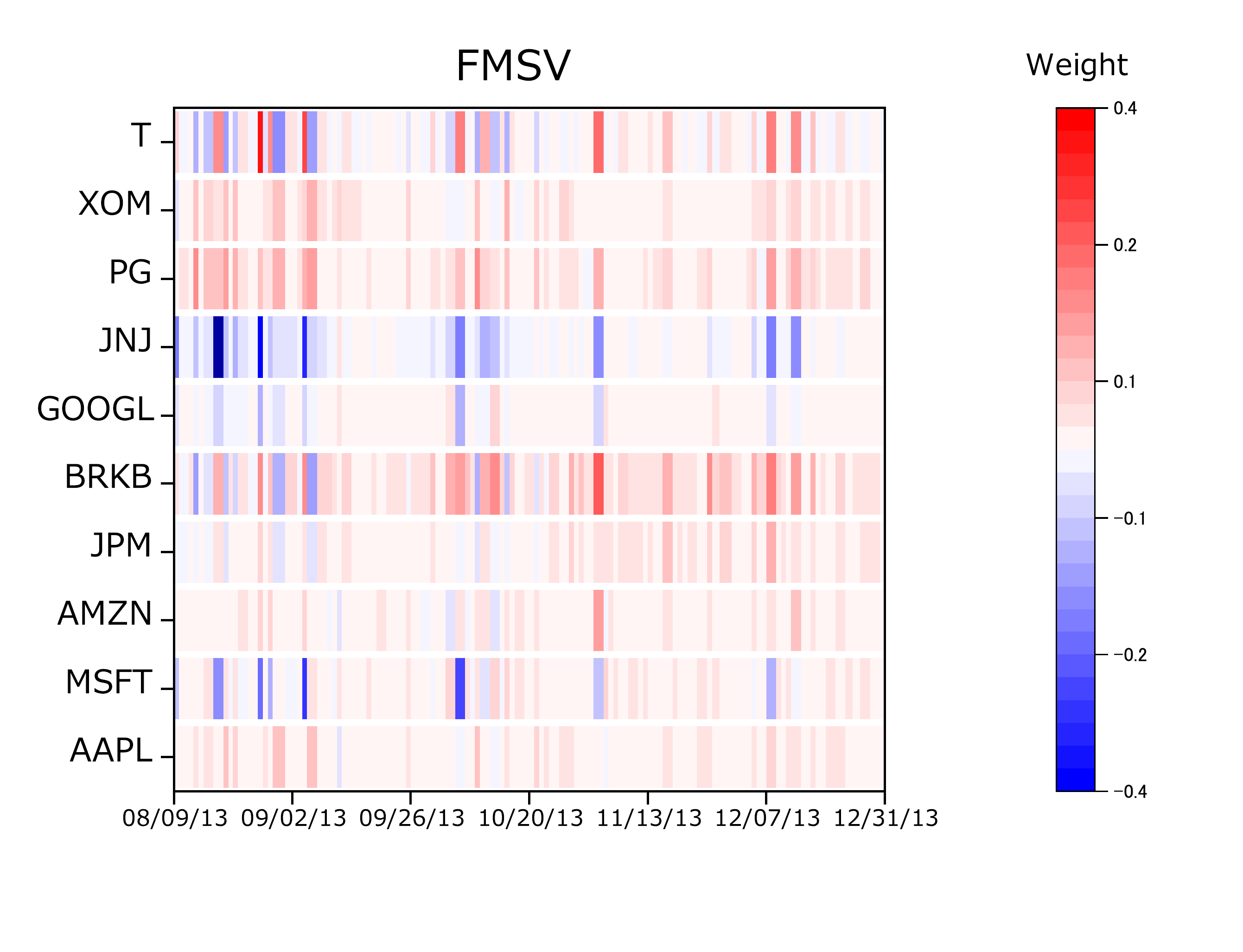}
\\
	\includegraphics[width=7.5cm]{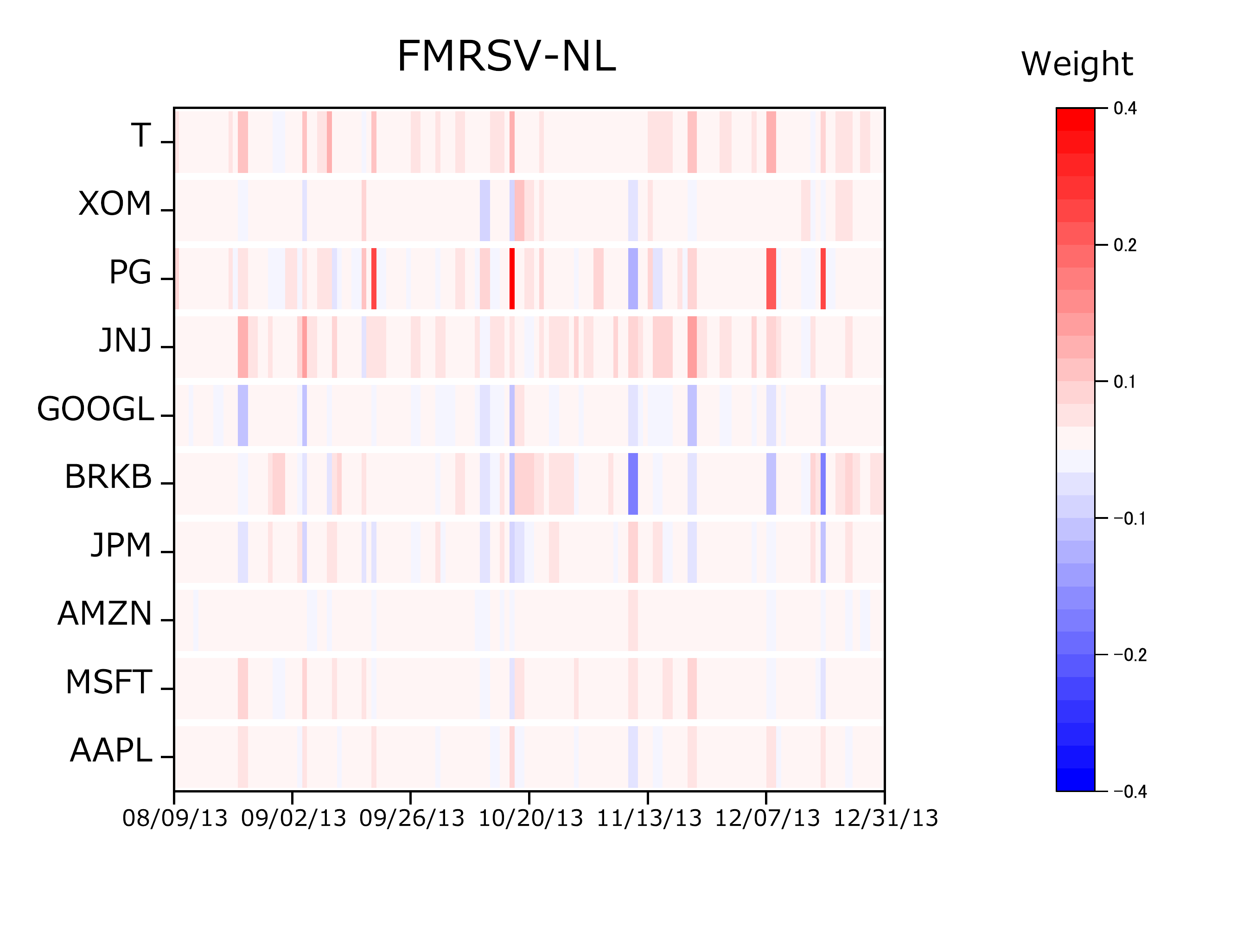}
	\includegraphics[width=7.5cm]{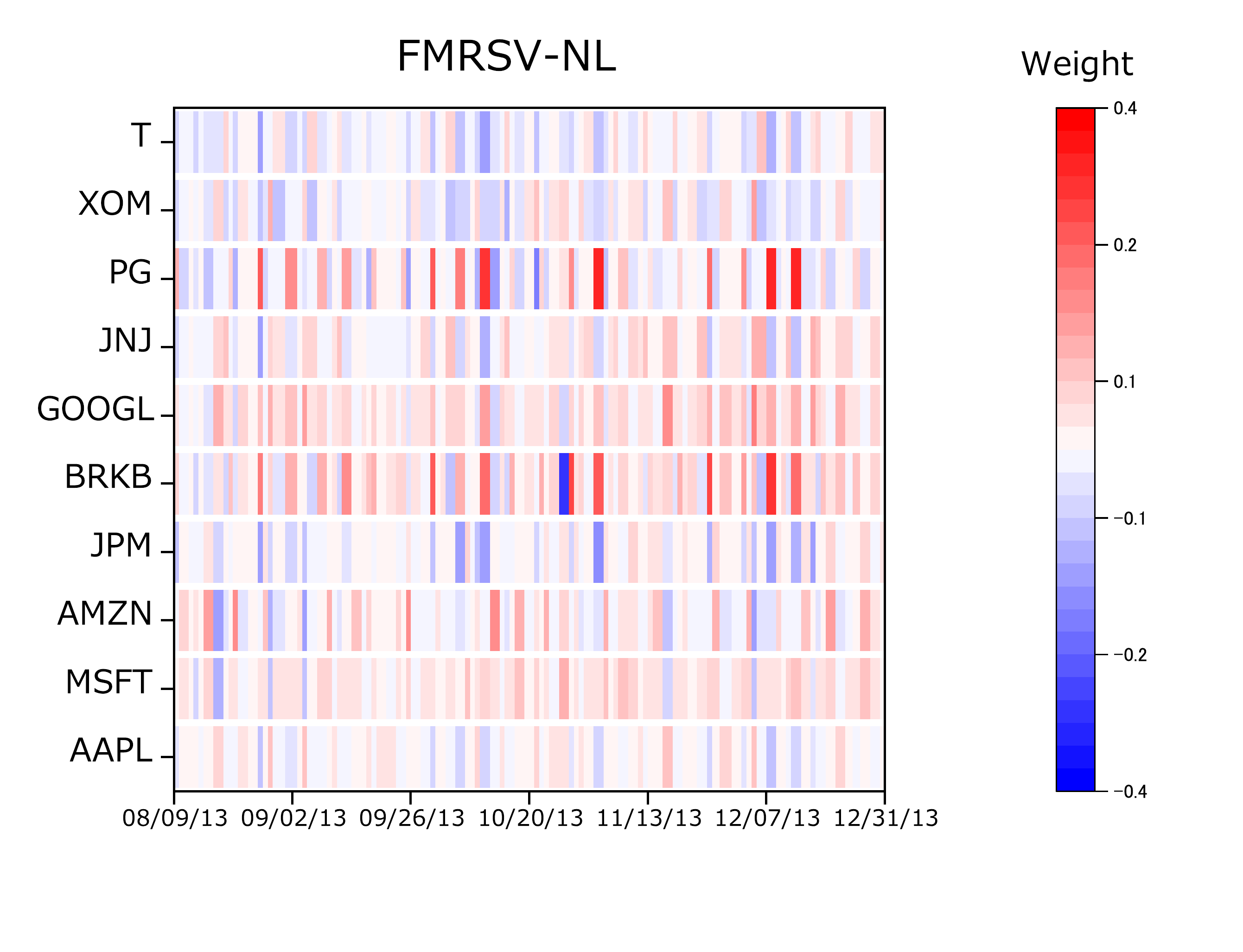}
\\
	\includegraphics[width=7.5cm]{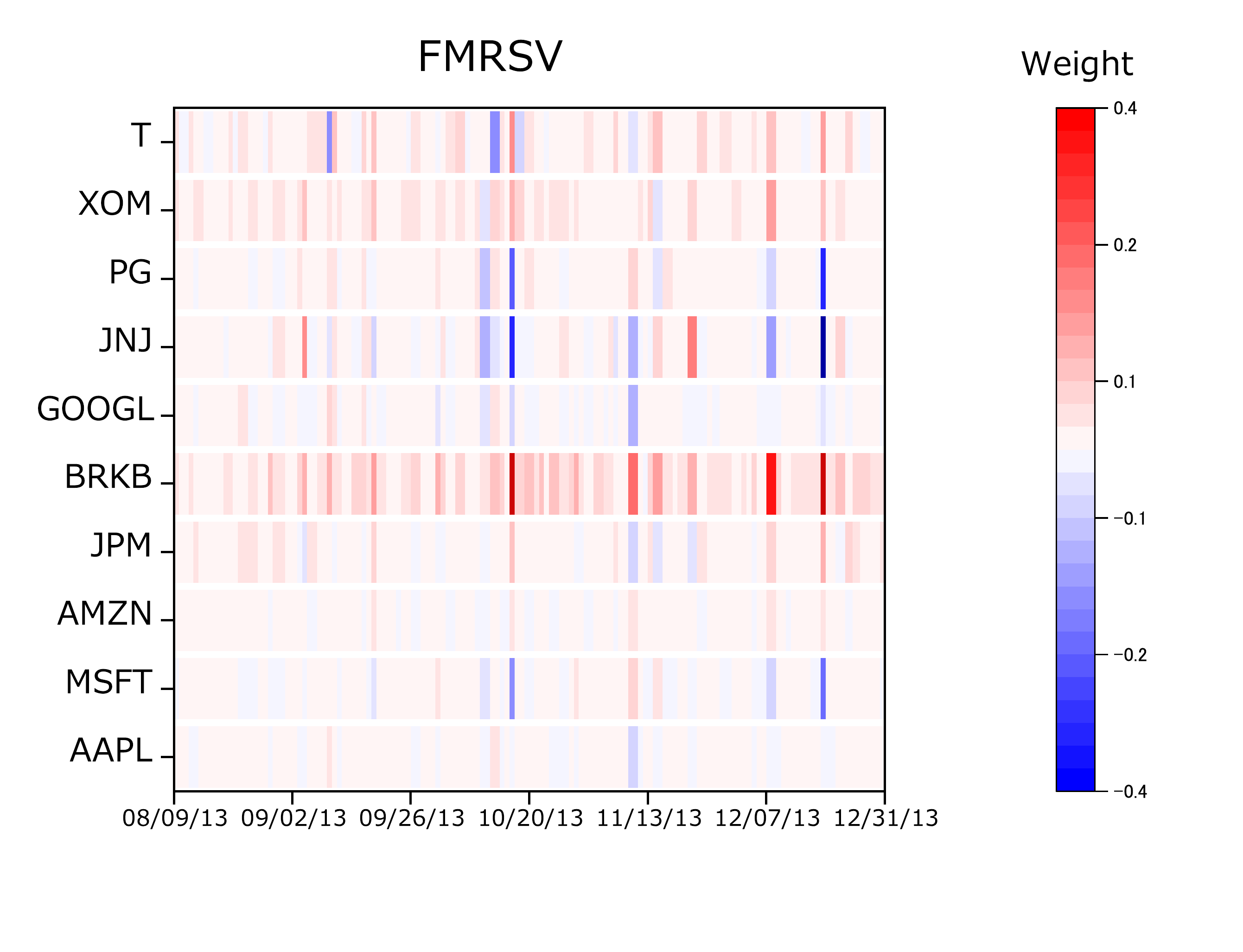}
	\includegraphics[width=7.5cm]{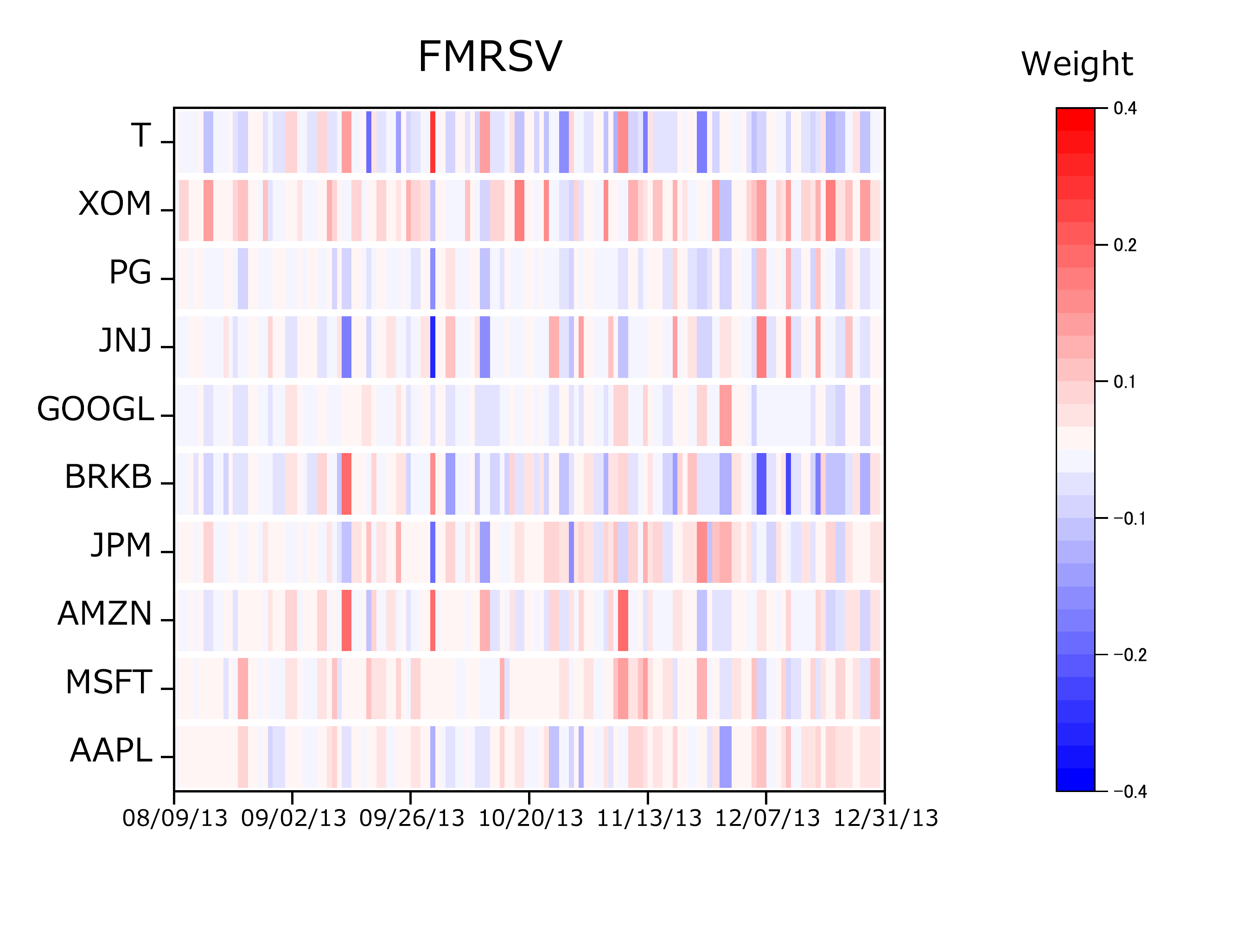} \\
	Left: Period I (8/9/2013--12/31/2013). Right: Period II (8/9/2019--12/31/2019).
	\caption{Split heatmaps of the portfolio weight $\bm{w}_t$ for the ten stocks with $\mu^*_p = 0.01$ (Period I) and $\mu^*_p = 0.015$ (Period II).}
	\label{fig:weight_2004-2013}
\end{figure}

Figure \ref{fig:weight_2004-2013} displays the split heatmaps of the portfolio weights for the three models during the two periods. We first compare three models in period I. In the FMSV model, the weights for T and BRKB tend to be positive and much larger than those of other stock returns, while the weights of JNJ are often found to be negative. Those weights are relatively unstable and sometimes become negative even for BRKB and T. For the FMRSV-NL and FMRSV models, most weights are positive for all ten stock returns and stable throughout period I. However, the FMRSV-NL model places relatively large weights on PG and JNJ, while the FMRSV model places heavy weights on BRKB.
In period II, the FMSV model places large positive weights on BRKB and negative weights on JNJ as in period I, but the weights for T are smaller and sometimes negative. They are large for BRKB and PG in the FMRSV-NL model, while they are more stable and larger for AMZN and XOM in the FMRSV model. The weights for T and BRKB seem unstable but become very small during the latter part of the forecasting period in the FMRSV model. 

\begin{figure}[H]
	\centering
	\includegraphics[width=7.5cm]{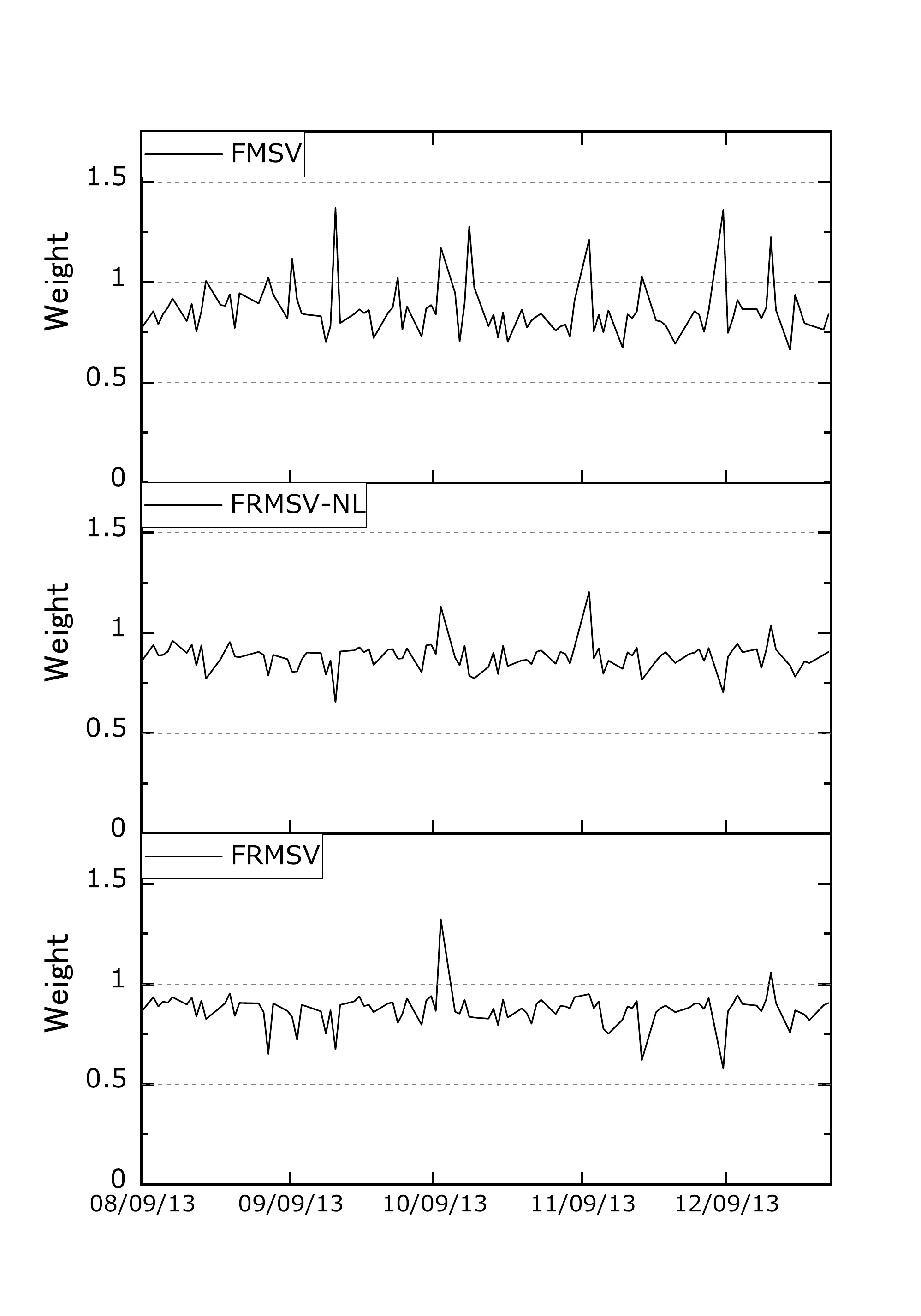}
	\includegraphics[width=7.5cm]{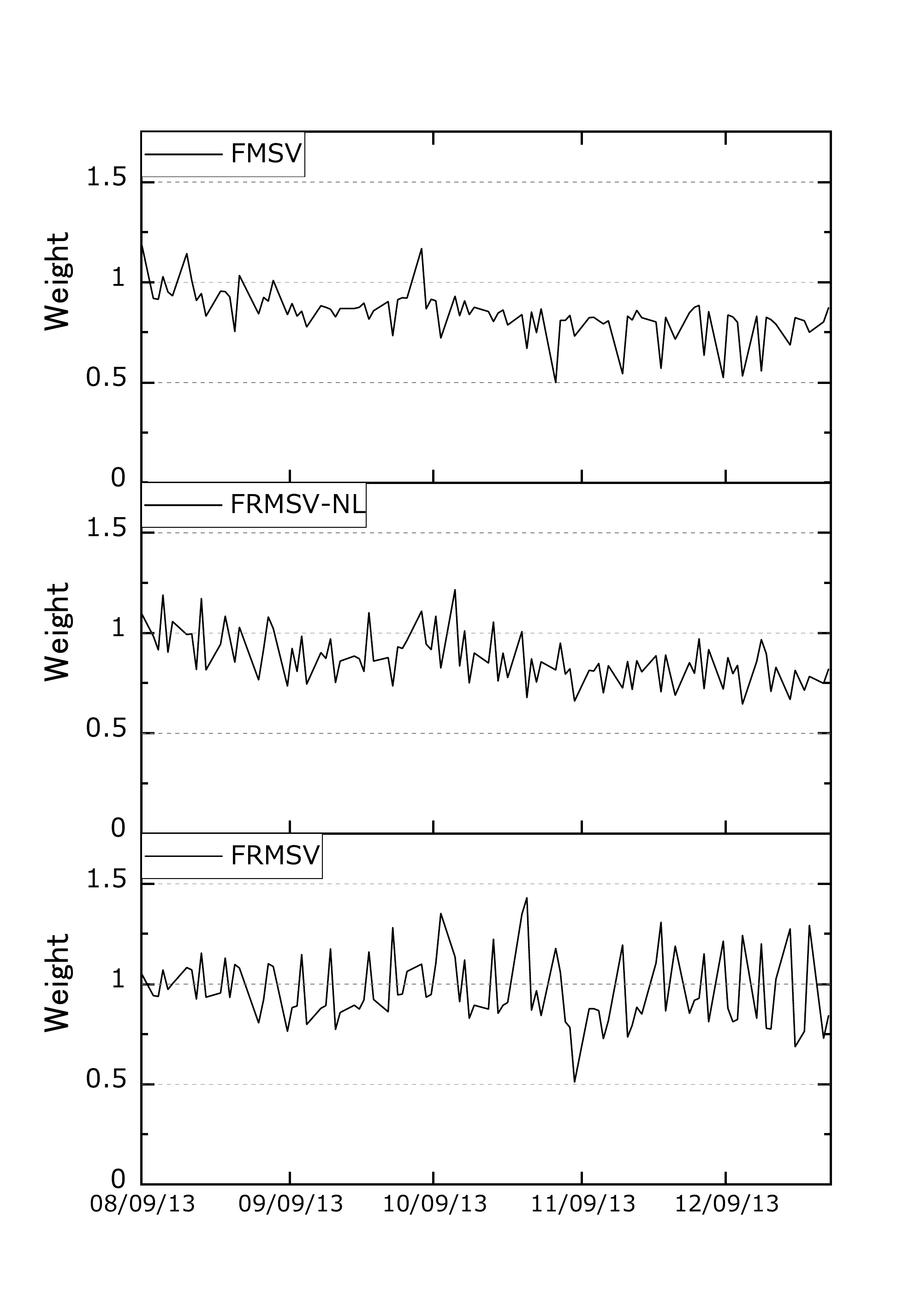}\\
	Left: Period I (8/9/2013--12/31/2013). Right: Period II (8/9/2019--12/31/2019).
	\caption{Time series plots of the portfolio weight $1-\bm{w}_t'\bm{1}_p$ for the federal funds rate with $\mu^*_p = 0.01$ (Period I) and $\mu^*_p = 0.015$ (Period II).}
	\label{fig:weight_ff}
\end{figure}

On the other hand, Figure \ref{fig:weight_ff} shows the time series plot of the weights of the risk-free asset. In period I, the weights for the federal funds rate are volatile in the FMSV model. They sometimes increase and decrease from over 1.3 to approximately 0.7. In the FMRSV-NL and FMRSV models, they are basically stable at approximately 0.9 except a couple of days.
In period II, the weights gradually decrease from approximately 1.0 to around 0.7 in the FMSV and FMRSV-NL models, while they are very volatile around approximately 1.0 in the FMRSV model.

In summary, the weights of the asset returns change over time and lead to better portfolio performance when we use the realized covariances, suggesting that including such an additional information is very effective. The importance of the leverage effect depends on the period, but the overall performance is more stable and better for the model with leverage. Using the information from both daily returns and realized covariances gives us more accurate and stable results in the estimation and in the portfolio performance based on the forecast.

\newpage

\section{Conclusion}
We propose a multivariate SV model with a dynamic factor structure and leverage effect, incorporating the realized measures of latent covariance and latent factors. Using the information of realized measures in addition to daily stock returns, we are able to estimate the model parameters and latent variables more accurately, and give more stable one-step ahead forecasts of the covariance matrices, which improves the portfolio performance as illustrated in our empirical studies. Taking account of the leverage effect from the information of daily returns is found to be important to obtain stable portfolio performance and becomes critical depending on the forecast period.

%


\section*{Acknowledgements}

The computational results were obtained by using Ox version 7 (\cite{Doornik(07)}). 
 This work was supported by JSPS KAKENHI Grant Numbers 19H00588, 20H00073.

\small
\bibliography{ref_FMRSV_paper}	
\normalsize
\clearpage
\small
\section*{Appendix}
\appendix
\section{MCMC algorithm}
\label{sec:generate-h2}
\noindent
{\it Generation of $\bm{h}^{(2)}$}. 
First we define $\bm{c}_t$ and $\boldV_{2t \vert h}$ such that
\begin{align}
	& \boldc_{t} =
	I(t<T)
	\left(
	\begin{array}{c}
	\frac{\rho_{p+1}\exp(h_{p+1,t}/2)}{\sigma_{p+1,\eta}}\left\{h_{p+1,t+1} - \mu_{p+1} - \phi_{p+1} (h_{p+1,t} - \mu_{p+1})\right\}  \label{eq:c_t}\\
	\vdots \\
	\frac{\rho_{p+q}\exp(h_{p+q,t}/2)}{\sigma_{p+q,\eta}}\left\{h_{p+q,t+1} - \mu_{p+q} - \phi_{p+q} (h_{p+q,t} - \mu_{p+q})\right\}  \\
	\end{array}
	\right), \\
	& \boldV_{2t \vert h} = 
		\left\{\mathbf{I}-I(t<T)\diag(\rho_{p+1}^2,\ldots,\rho_{p+q}^2)\right\}\mathbf{V}_{2t}.
\end{align}
Then, the log conditional posterior density of $\boldh^{(2)}$ given the other latent variables and parameters is
\begin{eqnarray}
\lefteqn{\log \pi(\boldh^{(2)} \vert \cdot)}&& \nonumber\\
	 &=& \mbox{const}+
	 \frac{s_0}{2}\sum_{t=1}^{T}\log \dtm{\boldB \boldV_{2t}\boldB' + \boldV_{1t}} -\frac{k_0}{2}\tr \left( \sum_{t=1}^{T} \boldB \boldV_{2t}\boldB' \boldW_t^{-1} \right) \nonumber\\
	&&-\frac{1}{2}\sum_{t=1}^{T}\left\{\log\dtm{\boldV_{2t\vert h}} 
	+ (\boldf_t - \boldgamma - \boldPsi (\boldf_{t-1} - \boldgamma) - \boldc_{t})'\boldV_{2t\vert h}^{-1}(\boldf_t - \boldgamma - \boldPsi (\boldf_{t-1} - \boldgamma) - \boldc_{t})\right\} \nonumber\\
	&&-\frac{1}{2} \sum_{t=1}^{T-1}\left\{\boldh_{t+1} - (\boldI - \boldPhi)\boldmu - \boldPhi\boldh_{t}\right\}'\boldSigma_{\eta\eta}^{-1}\{\boldh_{t+1} - (\boldI - \boldPhi)\boldmu - \boldPhi\boldh_{t}\}-\frac{1}{2} \{\boldh_{1} - \boldmu)'\boldSigma_{h,0}^{-1}
	(\boldh_{t} - \boldmu)  \nonumber\\
	& = & 
	\mbox{const} -\sum_{t=1}^{T-1}\sum_{i=1}^{q}\frac{\{h_{p+i,t+1}-h_{p+i,t}-(1-\phi_{p+i})\mu_{p+i}\}^2}{2\sigma_{\eta,p+i}^{2}}-\sum_{i=1}^{q}\frac{(1-\phi_{p+i}^2)(h_{p+i,1}-\mu_{p+i})^2}{2\sigma_{\eta,p+i}^2}
	\nonumber\\
	&& +\sum_{t=1}^{T}l_t^*, 
\end{eqnarray}
where
\begin{eqnarray*}
l_t^* &=& \frac{s_0}{2}\log\dtm{\boldB \boldV_{2t}\boldB' + \boldV_{1t}} 
-\frac{k_0}{2}\tr \left(\boldB \boldV_{2t}\boldB' \boldW_t^{-1} \right)
	-\frac{1}{2}\sum_{i=1}^{q}\left\{h_{p+i,t}+(f_{it}- \mu_{f,it})^2\sigma_{f,it}^{-2}\right\},\\
&&	\mu_{f,it} = \gamma_i + \psi_i(f_{i,t-1} - \gamma_i) + \frac{\rho_{p+i}\exp(h_{p+i,t}/2)}{\sigma_{p+i,\eta}}\left\{h_{p+i,t+1} - \mu_{p+i} - \phi_{p+i} (h_{p+i,t} - \mu_{p+i}) \right\}I(t<T), \\
&&	\sigma_{f,it}^2 = \left\{1-I(t<T)\rho_{p+i}^2\right\}\exp(h_{p+i,t}).
\end{eqnarray*}
When there is a correlation between $f_{it}$ and $h_{p+i,t+1}$, the approximated conditional  distribution of $(h_{p+i,s},\ldots,h_{p+i,s+m})$ does not have a diagonal covariance matrix, and the sampling algorithm for $\bm{h}^{(1)}$ does not apply to that for $\bm{h}^{(2)}$. Thus we take an alternative sampling algorithm based on \cite{omori2008block} to approximate the nonlinear Gaussian state space model using the linear Gaussian state space model. 
As in sampling $\bm{h}^{(1)}$, we consider sampling the disturbances $(\eta_{p+i,s-1},\ldots,\eta_{p+i,s+m-1})$ instead of the state variables $(h_{p+i,s},\ldots,h_{p+i,s+m})$ given the other state variables and parameters.
The logarithm of the conditional posterior density of $(\eta_{p+i,s-1},\ldots,\eta_{p+i,s+m-1})$ is given by
\begin{align}
&\log f(\eta_{p+i,s-1},\ldots,\eta_{p+i,s+m-1} \vert \cdot) 
	= \mbox{const} - \frac{1}{2} \sum_{t=s-1}^{t=s+m-1}\eta_{p+i,t}^2 + L,
\end{align}
where
\begin{align*}
	& L=\displaystyle \sum_{t=s}^{t=s+m} l_{it}^* - \frac{\{h_{p+i,s+m+1} - (1 - \phi_{p+i})\mu_{p+i} - \phi_{p+i}h_{p+i,s+m}\}^2}{2 \sigma_{\eta,p+i}^2}I(s+m<T),
\\
	& l_{it}^* = \frac{s_0}{2}\log\dtm{\boldB \boldV_{2t}\boldB' + \boldV_{1t}} 
-\frac{k_0}{2}\tr \left(\boldB \boldV_{2t}\boldB' \boldW_t^{-1} \right)
	-\frac{1}{2}\left\{h_{p+i,t}+(f_{it}- \mu_{f,it})^2\sigma_{f,it}^{-2}\right\}.
\end{align*}
In order to approximate $L$ using a logarithm of a normal probability density, we define $d_t,A_t, B_t$ and $\boldQ$  as follows.
\begin{align*}
	d_t = \frac{\partial L}{\partial h_{p+i, t}},\quad
	A_t = -E\left(\frac{\partial^2 L}{\partial h_{p+i,t}^2}\right), \quad
	B_t = -E\left(\frac{\partial^2 L}{\partial h_{p+i,t}\partial h_{p+i,t-1}}\right),
\end{align*}
\begin{align*}
	\boldQ = 
	\begin{bmatrix}
		A_{s}&B_{s+1}&0&\cdots&0 \\
		B_{s+1}&A_{s+1}&B_{s+2}&\cdots&0\\
		0&B_{s+2}&A_{s+2}&\ddots&\vdots\\
		\vdots&\ddots&\ddots&\ddots&B_{s+m}\\
		0&\cdots&0&B_{s+m}&A_{s+m}
	\end{bmatrix},
\end{align*}
where the expected value is taken with respect to $f_{it}$. 
Let $\boldb_i$ denote the $i$-th column vector ($i = 1,\ldots, q$) of $\boldB=(\bm{b}_1,\ldots,\bm{b}_q)$.
Using 
\begin{eqnarray*}
	\frac{\partial \log\dtm{\boldB \boldV_{2t}\boldB' + \boldV_{1t}}}
	     {\partial h_{p+i,t}}
	&=& \boldb'_i (\boldB \boldV_{2t}\boldB' + \boldV_{1t})^{-1} \boldb_i \exp(h_{p+i,t}), 
\end{eqnarray*}
(the proof is given by Proposition 3 in Appendix \ref{sec:prop3}), it can be shown that 
\begin{eqnarray*}
	d_t 
	&=&\frac{s_0}{2}\boldb'_i (\boldB \boldV_{2t}\boldB' + \boldV_{1t})^{-1} \boldb_i \exp(h_{p+i,t})
	-\frac{k_0}{2}\boldb_i'\boldW_t^{-1}\boldb_i\exp(h_{p+i,t})\\
	&&-\frac{1}{2} +  \frac{(f_{it} - \mu_{f,it})^2}{2 \sigma_{f,it}^2} + \frac{(f_{it} - \mu_{f,it})}{\sigma_{f,it}^2}\frac{\partial \mu_{f,t,i}}{\partial h_{p+i,t}} + \frac{(f_{i,t-1} - \mu_{f,i,t-1})}{\sigma_{f,i,t-1}^2}\frac{\partial \mu_{f,t-1,i}}{\partial h_{p+i,t}} \\
	&&+\frac{\phi_{p+i} (h_{p+i,t+1}
		-(1-\phi_{p+i})\mu_{p+i}
		-\phi_{p+i}h_{p+i,t})}{\sigma^2_{p+i,\eta}}I(t=s+m<T),
\end{eqnarray*}
where
\begin{eqnarray*}
\frac{\partial \mu_{f,t,i}}{\partial h_{p+i,t}} &=& 
		\frac{\rho_{p+i}}{\sigma_{p+i,\eta}} \left\{
		-\phi_{p+i} + \frac{h_{p+i,t+1} - (1-\phi_{p+i})\mu_{p+i} - \phi_{p+i} h_{p+i,t}}{2}\right\}\exp\left(\frac{h_{p+i,t}}{2}\right)I(t<T),\\
\frac{\partial \mu_{f,i,t-1}}{\partial h_{p+i,t}} &= & 
\frac{\rho_{p+i}}{\sigma_{p+i,\eta}}\exp\left(\frac{h_{p+i,t-1}}{2}\right)I(t>1).
\end{eqnarray*}
Further,
\begin{align*}
	A_t &=\frac{s_0}{2}\left\{\boldb_i'(\boldB \boldV_{2t}\boldB' + \boldV_{1t})^{-1}\boldb_i\right\}^2 \exp(2h_{2t,i}) \\
	& \quad +\frac{1}{2} + \sigma_{f,it}^{-2} \left(\frac{\partial \mu_{f,it}}{\partial h_{p+i,t}}\right)^2 + \sigma_{f,i,t-1}^{-2}\left(\frac{\partial \mu_{f,i,t-1}}{\partial h_{p+i,t}} \right)^2  
	+ I(t= s+m < T) \phi_{p+i}^2 \sigma^{-2}_{\eta,p+i}, \\
	B_t &= \sigma_{f,t-1}^{-2} \frac{\partial \mu_{f,i,t-1}}{\partial h_{p+i,t-1}}\frac{\partial \mu_{f,i,t-1}}{\partial h_{p+i,t}}, \quad t=s+1,\ldots,s+m, \quad B_{s} = 0,
\end{align*}
using
\begin{eqnarray*}
\lefteqn{\frac{\partial^2 \log\dtm{\boldB \boldV_{2t}\boldB' + \boldV_{1t}}}
	     {\partial h_{p+i,t}^2}} && \\
	&=& \boldb'_i (\boldB \boldV_{2t}\boldB' + \boldV_{1t})^{-1} \boldb_i \exp(h_{2t,i}) -
	\left\{\boldb_i'(\boldB \boldV_{2t}\boldB' + \boldV_{1t})^{-1}\boldb_i\right\}^2 \exp(2h_{2t,i}) ,
\end{eqnarray*}
(the proof is given by Proposition 3 in Appendix \ref{sec:prop3}) and
\begin{eqnarray*}
E(\boldW_t^{-1}) & = & \frac{s_0}{k_0} (\boldB\boldV_{2t}\boldB' + \boldV_{1t})^{-1}.
\end{eqnarray*}
Let $\bm{h}_{p+i}$ denote $(h_{p+i,s},\ldots,h_{p+i,s+m})$. Then, via the Taylor expansion of $L$ around the conditional mode, $\hat{\bm{h}}_{p+i}$, we obtain
\begin{align*}
	&\log f(\eta_{p+i,s-1},\ldots,\eta_{p+i,s+m-1} \vert \cdot) =\mbox{const} - \frac{1}{2} \sum_{t=s-1}^{t=s+m-1}\eta_{p+i,t}^2 + L \\
	&\approx \mbox{const} - \frac{1}{2} \sum_{t=s-1}^{t=s+m-1}\eta_{p+i,t}^2 + \hat{L} + \hat{\boldd}' (\boldh_{p+i} - \hat{\boldh}_{p+i}) - \frac{1}{2} (\boldh_{p+i} - \hat{\boldh}_{p+i})'\hat{\boldQ}(\boldh_{p+i} - \hat{\boldh}_{p+i}) \\
	&\equiv \mbox{const} + \log f^*(\eta_{p+i,s-1},\ldots,\eta_{p+i,s+m-1} \vert \cdot).
\end{align*}
where $\boldd = (d_{s},\ldots,d_{s+m})'$ and $\hat{L},\hat{\boldd}$, and $\hat{\boldQ}$ are the values of $L,\boldd$, and $\boldQ$ evaluated at $\boldh_{p+i} = \hat{\boldh}_{p+i}$. $f^*$ is the posterior density of the disturbances for the linear Gaussian state space model in (\ref{eq:ssm1}), (\ref{eq:ssm2}) and (\ref{eq:ssm3}). We sample $\eta_{p+i,s-1},\ldots,\eta_{p+i,s+m-1}$ as follows:
\begin{enumerate}
	\item
	Set some initial value of $\hat{\boldh}_{p+i}$.
	\item Compute $\hat{d}_t,\hat{A}_t$, and $\hat{B}_t$ at $\hat{\boldh}_{p+i}$ for $t = s,\ldots, s+m$.
	\item
	Initialize $D_{s} = \hat{A}_{s} $, $J_{s} = 0$, and $b_{s} = \hat{d}_{s}$ and derive $D_t,J_t$, and $b_t$ recursively for $t=s+1,\ldots,s+m$:
	\begin{align*}
		&D_t = \hat{A}_t - \hat{B}_t D_{t-1}^{-1} \hat{B}_t,
		 \quad J_t = \hat{B}_t K_{t-1}^{-1},
		 \quad  b_t = \hat{d}_t - J_{t} K_{t-1}^{-1}b_{t-1},
	\end{align*}
	where $K_t = \sqrt{D_t}$ and set $J_{s+m+1} = 0$.
	\item
	Define $\hat{y}_t = \hat{\gamma}_t + D_t^{-1}b_t$, where
	\begin{align*}
		\hat{\gamma}_t = \hat{h}_{p+i,t} + K_{t}^{-1}J_{t+1}\hat{h}_{p+i,t+1} \quad
		t = s,\ldots,s+m.
	\end{align*}
	\item
	Construct the approximated linear Gaussian state space model given by
	\begin{eqnarray}
	\hat{y}_{i,t} &=& z_t h_{p+i,t} + \mathbf{G}_t \bm{\xi}_t, \quad t=s,\ldots,s+m, \label{eq:ssm1}\\
	h_{p+i,t+1} &=& (1 - \phi_{p+i})\mu_{p+i} + \phi_{p+i} h_{p+i,t} + \mathbf{H}_t \bm{\xi}_t, \quad t=s,\ldots,s+m-1, \label{eq:ssm2}\\
		&&\bm{\xi}_t = (\xi_{1t},\xi_{2t})' \sim \Normal(\boldzero, \boldI_2),\notag\\	h_{p+i,s} &=& 
	\left\{
	\begin{array}{ll}
	(1 - \phi_{p+i})\mu_{p+i} + \phi_{p+i} h_{p+i,s-1} + \mathbf{H}_{s-1} \bm{\xi}_{s-1}, & \quad s>1, \\
	\mu_{p+i}+ \mathbf{H}_0 \bm{\xi}_0, & \quad s=1, 
	\end{array}
	\right.	\label{eq:ssm3}
	\end{eqnarray}
where
	\begin{eqnarray*}
		z_t &=& 1 + \phi_{p+i}K_t^{-1} J_{t+1},\quad
		\mathbf{G}_t = K_{t}^{-1}[1, \sigma_{\eta,p+i}J_{t+1}], \quad t=1,\ldots,T,\\
		\mathbf{H}_t &=&
		\left\{
		\begin{array}{ll}
		 (0, \sigma_{\eta,p+i}), & t=1,\ldots,T-1, \\ 
		\left(0, \frac{\sigma_{\eta,p+i}}{\sqrt{1-\phi_{p+i}^2}}\right), & t=0.
		\end{array}
		\right.
	\end{eqnarray*}
	In order to find the posterior mode, we implement the Kalman filter and a disturbance smoother for (\ref{eq:ssm1}) -- (\ref{eq:ssm3})  and update $\hat{\boldh}_{p+i}$. We repeat Steps 2-5 several times or until some convergence criterion is met. Otherwise, we go to Step 6. 

	\item
	Generate a candidate $(\eta_{p+i,s-1}^n,\ldots,\eta_{p+i,s+m-1}^n)$ via a simulation smoother using models (\ref{eq:ssm1}) -- (\ref{eq:ssm3}) given $\hat{\boldh}_{p+i}$. These samples are generated from $f^*$.
	\item
	Conduct the MH algorithm where we accept a candidate $(\eta_{p+i,s-1}^n,\ldots,\eta_{p+i,s+m-1}^n)$ with probability
	\begin{align*}
		\min\left\{
		1,
		\frac{f(\eta_{p+i,s-1}^n,\ldots,\eta_{p+i,s+m-1}^n|\cdot)f^*(\eta_{p+i,s-1}^o,\ldots,\eta_{p+i,s+m-1}^o|\cdot)}
		{f(\eta_{p+i,s-1}^o,\ldots,\eta_{p+i,s+m-1}^o|\cdot)f^*(\eta_{p+i,s-1}^n,\ldots,\eta_{p+i,s+m-1}^n|\cdot)}
		\right\}
		,
	\end{align*}
	where $(\eta_{p+i,s-1}^o,\ldots,\eta_{p+i,s+m-1}^o)$ is a current sample.
\end{enumerate}

\section{Propositions}
\subsection{Proposition 1}
\label{sec:prop1}
\begin{enumerate}
\item[(i)]
\begin{eqnarray}
\frac{\partial \log \dtm{\boldB \boldV_{2t}\boldB' + \boldV_{1t}}}{\partial \boldbeta_i} &=&
2 \boldV_{2t} \boldB' (\boldB \boldV_{2t}\boldB' + \boldV_{1t})^{-1} \bolde_i,
\end{eqnarray}
where $\bolde_i$ denotes a $p \times 1$ vector with the $i-$th element equal to one and zero elements.

\item[(ii)]
\begin{eqnarray}
\frac{\partial^2 \log \dtm{\boldB \boldV_{2t}\boldB' + \boldV_{1t}}}{\partial \boldbeta_i\partial \boldbeta_i'} &=&
	2 d_{ii}
	\left\{
	\boldV_{2t} - \boldV_{2t} \boldB' (\boldB\boldV_{2t}\boldB' + \boldV_{1t})^{-1} \boldB \boldV_{2t}
	\right\} \nonumber\\
	&&
	- \frac{1}{2} \left[\frac{\partial \log \dtm{\boldB \boldV_{2t}\boldB' + \boldV_{1t}}}{\partial \boldbeta_i}\right]
	\left[\frac{\partial \log \dtm{\boldB \boldV_{2t}\boldB' + \boldV_{1t}}}{\partial \boldbeta_i}\right]',\nonumber\\
\end{eqnarray}
where $d_{ii}$ is the $(i,i)$-th element of $(\boldB \boldV_{2t}\boldB' + \boldV_{1t})^{-1}$.
\end{enumerate}

\noindent
{\bf Proof:}\\
(i) Let $\boldX_t = \boldB \boldV_{2t}\boldB' + \boldV_{1t}$. Then, using a chain rule and $\partial \log|\mathbf{X}|/\partial \mbox{vec}(\mathbf{X})=\mbox{vec}(\mathbf{X}^{-1\prime})$ (see, e.g., 17.29 and 17.52 of \cite{Seber(08)}), we obtain the first derivative
	\begin{align*}
		\frac{\partial \log|\boldX_t|}{\partial \boldbeta_i'}
		&= \frac{\partial \log\dtm{\boldX_t}}{\partial \vecm(\boldX_t)'}
		\times \frac{\partial\vecm(\boldX_t)}{\partial \vecm(\boldB')'}
		\times \frac{\partial \vecm(\boldB')}{\partial \boldbeta_i'} \\
		&= \vecm(\boldX_t^{-1\prime})'
		\times \frac{\partial \vecm (\boldB \boldV_{2t} \boldB')}{\partial \vecm(\boldB')'}
		\times (\bolde_i \otimes \boldI_q) \\
		&= \vecm(\boldX_t^{-1})'
		\times \left\{
		(\boldI_{p^2} + \boldK_{pp})(\boldI_{p} \otimes \boldB \boldV_{2t}) 
		\right\}
		\times (\bolde_i \otimes \boldI_q)
		\\
		&= 2 \vecm(\boldX_t^{-1})'
		(\boldI_{p} \otimes \boldB \boldV_{2t})
		(\bolde_i \otimes \boldI_q)\\
		&= 2 \vecm(\boldX_t^{-1})'
		(\bolde_i \otimes \boldB \boldV_{2t}).
	\end{align*}
where $\boldK_{mn}$ is a vec-permutation matrix such that $\mbox{vec}(\mathbf{A}')=\mathbf{K}_{mn}\mbox{vec}(\mathbf{A})$ for an $m\times n$ matrix $\mathbf{A}$. The third equality follows from $\partial \mbox{vec}(\mathbf{X}'\mathbf{A}\mathbf{X})/\partial \mbox{vec}(\mathbf{X})'=(\mathbf{I}_{n^2}+\mathbf{K}_{nn})(\mathbf{I}_n\otimes\mathbf{X}'\mathbf{A})$ for an $m\times n$ matrix $\mathbf{X}$ and an $m\times m$ symmetric matrix $\mathbf{A}$ (e.g., 17.30(f) of \cite{Seber(08)}). Therefore,
	\begin{align*}
		\frac{\partial \log \dtm{\boldX_{t}}}{\partial \boldbeta_i}
		&= 2 (\bolde_i^\prime \otimes \boldV_{2t}\boldB')
		\vecm(\boldX_t^{-1}) 
		= 2
		\boldV_{2t} \boldB' \boldX_{t}^{-1}\bolde_i,
		\end{align*}
	using $\mbox{vec}(\mathbf{AXC})=(\mathbf{C}'\otimes \mathbf{A})\mbox{vec}(\mathbf{X})$ (e.g., 11.16(b) of \cite{Seber(08)}).\\

\noindent 
(ii) From (i) and a product rule (e.g., 17.30(h) \cite{Seber(08)}),
\begin{eqnarray*}
\frac{\partial^2 \log \dtm{\boldX_{t}}}{\partial \boldbeta_i\partial \boldbeta_i'}
	 &=&
		2 \times \frac{\partial \boldV_{2t} \boldB' \boldX_{t}^{-1} \bolde_i}{\partial \boldbeta_i'}
		\\
		&=& 2 \times \left[
		\left\{
		(\bolde_i'\boldX_{t}^{-1}) \otimes \boldI_q
		\right\}
		\frac{\partial \vecm(\boldV_{2t}\boldB')}{\partial \boldbeta_i'}
		+ (1 \otimes \boldV_{2t}\boldB')
		\frac{\partial \vecm(\boldX_{t}^{-1}\bolde_i )}{\partial \boldbeta_i'}
		\right]
		.
\end{eqnarray*}
Since
		\begin{align*}
			\frac{\partial \vecm(\boldV_{2t}\boldB')}{\partial \boldbeta_i'}
			&= \frac{\partial \vecm(\boldV_{2t}\boldB')}{\partial \vecm(\boldB')'}
			\times \frac{\partial\vecm(\boldB')}{\partial\boldbeta_i'} 
			= (\boldI_p \otimes \boldV_{2t})(\bolde_i \otimes \boldI_q)
			= \bolde_i \otimes \boldV_{2t},
		\end{align*}
using $\partial \mbox{vec}(\mathbf{AXB})/\partial \vecm(\mathbf{X})' = \mathbf{B}' \otimes \mathbf{A}$ (e.g. 17.30(b) of \cite{Seber(08)}), the first term in the bracket is 
		\begin{align*}
			&\left\{
			(\bolde_i'\boldX_t^{-1}) \otimes \boldI_q
			\right\}
			\frac{\partial \vecm(\boldV_{2t}\boldB')}{\partial \boldbeta_i'}= (\bolde_i' \boldX_{t}^{-1} \bolde_i) \otimes \boldV_{2t}
			= d_{ii} \boldV_{2t}.
		\end{align*}
For the second term in the bracket, via a product rule and a chain rule,
		\begin{eqnarray}
			(1 \otimes \boldV_{2t}\boldB')
			\frac{\partial \vecm(\boldX_{t}^{-1}\bolde_i )}{\partial \boldbeta_i'}
			&=&\boldV_{2t} \boldB' \left\{
			(\bolde_i' \otimes \boldI_p) \frac{\partial\vecm(\boldX_{t}^{-1})}{\partial \boldbeta_i'} 
			\right\} \nonumber\\
			&=& \boldV_{2t}\boldB'(\bolde_i' \otimes \boldI_p) \times \frac{\partial \vecm(\boldX_t^{-1})}{\partial \vecm(\boldX_t)'}
			\frac{\partial \vecm(\boldX_t)}{\partial \boldbeta_i'} \nonumber \\
			&=& \boldV_{2t} \boldB' (\bolde_i' \otimes \boldI_p)
			\left[
			\left\{
			-(\boldX_t^{-1})' \otimes \boldX_t^{-1}
			\right\}
			\frac{\partial \vecm(\boldB\boldV_{2t}\boldB')}{\boldbeta_i'}
			\right]. \hspace{3mm}\mbox{}\label{eq:2nd_term}
		\end{eqnarray}
Noting that $\mathbf{K}_{pm}(\bm{b} \otimes \mathbf{A}) = \mathbf{A}\otimes \bm{b}$ (e.g.  11.19(c)(ii) of \cite{Seber(08)}),  we substitute 
		\begin{align*}
			\frac{\vecm(\boldB\boldV_{2t}\boldB')}{\boldbeta_i'}
			&= (\boldI_{p^2} + \boldK_{pp})(\boldI_p \otimes \boldB \boldV_{2t}) (\bolde_i \otimes \boldI_q) 
			= (\boldI_{p^2} + \boldK_{pp})(\bolde_i \otimes \boldB \boldV_{2t}) \\
			&= \bolde_i \otimes \boldB\boldV_{2t} + \boldB \boldV_{2t} \otimes \bolde_i,
		\end{align*}
into Equation (\ref{eq:2nd_term}). Then (\ref{eq:2nd_term}) reduces to
		\begin{align*}
			& -\boldV_{2t}\boldB'
			\left\{
			(\bolde_i' \boldX_t^{-1}\bolde_i)\otimes \boldX_t^{-1} \boldB \boldV_{2t}
			+(\bolde_i' \boldX_t^{-1} \boldB \boldV_{2t}) \otimes (\boldX_t^{-1} \bolde_i)
			\right\}\\
			&= -d_{ii} \boldV_{2t} \boldB' \boldX_{t}^{-1} \boldB \boldV_{2t}
			- \left(
			\bolde_i' \boldX_{t}^{-1}\boldB \boldV_{2t} \right) \otimes 
			\left(
			\boldV_{2t} \boldB' \boldX_{t}^{-1} \bolde_i
			\right) \\
			&= -d_{ii} \boldV_{2t} \boldB'  \boldX_{t}^{-1} \boldB \boldV_{2t}
			- \left(
			\boldV_{2t} \boldB' \boldX_{t}^{-1} \bolde_i
			\right)
			\left(
			\boldV_{2t} \boldB' \boldX_{t}^{-1} \bolde_i
			\right)' ,
		\end{align*}
and the result follows.
\begin{flushright}
$\square$
\end{flushright}

\subsection{Proposition 2}
\label{sec:prop2}
\begin{enumerate}
\item[(i)]
\begin{eqnarray}
\frac{\partial \log\dtm{\boldB \boldV_{2t}\boldB' + \boldV_{1t}}}{\partial h_{it}}
&=& d_{ii} \exp(h_{it}), \quad i=1,\ldots,p,
\end{eqnarray}
\item[(ii)]
\begin{eqnarray}
\frac{\partial^2 \log\dtm{\boldB \boldV_{2t}\boldB' + \boldV_{1t}}}{\partial h_{it}^2}= d_{ii} \exp(h_{it}) - d_{ii}^2\exp(2h_{it}), \quad i=1,\ldots,p.
\end{eqnarray}
where $d_{ii}$ is the $(i,i)$-th element of $(\boldB \boldV_{2t}\boldB' + \boldV_{1t})^{-1}$. 
\end{enumerate}

\noindent
{\bf Proof:}\\
(i) Let $\boldX_t = \boldB \boldV_{2t}\boldB' + \boldV_{1t}$. Then, using a chain rule as in the proof of Proposition 1,
	\begin{align*}
		\frac{\partial \log|\boldX_t|}{\partial h_{it}}
		&= \frac{\partial \log\dtm{\boldX_t}}{\partial \vecm(\boldX_t)'}
		\times \frac{\partial\vecm(\boldX_t)}{\partial h_{it}}
\\
		&= \vecm(\boldX_t^{-1\prime})'
		\times \vecm (\bm{e}_i\bm{e}_i')\times \exp(h_{it})
		\\
		&= \mbox{tr}(\boldX_t^{-1\prime}\bm{e}_i\bm{e}_i')\times \exp(h_{it})
		=d_{ii}\exp(h_{it}).
	\end{align*}
\noindent
(ii) Since
	\begin{align*}
        \frac{\partial d_{ii}}{\partial h_{it}}
		&=\vecm (\bm{e}_i\bm{e}_i')' \frac{\partial \vecm(\boldX_t^{-1})}{\partial h_{it}}
\\
		&=\vecm (\bm{e}_i\bm{e}_i') (-\mathbf{X}_t^{-1\prime}\otimes\mathbf{X}_t^{-1})
		\vecm (\bm{e}_i\bm{e}_i')\exp(h_{it})
\\
		&=-(\bm{e}_i'\otimes \bm{e}_i') (\mathbf{X}_t^{-1}\otimes\mathbf{X}_t^{-1})
		(\bm{e}_i\otimes \bm{e}_i)\exp(h_{it})
\\
		&=-(\bm{e}_i'\mathbf{X}_t^{-1}\bm{e}_i\otimes \bm{e}_i'\mathbf{X}_t^{-1}\bm{e}_i)\exp(h_{it})=-d_{ii}^2\exp(h_{it}),
	\end{align*}
using $\partial \vecm(\boldX^{-1})/\partial \vecm(\boldX)'=-\mathbf{X}^{-1\prime}\otimes\mathbf{X}^{-1}$ (e.g. 17.30(d) of \cite{Seber(08)}), we obtain
	\begin{align*}
		\frac{\partial^2 \log|\boldX_t|}{\partial h_{it}^2}
		&= \frac{\partial d_{ii}}{\partial h_{it}}\exp(h_{it})+d_{ii}\exp(h_{it})
		=d_{ii}\exp(h_{it})-d_{ii}^2\exp(2h_{it}),
	\end{align*}
and the result follows.
\begin{flushright}
$\square$
\end{flushright}
\subsection{Proposition 3}
\label{sec:prop3}
\begin{enumerate}
\item[(i)]
\begin{eqnarray}
	\frac{\partial \log\dtm{\boldB \boldV_{2t}\boldB' + \boldV_{1t}}}
	     {\partial h_{p+i,t}}
	= \boldb'_i (\boldB \boldV_{2t}\boldB' + \boldV_{1t})^{-1} \boldb_i \exp(h_{p+i,t}), 
	\quad i=1,\ldots,q,
\end{eqnarray}
\item[(ii)]
\begin{eqnarray}
\lefteqn{\frac{\partial^2 \log\dtm{\boldB \boldV_{2t}\boldB' + \boldV_{1t}}}
	     {\partial h_{p+i,t}^2}} &&\nonumber \\
	&=& \boldb'_i (\boldB \boldV_{2t}\boldB' + \boldV_{1t})^{-1} \boldb_i \exp(h_{2t,i}) -
	\left\{\boldb_i'(\boldB \boldV_{2t}\boldB' + \boldV_{1t})^{-1}\boldb_i\right\}^2 \exp(2h_{2t,i}) , \nonumber\\
&& i=1,\ldots,q.
\end{eqnarray}
\end{enumerate}

\noindent
{\bf Proof:}\\
(i) Let $\boldX_t = \boldB \boldV_{2t}\boldB' + \boldV_{1t}$. Then, as in the proof of Proposition 2,
	\begin{align*}
		\frac{\partial \log|\boldX_t|}{\partial h_{p+i,t}}
		&= \frac{\partial \log\dtm{\boldX_t}}{\partial \vecm(\boldX_t)'}
		\times \frac{\partial\vecm(\boldX_t)}{\partial h_{p+i,t}}
\\
		&= \vecm(\boldX_t^{-1\prime})'
		\times \frac{\partial\vecm{(\boldB \boldV_{2t}\boldB')}}{\partial h_{p+i,t}}
\\
		&= \vecm(\boldX_t^{-1})'
		\times \vecm (\bm{b}_i\bm{b}_i')\times \exp(h_{p+i,t})
		\\
		&= \mbox{tr}(\boldX_t^{-1}\bm{b}_i\bm{b}_i')\times \exp(h_{p+i,t})
		=\bm{b}_i'\boldX_t^{-1}\bm{b}_i\exp(h_{p+i,t}).
	\end{align*}
\noindent
\noindent
(ii) Similar to the proof in Proposition 2, using
	\begin{align*}
        \frac{\partial \bm{b}_i'\boldX_t^{-1}\bm{b}_i}{\partial h_{p+i,t}}
		&=\vecm (\bm{b}_i\bm{b}_i')' \frac{\partial \vecm(\boldX_t^{-1})}{\partial h_{p+i,t}}
\\
		&=\vecm (\bm{b}_i\bm{b}_i') (-\mathbf{X}_t^{-1\prime}\otimes\mathbf{X}_t^{-1})
		\vecm (\bm{b}_i\bm{b}_i')\exp(h_{p+i,t})
\\
		&=\{\bm{b}_i'\boldX_t^{-1}\bm{b}_i\}^2\exp(h_{p+i,t}),
	\end{align*}
and the result follows.
\begin{flushright}
$\square$
\end{flushright}

\end{document}


\title{Dynamic factor, leverage and realized covariances \\ in multivariate stochastic volatility: Supplementary material}
\author{Yuta Yamauchi and Yasuhiro Omori}
\maketitle

\customlabel{eq:c_t}{32}
\setcounter{equation}{45}

\appendix
\section{MCMC algorithm for sampling other parameters}
\subsection{Joint posterior density}
The logarithm of the joint posterior density of $\bm{\theta}$ is given by
\begin{align*}
\lefteqn{\log\pi(\bm{\theta}|\bm{x}, \bm{y})} & \\
	&= \mbox{const}+\sum_{t=1}^{T} \frac{s_0}{2}\log \dtm{\boldB \boldV_{2t}\boldB' + \boldV_{1t}} -\frac{k_0}{2}\sum_{t=1}^{T} \tr\left\{ \left(\boldB \boldV_{2t}\boldB' +\boldV_{1t}\right)\boldW_t^{-1} \right\} \\
	&\quad -\frac{1}{2}\sum_{t=1}^{T}\log\dtm{\boldV_{1t}}
	-\frac{1}{2} \sum_{t=1}^{T} (\boldy_t - \boldB \boldf_t)'\boldV_{1t}^{-1}(\boldy_t - \boldB \boldf_t) \\
	&\quad -\frac{T-1}{2}\log\dtm{\boldSigma_{\eta\eta}} 
	-\frac{1}{2} \sum_{t=1}^{T-1} \{\boldh_{t+1} - (\boldI - \boldPhi)\boldmu - \boldPhi\boldh_{t}\}'
	\boldSigma_{\eta\eta}^{-1}
	\{\boldh_{t+1} - (\boldI - \boldPhi)\boldmu - \boldPhi\boldh_{t}\} \\
	&\quad -\frac{1}{2}\log\dtm{\boldSigma_{h,0}} 
	-\frac{1}{2}(\boldh_{1} -\boldmu)'\boldSigma_{h,0}(\boldh_{1}- \boldmu)\\
	&\quad -\frac{1}{2}\sum_{t=1}^{T}\log\dtm{\boldV_{2t\vert h}}
	-\frac{1}{2} \sum_{t=1}^{T} (\boldf_t - \boldgamma - \boldPsi (\boldf_{t-1} - \boldgamma) - \boldc_{t})'\boldV_{2t\vert h}^{-1}(\boldf_t - \boldgamma - \boldPsi (\boldf_{t-1} - \boldgamma) - \boldc_{t}) \\
	&\quad -\frac{1}{2}\sum_{t=1}^{T}\log\dtm{\boldSigma_{\nu}}
	-\frac{1}{2}
	\sum_{t=1}^{T}(\boldx_t-\boldA \boldf_t)' \boldSigma_{\nu}^{-1}(\boldx_t-\boldA \boldf_t) + \log\pi(\bm{\theta}),
	\end{align*}
where $\pi(\bm{\theta})$ denotes a prior density of $\bm{\theta}$.

\subsection{Generation of $\boldf_t$}
To sample $(\boldf_1,\ldots, \boldf_T)$ simultaneously, we consider the following state space equations:
\begin{align*}
	& \left(\begin{array}{c} \boldx_t \\ \boldy_t \end{array}\right)
	= \left(\begin{array}{c} \boldA \\ \boldB \end{array}\right) \boldf_t 
	+ \left(\begin{array}{c} \boldnu_t \\ \boldV_{1t}^{1/2} \boldepsilon_{1t} \end{array}\right),
	\\
	& \boldf_t = (\mathbf{I}_q-\mathbf{\Psi})\boldgamma + \boldc_t + \mathbf{\Psi}\boldf_{t-1} + \boldV_{2t\vert h}^{1/2} \boldepsilon_{2t}, \quad \mathbf{\Psi}=\mbox{diag}(\bm{\psi}),
\end{align*}
where $\bm{c}_{t}$ is defined in Equation (\ref{eq:c_t}).
We implement the simulation smoother for the linear and Gaussian state space model.
\subsection{Generation of  $\boldalpha$}
The conditional posterior density of $\boldalpha$ is given by
\begin{align}
\pi(\boldalpha|\cdot) 
	& \propto \exp\left[-\frac{1}{2}\sum_{j=2}^q(\boldalpha_j -\boldm_{\alpha_j})'\boldS_{\alpha_j}^{-1}(\boldalpha_j -\boldm_{\alpha_j})
	-\frac{1}{2}
	\sum_{t=1}^{T}(\boldA \boldf_t - \boldx_t)' \boldSigma_{\nu}^{-1}(\boldA \boldf_t - \boldx_t)
	\right] \nonumber \\
	& \propto \prod_{j=2}^q\exp\left[-\frac{1}{2}(\boldalpha_j -\boldm_{\alpha_j})'\boldS_{\alpha_j}^{-1}(\boldalpha_j -\boldm_{\alpha_j})
	-\frac{1}{2\sigma_{\nu,j}^2}
	\sum_{t=1}^{T}\{\boldalpha_j'\boldf_{jt}^* -(x_{jt}-f_{jt})\}^2
	\right]
\end{align}
where $\boldf_{jt}^*=(f_{1t},f_{2t},\ldots,f_{j-1,t})'$.
Noting that $\bm{\alpha}_j$s are conditionally independent, 
we generate $\boldalpha_j$ ($j=2,\ldots,q$) from $\Normal(\hat{\boldm}_{\alpha_j}, \hat{\boldSigma}_{\alpha_j})$ where
\begin{align*}
	&\hat{\boldm}_{\alpha_j} = \hat{\boldSigma}_{\alpha} \left\{
	\sum_{t=1}^{T}\left(
	\boldf^*_{jt}\sigma_{\nu,j}^{-2}(x_{jt} - f_{jt})
	\right)+
	\boldS_{\alpha_j}^{-1}\boldm_{\alpha_{j}}
	\right\}, \quad
	\hat{\boldSigma}_{\alpha_j}^{-1} = \sum_{t=1}^{T} \sigma_{\nu,j}^{-2}\boldf_{jt}^*\boldf_{jt}^{*\prime} + \boldS_{\alpha_j}^{-1}.
\end{align*}
\subsection{Generation of $\boldmu$}
The conditional posterior density of $\boldmu$ is given by
\begin{align}
\pi(\boldmu|\cdot)
	& \propto \exp\left[-\frac{1}{2}(\boldmu -\boldm_{\mu})'\boldS_\mu^{-1}(\boldmu -\boldm_{\mu})-\frac{1}{2}(\boldh_1 -\boldmu)'\boldSigma_{h,0}^{-1}(\boldh_1 -\boldmu)\right]
   \nonumber\\
	& \times \exp\left[-\frac{1}{2}\sum_{t=1}^{T-1}\{\boldh_{t+1} - (\boldI - \boldPhi) \boldmu -  \boldPhi\boldh_{t} - \boldc_t^*\}'
	\boldSigma_{\eta\vert\epsilon}^{-1}
	\{\boldh_{t+1} - (\boldI - \boldPhi) \boldmu -  \boldPhi\boldh_{t} - \boldc_t^*\}'\right],
\end{align}
where $\bm{c}_t^*=(c_{1t}^*,\ldots,c_{p+q,t}^*)'$ (see \ref{sec:mcmc_phi} for the definition of $c_{it}^*$). 
We generate $\boldmu$ from $\Normal(\hat{\boldm}_\mu,\hat{\boldSigma}_\mu)$ where
\begin{align*}
	&\hat{\boldm}_{\mu} = \hat{\boldSigma}_{\mu}\times\left[\boldS_\mu^{-1}\boldm_{\mu} + \boldSigma_{h,0}^{-1}\boldh_1 + (\boldI - \boldPhi)\boldSigma_{\eta\vert\epsilon}^{-1}\sum_{t=1}^{T-1}\left\{\boldh_{t+1} - \boldPhi\boldh_t - \boldc_t^* \right\} \right]\\
	&\hat{\boldSigma}_{\mu}^{-1} = \boldS_{\mu}^{-1} + \boldSigma_{h,0}^{-1} + (T-1)(\boldI_{p+q} - \boldPhi)\boldSigma_{\eta\vert\epsilon}^{-1}(\boldI_{p+q} - \boldPhi), \\
& \hspace{10mm}\boldSigma_{\eta\vert\epsilon}=\mbox{diag}\left(\sigma_{\eta,1}^2,\ldots,\sigma_{\eta,p}^2, \sigma_{\eta,p+1}^2(1-\rho_{p+1}^2),\ldots,\sigma_{\eta,p+q}^2(1-\rho_{p+q}^2)\right).
\end{align*}
\subsection{Generation of $\boldgamma$}
The conditional posterior density of $\boldgamma$ is given by
\begin{align}
\pi(\boldgamma|\cdot) 
	& \propto \exp\left[-\frac{1}{2}(\boldgamma -\boldm_{\gamma})'\boldS_\gamma^{-1}(\boldgamma -\boldm_{\gamma})\right]\nonumber \\
	& \times \exp\left[-\frac{1}{2}\sum_{t=2}^{T}\{\boldf_{t} - (\boldI - \boldPsi) \boldgamma -  \boldPsi\boldf_{t-1} - \boldc_t\}'
	\boldV_{2t\vert h}^{-1}
	\{\boldf_{t} - (\boldI - \boldPsi) \boldgamma -  \boldPsi\boldf_{t-1} - \boldc_t\}\right],
\end{align}
where $\bm{c}_{t}$ is defined in Equation (\ref{eq:c_t}).
We generate $\boldgamma$ from $\Normal(\hat{\boldm}_\gamma,\hat{\boldSigma}_\gamma)$ where
\begin{align*}
	&\hat{\boldm}_{\gamma} = \hat{\boldSigma}_{\gamma}\times\left[\boldS_{\gamma}^{-1}\boldm_{\gamma} + (\boldI_q - \boldPsi)\sum_{t=2}^{T}\boldV_{2t\vert h}^{-1}\left\{\boldf_{t} - \boldPsi\boldf_{t-1} - \boldc_t \right\} \right]\\
	&\hat{\boldSigma}_{\gamma}^{-1} = \boldS_{\gamma}^{-1} + \sum_{t=2}^{T}(\boldI_{q} - \boldPsi)\boldV_{2t\vert h}^{-1}(\boldI_{q} - \boldPsi).
\end{align*}
\subsection{Generation of $\boldphi$}
\label{sec:mcmc_phi}
The posterior distributions of $\boldphi=(\phi_1,\ldots,\phi_{p+q})'$ are conditionally independent and the conditional posterior density of $\phi_i$ is given by
\begin{align}
\pi(\phi_i|\cdot) & \propto &  k(\phi_i)\times
	\exp\left[-\frac{1}{2\sigma_{\eta,i}^2(1-\rho_i^2)}\sum_{t=1}^{T-1}\{h_{i,t+1} - \mu_i -  \phi_i(h_{it} - \mu_i) - c_{it}^*\}^2
\right],
\end{align}
where $\rho_1=\ldots=\rho_p=0$, $c_{1t}^*=\ldots=c_{pt}^*=0$ and 
\begin{align*}
&	k(\phi_i) = \sqrt{1-\phi_i^2}\left(\frac{1+\phi_i}{2}\right)^{a_\phi - 1}\left(\frac{1+\phi_i}{2}\right)^{b_\phi - 1}, \quad i=1,\ldots,p+q,
\\
	&c_{p+i,t}^* =\rho_{p+i}\sigma_{\eta,p+i}\exp(-h_{p+i,t}/2)\{f_{it}-\gamma_i-\psi_i(f_{i,t-1}-\gamma_i)\}, & i = 1,\ldots,q,
\end{align*}
Let $\phi_i^o$ denote a current sample of $\phi_i$. Then we generate $\phi_i^{n}$ from $\Normal(m_{\phi,i},\sigma^2_{\phi,i})$, where
\begin{align*}
	&m_{\phi,i} = \frac{\sum_{t=1}^{T-1}(h_{i,t+1} - \mu_i-c_{it}^*)(h_{it} - \mu_i)}{\sum_{t=2}^{T-1}(h_{it} - \mu_i)^2}, \quad
	\sigma_{\phi,i}^2 = \frac{\sigma_{\eta,i}^{2}(1-\rho_i^2)}{\sum_{t=2}^{T-1}(h_{it} - \mu_i)^2},
\end{align*}
and accept it with probability $\min\{1, k(\phi_i^n)/k(\phi_i^o)\}$.
\subsection{Generation of $\boldpsi$}
The posterior distributions of $\boldpsi=(\psi_1,\ldots,\psi_{q})'$ are conditionally independent and the conditional posterior density of $\psi_i$ is given by
\begin{align}
\pi(\psi_i|\cdot) & \propto k(\psi_i)\times\exp\left[-\frac{1}{2}\sum_{t=2}^{T}
	\frac{\{f_{it} - \gamma_i -  \psi_i(f_{i,t-1} - \gamma_i) - c_{it}\}^2}
	{\{1-\rho_{p+i}^2I(t<T)\}\exp(h_{p+i,t})}\right]
\end{align}
where $c_{it}$ is defined in Equation (\ref{eq:c_t}), and 
\begin{align*}
	k(\psi_i) = \left(\frac{1+\psi_i}{2}\right)^{a_\psi - 1}\left(\frac{1-\psi_i}{2}\right)^{b_\psi - 1}, \quad i=1,\ldots,q.
\end{align*}
Let $\psi_i^o$ denote a current sample of $\psi_i$. Then we generate $\psi_i^{n}$ from $\Normal(m_{\psi,i},\sigma^2_{\psi,i})$, where
\begin{align*}
	&m_{\psi,i} = \sigma_{\psi,i}^{2}
	\sum_{t=2}^{T}
\frac{(f_{i,t} - \gamma_i-c_{it})(f_{i,t-1} - \gamma_i)}{\{1-\rho_{p+i}^2I(t<T)\}\exp(h_{p+i,t})}, \quad
	\sigma_{\psi,i}^{-2} =  \sum_{t=2}^{T}\frac{(f_{i,t-1} - \gamma_i)^2}{\{1-\rho_{p+i}^2I(t<T)\}\exp(h_{p+i,t})},
\end{align*}
and accept it with probability $\min\{1, k(\psi_i^n)/k(\psi_i^o)\}$.
\subsection{Generation of $\bm{\rho}$}
The conditional posterior density of $\rho_{k}$ $(k=p+1,\ldots,p+q)$ is given by
\begin{eqnarray*}
\pi(\rho_{k}|\cdot) & \propto &
\left(\frac{1+\rho_{k}}{2}\right)^{a_\rho - 1}\left(\frac{1+\rho_{k}}{2}\right)^{b_\rho - 1} (1-\rho_{k}^2)^{-(T-1)/2}\\
&& \times
	\exp\left[-\frac{1}{2\sigma_{\eta,k}^2(1-\rho_{k}^2)}
	\sum_{t=1}^{T-1}\{h_{k,t+1} - \mu_k - \phi_{k}(h_{k,t} - \mu_k) - c_{kt}^*\}^2
   \right], 
\end{eqnarray*}
where  $c_{kt}^*$ is defined in \ref{sec:mcmc_phi}.
In order to sample $\rho_{k}$, we consider Fisher transformation, $g_{k} = \log(1 + \rho_{k}) - \log(1 - \rho_{k})$ (or $\rho_{k}=\{\exp(g_{k})-1\}/\{\exp(g_{k})+1\})$. Let $l(g_k)$ denote the logarithm of the posterior density of $g_{k}$ given by
\begin{align*}
	l(g_{k}) &\equiv \log \pi(\rho_{k} \vert \cdot) + \log(J(g_{k})), \quad k=p+1,\ldots,p+q,
\end{align*}
where $J(g_{k})$ is Jacobian such that
\begin{align*}
	&J(g_{k}) = \frac{2\exp(g_{k})}{(\exp(g_{k}) + 1)^2}.
\end{align*}
First, we compute the mode $\hat{g}_k$ of $l(g_k)$ numerically. Then we consider 
Taylor expansion of $l(g_k)$ around $\hat{g}_k$ :
\begin{align*}
	l(g_k) \approx l(\hat{g}_k) + m_{g_k}(g_k - \hat{g}_k) - \frac{(g_k - \hat{g}_k)^2}{2S_{g_k}}\equiv l^*(g_k),
\end{align*}
where
\begin{align*}
	&m_{g_k} = \left.\frac{\partial l(g_k)}{\partial g_k} \right|_{g_k = \hat{g}_k}, \quad
	S_{g_k}^{-1} = -\left.\frac{\partial^2 l(g_k)}{\partial g_k^2} \right|_{g_k = \hat{g}_k}.
\end{align*}
To conduct MH algorithm, we propose a candidate $g_k^n$ from $\Normal(\hat{g}_k + S_{g_k}m_{g_k}, S_{g_k})$ and accept it with probability $\{1,\exp(l(g_k^n)-l(g_k^o)-l^*(g_k^n)+l^*(g_k^o) )\}$ where $g_k^o$ is a current sample.
\subsection{Generation of $\bm{\sigma}_{\eta}$}
First we define $\tilde{n}_{\eta,i} = n_{\eta} +T$ and
\begin{align*}
	&\tilde{d}_{\eta,i} = d_\eta + \sum_{t=1}^{T-1} \left\{h_{i,t+1} - (1-\phi_i) \mu_i - \phi h_{i,t}\right\}^2+(1-\phi_i^2)(h_{i1}-\mu_i)^2,
\end{align*}
for $i=1,\ldots,p+q$.\\

\noindent
{\it Generation of $\sigma_{\eta,1}^2,\ldots,\sigma_{\eta,p}^2$}. We generate $\sigma_{\eta,i}^2$ from the conditional posterior distribution, $\sigma_{\eta,i}^2 \sim \IG(\tilde{n}_{\eta,i}/2,\tilde{d}_{\eta,i}/2)$, $i=1,\ldots,p$.\\

\noindent
{\it Generation of $\sigma_{\eta,p+1}^2,\ldots,\sigma_{\eta,p+q}^2$}.
The conditional posterior density for $\sigma_{\eta,p+i}^2$  is given by
\begin{eqnarray}
\pi(\sigma_{\eta,p+i}^2|\cdot)&\propto&
	 (\sigma_{\eta,p+i}^2)^{-\frac{\tilde{n}_{\eta,i}}{2}-1}
	\exp\left[-\frac{\tilde{d}_{\eta,i}}{2\sigma_{\eta,p+i}^2}+k(\sigma_{\eta,p+i}^{2})\right],
	\quad i=1,\ldots,q,\\
&& \hspace{10mm}k(\sigma_{\eta,p+i}^{2}) =	-\frac{1}{2}
	\sum_{t=1}^{T-1}\frac{(f_{it} - \gamma - \psi_i(f_{i,t-1} - \gamma) - c_{it})^2}{(1-\rho_{p+i}^2)\exp(h_{p+i,t})},\nonumber
\end{eqnarray}
where  $c_{it}$ is defined in Equation (\ref{eq:c_t}).
We conduct MH algorithm to sample from its conditional posterior distribution. Let $\sigma_{\eta,p+i}^{2,o}$ denote a current sample, and we propose a candidate $\sigma_{\eta,p+i}^{2,n} \sim \IG(\tilde{n}_{\eta,i}/2,\tilde{d}_{\eta,i}/2)$. Accept it with probability $\min\{1,\exp(k(\sigma_{\eta,p+i}^{2,n})-k(\sigma_{\eta,p+i}^{2,o}))\}$.
\subsection{Generation of $\bm{\sigma}_{\nu}$}
Let $\tilde{x}_{jt}$ denote the $j$-th element of $\bm{x}_t-\mathbf{A}\bm{f}_t$.
We generate $\sigma_{\nu,j}^2$ from the conditional posterior distribution, $\sigma_{\nu,j}^2 \sim \IG(\tilde{n}_{\nu,j}/2,\tilde{d}_{\nu,j}/2)$, $j=1,\ldots,q$ where $\tilde{n}_{\nu,j} = n_{\nu} +T$ and $\tilde{d}_{\nu,j} = d_\nu + \sum_{t=1}^{T}\tilde{x}_{jt}^2$.

\subsection{Generation of $\delta$}
The conditional posterior density of $\delta$ is
\begin{align*}
\pi(\delta \vert \cdot)
&=\frac{1}{2^\frac{(\delta+p+3)p}{2}\Gamma_p(\frac{\delta+p+3}{2})}\dtm{(\delta+2) (\boldB\boldV_{2t}\boldB' + \boldV_{1t})}^{\frac{\delta+p+3}{2}}
\times \dtm{\boldW_t}^{-\frac{\delta+2p+4}{2}} \\
&\quad\times \exp\left[-\frac{1}{2} \tr \left\{(\delta+2) (\boldB\boldV_{2t}\boldB' + \boldV_{1t}\right)\boldW_t^{-1} \} \right].
\end{align*}
We generate $\log \delta$ using random-walk MH algorithm. Given the current value of $\log \delta^o$, we generate $\log \delta^n \sim N(\log \delta^o,\sigma_{\delta}^2)$. The $\sigma_{\delta}$ is a tuning parameter to optimize the sampling efficiency. In our empirical study,  we set $\sigma_{\delta}=0.001$ and the acceptance rate was 0.898.
\section{Figures}
\subsection{The effect of the realized measures on the estimated dynamic correlations}
\begin{figure}[H]
	\centering
	\includegraphics[width=10cm]{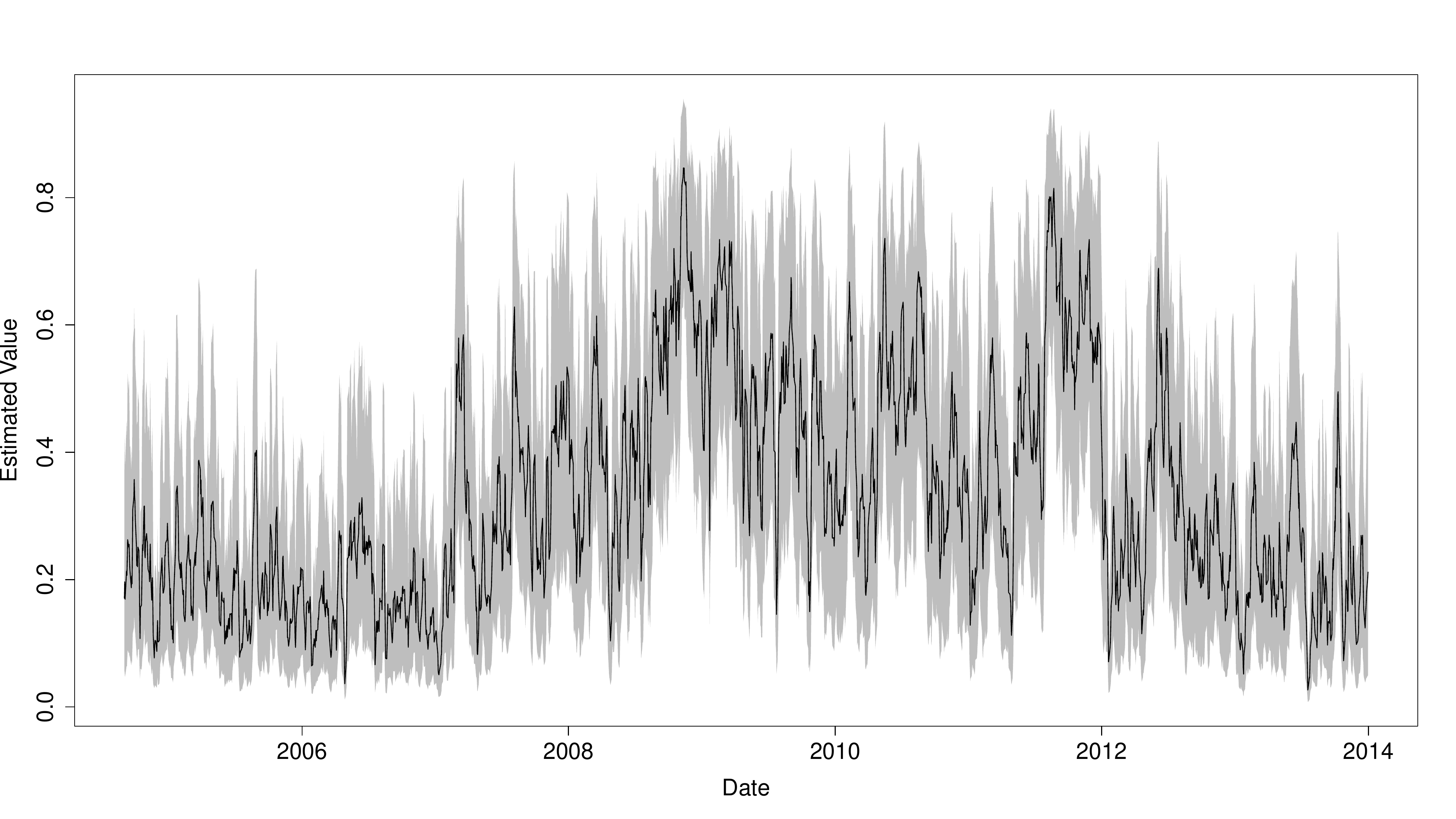}
	\caption{FMSV model. The posterior mean (solid line) and 95\% credible interval (shaded area) of estimated dynamic correlation coefficients between AAPL and MSFT.}
	\label{fig:corr_norcov}
\end{figure}

\begin{figure}[H]
	\centering
	\includegraphics[width=10cm]{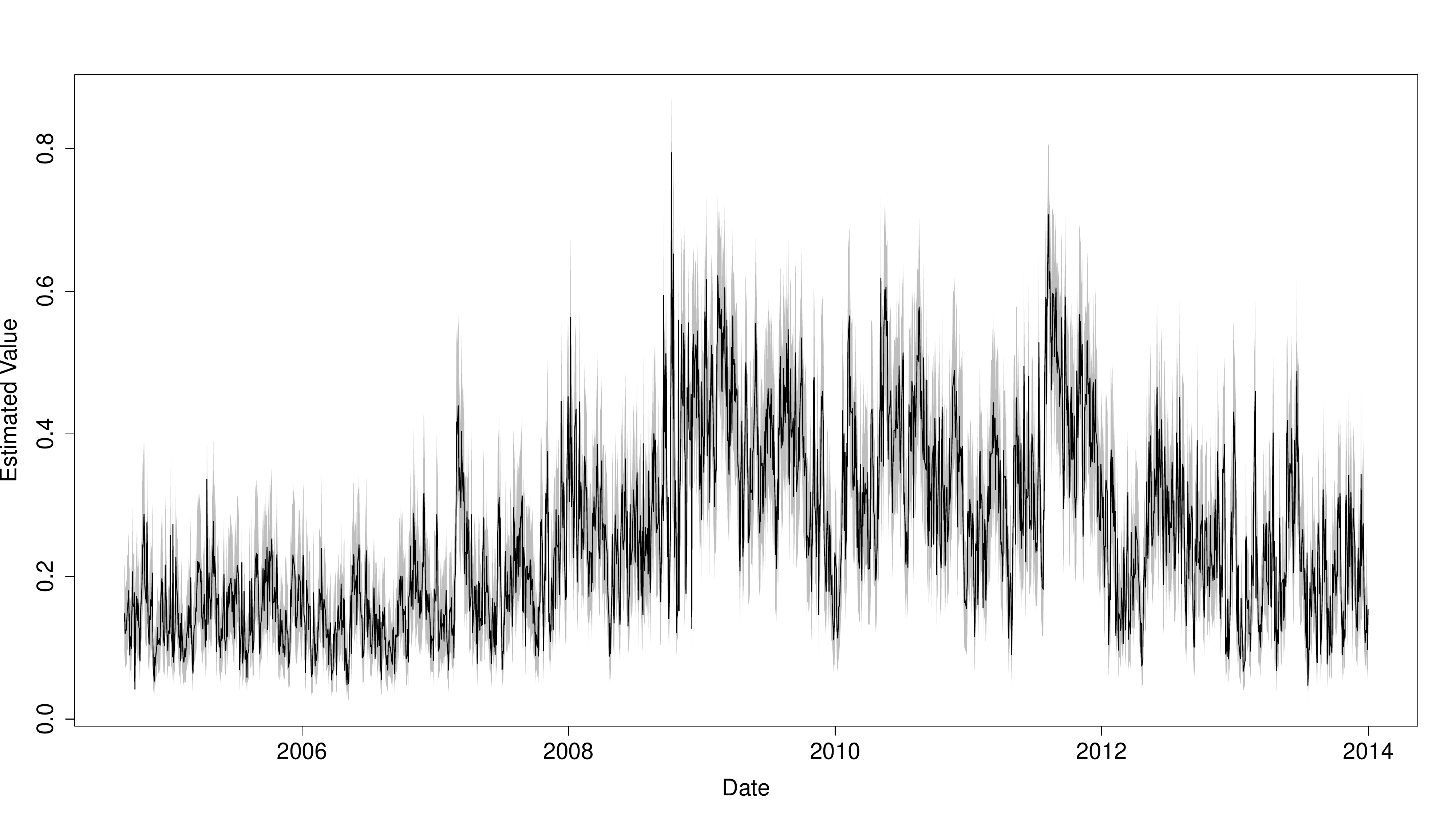}
	\caption{FMRSV model. The posterior mean (solid line) and 95\% credible interval (shaded area) of estimated dynamic correlation coefficients between AAPL and MSFT.}
	\label{fig:corr}
\end{figure}

Figures \ref{fig:corr_norcov} and \ref{fig:corr}  show the estimated dynamic correlation coefficients between AAPL and MSFT. There are rises in the correlation coefficients among stocks around the period of the global financial crisis.

\begin{figure}[H]
	\centering
	\includegraphics[width=10cm]{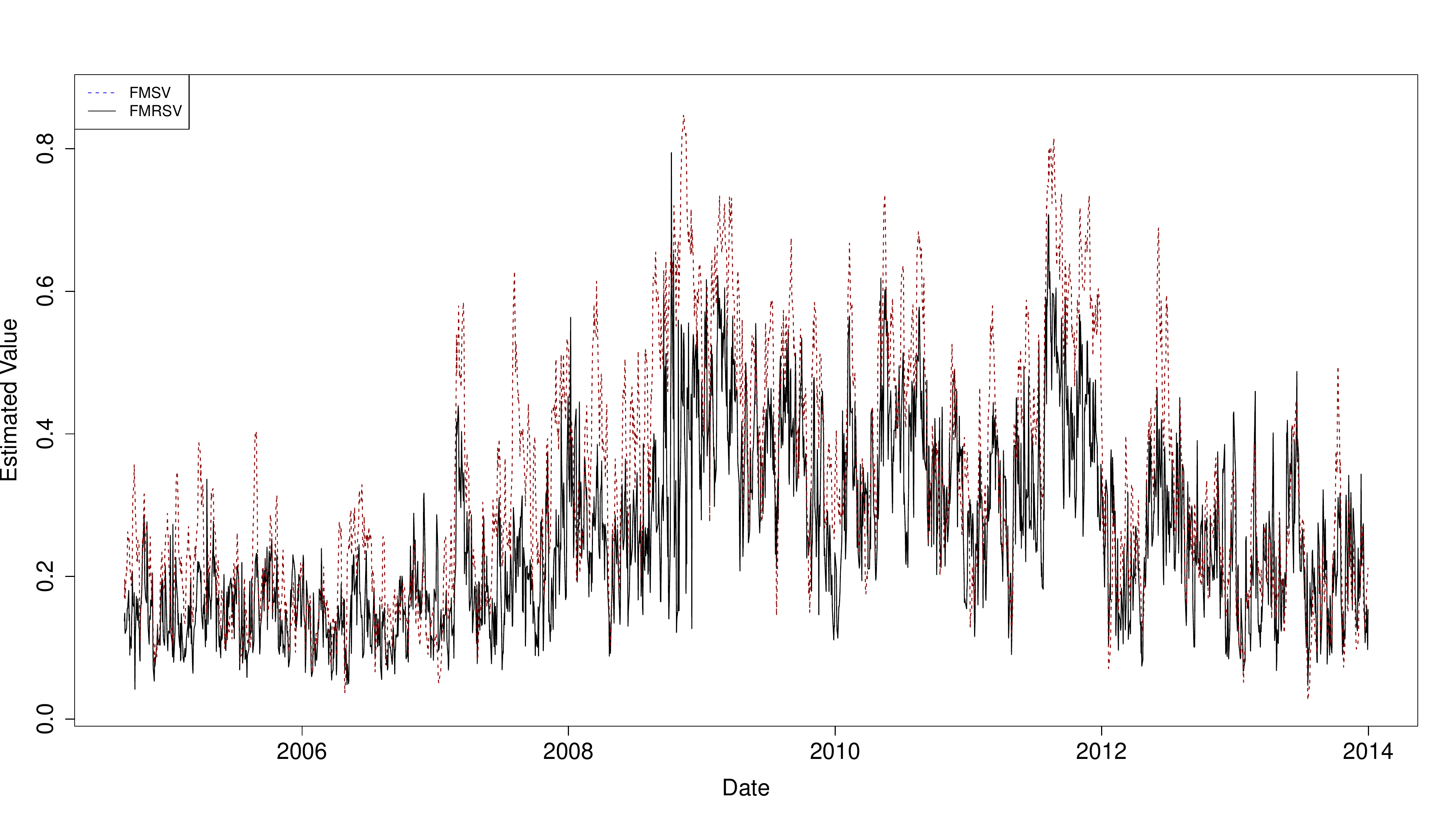}
	\caption{FMSV (black solid line) and FMRSV (red dashed line) models. The posterior mean of estimated dynamic correlation coefficients between AAPL and MSFT.}
	\label{fig:corr_lines}
\end{figure}

%
\noindent
Figure \ref{fig:corr_lines} shows the correlation coefficients in the FMSV and FMRSV models simultaneously. The estimates of the FMSV model are sometimes larger than those of the FMRSV model.

\begin{figure}[H]
	\centering
	\includegraphics[width=12cm]{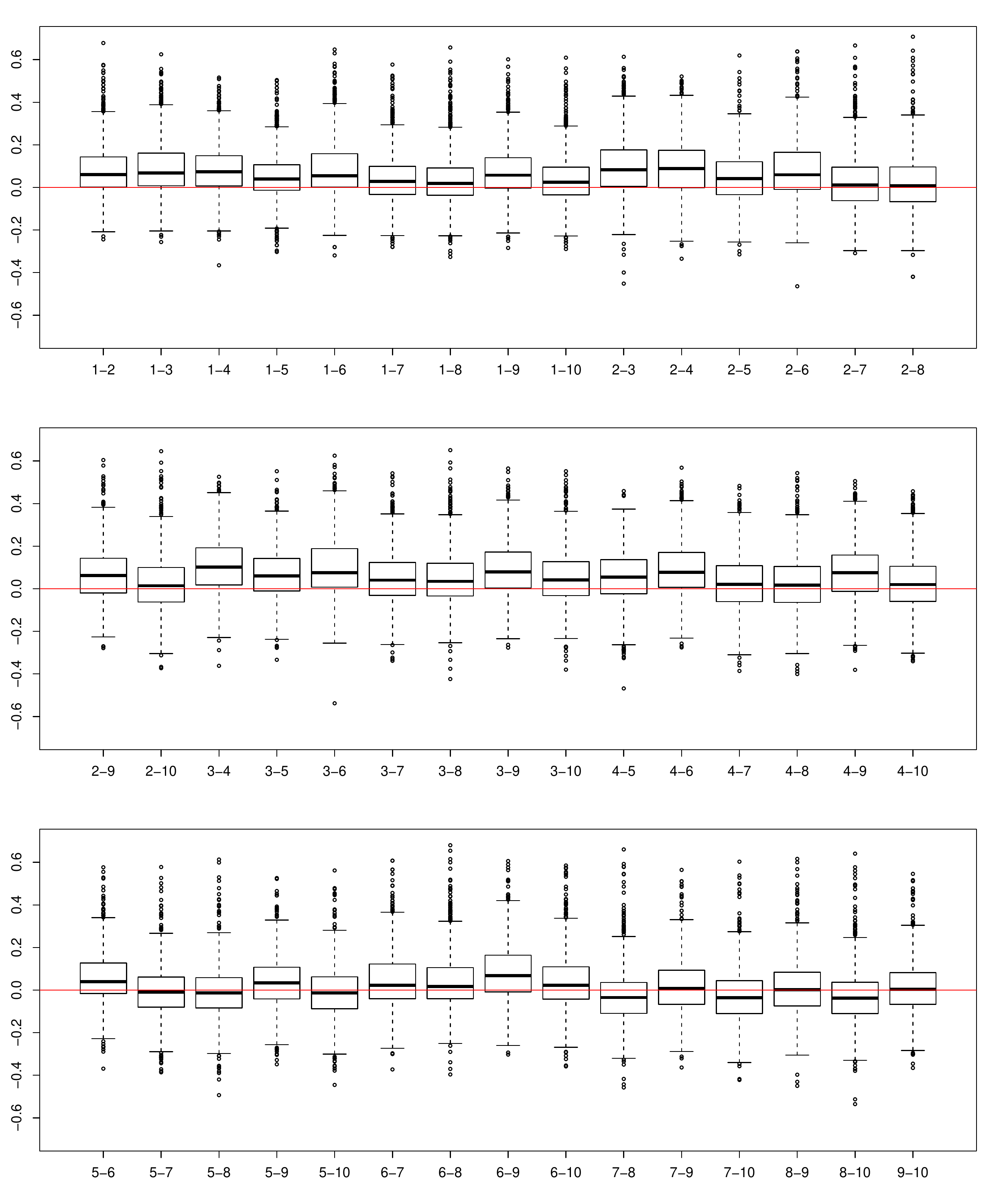}
	\caption{Boxplots of the difference between dynamic correlation estimates (the estimate of FMSV model minus the estimate of FMRSV model).}
	\label{fig:corr_boxplot}
\end{figure}

\noindent
Further, Figure \ref{fig:corr_boxplot} shows the boxplots of the difference between the estimated dynamic correlations in the FMSV and FMRSV models. The estimates of the FMSV model tend to be larger than those of the FMRSV model.

\section{Simulation study}
We generated the artificial data for the case $p=9$ and $q=2$ to illustrate our estimation method. The true values of parameters are given in Tables 1 and 2.
The MCMC simulation is iterated to obtain 10,000 posterior samples after discarding 2,000 samples as the burn-in period.
The prior distributions are assumed as follows:
\begin{align*}
	& \mu_i \sim \Normal(0, 10000), \quad \quad \frac{1+\phi_i}{2} \sim \text{Beta}(1,1), \quad \sigma_{\eta,i}^2 \sim \IG\left(\frac{0.1}{2}, \frac{0.1}{2}\right),\quad i=1,\ldots 11, \\\
	&\gamma_j \sim \Normal(0, 10000), \quad \frac{1+\psi_j}{2} \sim \text{Beta}(1,1),\quad \frac{1+\rho_{9+j}}{2} \sim \text{Beta}(1,1), \quad \sigma_{\nu,j}^2 \sim \IG\left(\frac{0.1}{2}, \frac{0.1}{2}\right),\quad j=1,2,\\
	&\alpha_{21} \sim \Normal(0, 10000), \quad \beta_{ij} \sim \Normal(0, 10000), \quad i=1,\ldots 9, \quad j=1,2, \quad \pi(\delta) \propto I(\delta>0).
\end{align*}
As shown in Tables 1 and 2, the parameter estimates are close to true values, which are included in almost all of the 95\% credible intervals. The inefficiency factors for $\sigma_{\eta,10}$ and $\delta$ seem to be relatively large, but overall vary from  1 to 200.

\scriptsize
\begin{table}[H]
	\newcolumntype{.}{D{.}{.}{-1}}
\begin{center}
	\begin{tabular}{llllr}
		\hline\hline
		\multicolumn{1}{l}{Par.}&\multicolumn{1}{c}{Mean}&\multicolumn{1}{c}{True}&\multicolumn{1}{c}{95\% interval}&\multicolumn{1}{c}{IF}\tabularnewline
		\hline
		$\sigma_{\eta,1}$&$0.100$&$0.1$&$[0.0821,0.120]$&$63$\tabularnewline
		$\sigma_{\eta,2}$&$0.113$&$0.1$&$[0.0945,0.136]$&$68$\tabularnewline
		$\sigma_{\eta,3}$&$0.102$&$0.1$&$[0.0881,0.120]$&$49$\tabularnewline
		$\sigma_{\eta,4}$&$0.113$&$0.1$&$[0.0942,0.134]$&$53$\tabularnewline
		$\sigma_{\eta,5}$&$0.105$&$0.1$&$[0.0880,0.124]$&$71$\tabularnewline
		$\sigma_{\eta,6}$&$0.108$&$0.1$&$[0.0916,0.128]$&$85$\tabularnewline
		$\sigma_{\eta,7}$&$0.106$&$0.1$&$[0.0891,0.126]$&$55$\tabularnewline
		$\sigma_{\eta,8}$&$0.109$&$0.1$&$[0.0927,0.128]$&$72$\tabularnewline
		$\sigma_{\eta,9}$&$0.112$&$0.1$&$[0.0957,0.132]$&$52$\tabularnewline
		$\sigma_{\eta,10}^{\dagger}$&$0.128$&$0.1$&$[0.0890,0.182]$&$361$\tabularnewline
		$\sigma_{\eta,11}^{\dagger}$&$0.135$&$0.1$&$[0.0997,0.176]$&$215$\tabularnewline
		$\sigma_{\nu,1}^{\dagger}$&$0.143$&$0.1$&$[0.0975,0.181]$&$191$\tabularnewline
		$\sigma_{\nu,2}^{\dagger}$&$0.0911$&$0.1$&$[0.0724,0.110]$&$153$\tabularnewline
		$\delta$&$8.028$&$8$&$[7.882,8.192]$&$307$\tabularnewline
		\hline
\end{tabular}\end{center}
	\caption{}
	\label{table:sim_q2_2}
	$\dagger$: parameters related to the market factor.
\end{table}

\scriptsize
\begin{table}[H]
\newcolumntype{.}{D{.}{.}{-1}}
\begin{center}
	\begin{tabular}{llllrllllr}
		\hline\hline
		\multicolumn{1}{l}{Par.}&\multicolumn{1}{c}{Mean}&\multicolumn{1}{c}{True}&\multicolumn{1}{c}{95\% interval}&\multicolumn{1}{c}{IF}&\multicolumn{1}{c}{Par.}&\multicolumn{1}{c}{Mean}&\multicolumn{1}{c}{True}&\multicolumn{1}{c}{95\% interval}&\multicolumn{1}{c}{IF}\tabularnewline
		\hline
		$\alpha_{21}$&$0.538$&$0.5$&$[0.480,0.581]$&$183$&$\mu_{1}$&$-0.989$&$-1$&$[-1.035,-0.944]$&$9$\tabularnewline
		$\beta_{11}$&$1.075$&$1$&$[1.006,1.136]$&$156$&$\mu_{2}$&$-0.980$&$-1$&$[-1.021,-0.939]$&$13$\tabularnewline
		$\beta_{12}$&$0.969$&$1$&$[0.943,0.995]$&$35$&$\mu_{3}$&$-0.942$&$-1$&$[-0.991,-0.893]$&$8$\tabularnewline
		$\beta_{21}$&$1.040$&$1$&$[0.969,1.099]$&$148$&$\mu_{4}$&$-1.011$&$-1$&$[-1.054,-0.968]$&$11$\tabularnewline
		$\beta_{22}$&$0.994$&$1$&$[0.968,1.019]$&$30$&$\mu_{5}$&$-0.982$&$-1$&$[-1.029,-0.934]$&$8$\tabularnewline
		$\beta_{31}$&$1.048$&$1$&$[0.980,1.107]$&$134$&$\mu_{6}$&$-0.993$&$-1$&$[-1.040,-0.948]$&$10$\tabularnewline
		$\beta_{32}$&$1.002$&$1$&$[0.975,1.029]$&$16$&$\mu_{7}$&$-0.992$&$-1$&$[-1.035,-0.950]$&$10$\tabularnewline
		$\beta_{41}$&$1.058$&$1$&$[0.990,1.116]$&$158$&$\mu_{8}$&$-0.982$&$-1$&$[-1.028,-0.936]$&$8$\tabularnewline
		$\beta_{42}$&$0.986$&$1$&$[0.960,1.010]$&$28$&$\mu_{9}$&$-0.980$&$-1$&$[-1.025,-0.935]$&$9$\tabularnewline
		$\beta_{51}$&$1.035$&$1$&$[0.966,1.093]$&$154$&$\mu_{10}^{\dagger}$&$-0.990$&$-1$&$[-1.071,-0.913]$&$73$\tabularnewline
		$\beta_{52}$&$0.992$&$1$&$[0.967,1.018]$&$26$&$\mu_{11}^{\dagger}$&$-0.546$&$-0.5$&$[-0.617,-0.475]$&$73$\tabularnewline
		$\beta_{61}$&$1.034$&$1$&$[0.965,1.095]$&$149$&$\phi_{1}$&$0.892$&$0.9$&$[0.846,0.929]$&$50$\tabularnewline
		$\beta_{62}$&$0.989$&$1$&$[0.963,1.014]$&$25$&$\phi_{2}$&$0.865$&$0.9$&$[0.812,0.906]$&$45$\tabularnewline
		$\beta_{71}$&$1.060$&$1$&$[0.993,1.120]$&$155$&$\phi_{3}$&$0.899$&$0.9$&$[0.864,0.929]$&$24$\tabularnewline
		$\beta_{72}$&$0.983$&$1$&$[0.957,1.008]$&$33$&$\phi_{4}$&$0.872$&$0.9$&$[0.827,0.912]$&$34$\tabularnewline
		$\beta_{81}$&$1.055$&$1$&$[0.988,1.114]$&$145$&$\phi_{5}$&$0.893$&$0.9$&$[0.852,0.927]$&$45$\tabularnewline
		$\beta_{82}$&$0.986$&$1$&$[0.960,1.011]$&$26$&$\phi_{6}$&$0.888$&$0.9$&$[0.848,0.922]$&$49$\tabularnewline
		$\beta_{91}$&$1.071$&$1$&$[1.001,1.132]$&$152$&$\phi_{7}$&$0.878$&$0.9$&$[0.834,0.915]$&$35$\tabularnewline
		$\beta_{92}$&$0.979$&$1$&$[0.952,1.005]$&$34$&$\phi_{8}$&$0.885$&$0.9$&$[0.844,0.919]$&$47$\tabularnewline
		$\psi_{1}^{\dagger}$&$0.299$&$0.3$&$[0.254,0.343]$&$6$&$\phi_{9}$&$0.882$&$0.9$&$[0.841,0.917]$&$31$\tabularnewline
		$\psi_{2}^{\dagger}$&$0.282$&$0.3$&$[0.238,0.326]$&$2$&$\phi_{10}^{\dagger}$&$0.841$&$0.9$&$[0.609,0.934]$&$240$\tabularnewline
		$\gamma_{1}^{\dagger}$&$0.0631$&$0.05$&$[0.0249,0.101]$&$1$&$\phi_{11}^{\dagger}$&$0.851$&$0.9$&$[0.753,0.917]$&$178$\tabularnewline
		$\gamma_{2}^{\dagger}$&$0.0619$&$0.05$&$[0.0143,0.109]$&$2$&$\rho_{10}^{\dagger}$&$-0.222$&$-0.2$&$[-0.405,-0.042]$&$54$\tabularnewline
		&&&&&$\rho_{11}^{\dagger}$&$0.121$&$0$&$[-0.0213,0.260]$&$49$\tabularnewline
		\hline
\end{tabular}\end{center}
	\caption{}
	\label{table:sim_q2_1}
	$\dagger$: parameters related to the market factor.
\end{table}